\documentclass[useAMS,usenatbib]{mn2e}

\usepackage{graphicx}
\usepackage{graphics}
\usepackage{color}
\usepackage{url}
\usepackage[caption=false]{subfig}
\usepackage{verbatim}
\usepackage{array}
\usepackage{amssymb}
\usepackage{amsmath}
\usepackage{longtable}
\usepackage{pbox}

\newcommand{\kepler}{\it Kepler}

\newcommand{\lamost}{\sc lamost}
\newcommand{\gaia}{\it Gaia}
\newcommand{\Teff}{$T_{\text{eff}}$}
\newcommand{\logg}{$\log{g}$}
\newcommand{\FeH}{[Fe/H]}
\newcommand{\mudeltamu}{$\mu/\Delta \mu$}
\newcommand{\No}{N$^{\text{o}}$}

\newcommand{\ion}[2]{{#1~\uppercase \expandafter{\romannumeral #2}}}
\newcolumntype{C}[1]{>{\centering\arraybackslash}m{#1}}
\newcolumntype{L}[1]{>{\arraybackslash}m{#1}}

\begin{document}

\title[]{On the Identification of Wide Binaries in the {\kepler} Field}

\author[D.~Godoy-Rivera et al.]{Diego~Godoy-Rivera$^{1}$\thanks{e--mail:godoyrivera.1@osu.edu} and Julio~Chanam\'e$^{2,3}$\\
  $^{1}$  Department of Astronomy, The Ohio State University, 140 West 18th Avenue, Columbus, OH 43210, USA \\
  $^{2}$  Instituto de Astrof\'isica, Pontificia Universidad Cat\'olica de Chile, Av. Vicu\~na Mackenna 4860, 782-0436 Macul, Santiago, Chile \\
  $^{3}$ Millennium Institute of Astrophysics, Santiago, Chile\\
}

\maketitle

\begin{abstract}
We perform a search for wide binaries in the {\kepler} field with the prospect of providing new constraints for gyrochronology. First, we construct our base catalog by compiling astrometry for the stars observed by {\kepler}, and supplement it with parallaxes, radial velocities (RVs), and metallicities. We then mine our base catalog for wide binary candidates by matching the stars' proper motions, as well as parallaxes, RVs, and metallicities, if available. We mitigate the presence of chance alignments among our candidates by performing a comprehensive data-based contamination analysis in the proper motion versus angular separation phase space. Our final sample contains 55 binary candidates. A crossmatch of our pairs with the Second Data Release (DR2) from {\gaia} validates our candidates and confirms the reliability of our search method, particularly for $\varpi \gtrsim 2$ mas. Due to the implicit {\kepler} selection function and image scale per pixel, our binary search is incomplete for angular separations of $\Delta \theta \lesssim 20\arcsec$. We crossmatch our candidates with rotation period and asteroseismic ages catalogs, and find that our binary candidates do not follow a simple period-color relation, in agreement with previous studies. Two pairs have an age estimate for one component star and rotation period for its companion, positioning them as potentially new gyrochronology constraints at old ages. This is the first study that uses RVs and metallicities as criteria, rather than as a confirmation, in a binary search.
\end{abstract}

\section{Introduction}
\label{sec:intro}

Binary stars are common, and about half of all main sequence (MS) stars exist in binary systems \citep{fischer92,raghavan10}. These systems show a wide distribution of orbital periods and semimajor axes \citep{duquennoy91b,raghavan10}. At the wide separation end, the orbital periods are extremely long ($P_{\text{orb}}\sim10^3-10^6$ yr or more), making the observation of the members of a wide binary orbiting each other a difficult task. Nevertheless, their gravitationally bound nature does provide us with clues on how to find them.

For true, gravitationally bound wide binaries, the component stars are expected to have the same 3D-positions and 3D-velocities, down to the level of their orbital sizes and velocities. In terms of typical observables, this translates into pairs of stars with matching positions, proper motions, parallaxes, and radial velocities (RVs). Thus, wide binary studies have been based on either statistical \citep{bahcall81,longhitano10}, or kinematic, common-proper-motion searches \citep{chaname04,lepine07}. While some studies have also included photometric \citep{quinn09b,sesar08} and trigonometric \citep{andrews17,oelkers17,oh17a,shaya11} parallaxes in their searches, including RV-consistency as a search criterion remains unexplored.

An extra dimension could be added to the 6D-position and velocity phase space, as the components of binaries are expected to come from the same parental material \citep{kouwenhoven10}. This translates to stars with similar chemical abundances, and observational evidence supporting this has been reported in the literature \citep{andrews18,desidera04,desidera06}. Thus, the matching of chemical abundances arises as a potential constraint in searches for binary stars.

Another application that can be derived from the expected similarities in abundances, is to use wide binaries as calibrators for chemical tagging studies \citep{andrews18}. Chemical tagging suggests that, groups of stars that were born kinematically consistent but have since slowly dispersed in phase space, could be traced back and recognized as such by means of detailed chemical abundance matching \citep{freeman02}. This idea has been tested on clusters \citep{blancocuaresma15,hogg16,ness18}, but using binaries could allow the expansion of this technique to unexplored metallicity regimes.

This only adds to the large number of applications that wide binaries have in astrophysics \citep{chaname07}. Some of these include to provide constraints on the properties of dark matter \citep{penarrubia16,quinn09a,yoo04}, on the initial-to-final mass relation of white dwarfs \citep{catalan08,andrews15}, and on Type Ia supernovae progenitors \citep{thompson11,katz12}.

One further application of wide binaries, so far largely unexploited, is their potential contribution to gyrochronology \citep{barnes07,chaname12}. As they age, cool MS stars lose angular momentum via magnetized winds \citep{kawaler88,parker58,schatzman62,weber67}, and this is expressed as a spin down in their rotation periods \citep{skumanich72,barnes10}. \citet{barnes03} proposed the idea of using the stars' rotation rates (and masses) to obtain age estimates. This technique would be of particular use for isolated, field stars \citep{barnes07,mamajek08}.

To calibrate the gyrochronology relations, however, independent estimates of the stars' ages and rotation periods are needed. The existing constraints come mainly from open clusters studies \citep{barnes03,barnes15,meibom09,meibom11}, and have been recently extended up to $\sim 4$ Gyr (\citealt{barnes16}; but see also \citealt{epstein14}, \citealt{vansaders16}, \citealt{vansaders18}). 

The {\kepler} mission \citep{borucki10} has played a fundamental role in the acquisition of these constraints \citep{angus15,garcia14}, granting observations that could have not been done from the ground. Given their exquisite photometric precision, the {\kepler} observations have allowed rotation periods to be determined for several tens of thousands of stars (e.g., \citealt{mcquillan14}), as well as asteroseismic studies to be carried out for several hundred targets (e.g., \citealt{chaplin14}).

As first proposed by \citet{chaname12}, wide binaries can contribute to this growing literature of gyrochronology constraints, particularly in regimes of age and metallicity unexplored by clusters. This idea makes use of the coeval nature of the components of a binary. If, by some technique (e.g., asteroseismology), the age for one component star is obtained, the age for the entire system is simultaneously derived. When complemented with a rotation period measurement of an FGK-type component, wide binaries can provide age-rotation constraints difficult to obtain otherwise. Accordingly, a sample of {\kepler} wide binaries holds immense interest.

In this paper we search for wide binaries composed of stars observed by {\kepler}. In \S \ref{sec:data} we describe the data we use and characterize our base catalog. We construct different data subsamples and describe the proper motion parameters used in \S \ref{sec:define_subsamples}. In \S \ref{sec:search_algorithm} we describe our search algorithm and explain the basis of our contamination analysis. Sections \S \ref{sec:search_subs1}, \ref{sec:search_subs2}, \ref{sec:search_subs3}, \ref{sec:search_subs4} and \ref{sec:search_all_subs} are dedicated to explain our candidate selection process. We compare our candidates with previous studies in \S \ref{sec:comparison_with_others}, and examine them in the context of age-rotation relations in \S \ref{sec:age_rotation}. We conclude in \S \ref{sec:conclusions}.
\section{Data} 
\label{sec:data}

\subsection{Catalog Compilation}

To perform the desired search, we compile a list of stars in the {\kepler} field that were actually observed by the spacecraft, for which enough information was available (including astrometry and photometry). To accomplish this, we have collected data from a number of different catalogs, which we detail below.
\subsubsection{{\kepler} Input Catalog (KIC)}
\label{subsubsec:data_kic}

Before {\kepler} began its observations, the {\kepler} team observed the original {\kepler} field and its surroundings, and constructed the {\kepler} Input Catalog (KIC; \citealt{brown11}). This catalog contains $\sim 1.3 \times 10^7$ stars, and its purpose was to pre-select optimal targets following {\kepler}'s original purpose \citep{borucki10}.  As detailed in \citet{brown11}, a photometric catalog in Sloan Digital Sky Survey (SDSS; \citealt{york00})-like $griz$ bands was constructed, and the most important stellar parameters were derived (including {\Teff} and {\logg}).

We used the KIC as the source of the stars' RA ($\alpha$) and Dec ($\delta$) coordinates. Additionally, part of our wide binary search used the KIC photometry in order to derive photometric distances (see \S \ref{subsubsec:distances_photometric}).
\subsubsection{Catalog of \citet{huber14} 
\label{subsubsec:data_h14}}

In order to only include the stars that were actually observed by {\kepler}, we chose the catalog of \citet{huber14} as the starting point. This catalog contains a list of 196,468 stars observed by {\kepler} during Quarters 1 to 16. After compiling input parameters for these stars from a variety of observing techniques (including asteroseismology, spectroscopy, photometry), \citet{huber14} homogeneously fitted them to a grid of isochrones and derived improved stellar parameters. 

While this catalog meant an improvement in many regards, the input {\logg} values for $\sim$ 70\% of the stars were still based on the KIC values.
\subsubsection{UCAC4}
\label{subsubsec:data_ucac4}

The importance of proper motions in our catalog compilation cannot be overstated. Given its completeness down to magnitude $R=16$, we decided to use the UCAC4 catalog \citep{zacharias13} as the source of our proper motions. This limit virtually includes all the {\kepler} targets, therefore by using UCAC4 we secured that most of the stars would have proper motion measurements.
\subsubsection{Catalog Cross-match}
\label{subsubsec:data_crossmatch}

\begin{figure}
\centering
\subfloat{{\includegraphics[width=1.0\linewidth]{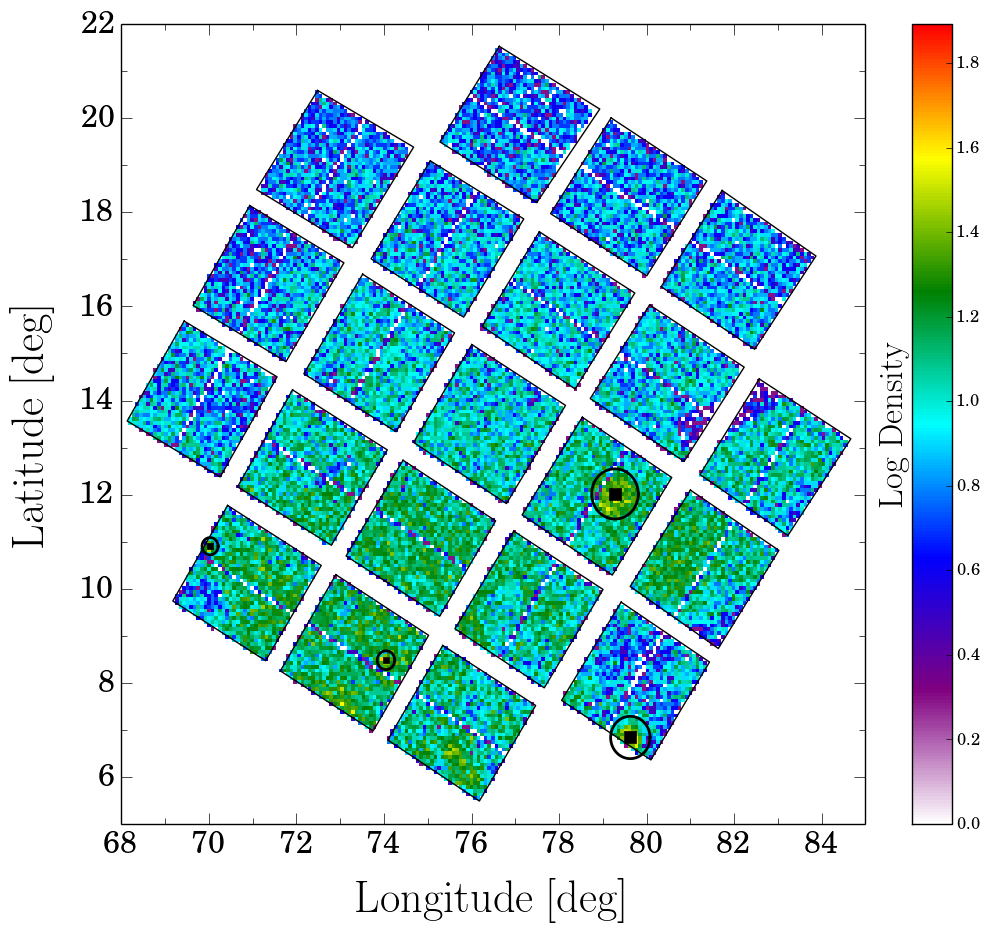}}}
\caption{Density map of the spatial distribution of our base catalog, containing the 182,821 sources identified in the cross-match of the catalogs of \citet{huber14}, KIC, and UCAC4. The positions of four known open clusters in the {\kepler} field (NGC 6791, NGC 6811, NGC 6819 and NGC 6866) are shown as the black squares, and overdensities of stars can be seen around some of them. The exclusion area defined for the clusters (see \S \ref{subsec:search_algorithm}) are shown as black circumferences around them. Overall, the density of stars increases heavily towards the Galactic plane.}
\label{fig:spatial_density}
\end{figure}

We compiled a list of the KIC IDs of the {\kepler} targets from \citet{huber14}, and using VizieR\footnote{http://vizier.u-strasbg.fr/viz-bin/VizieR} we looked for these stars in the KIC and UCAC4 catalogs. We discarded the stars that were found in UCAC4 but did not have proper motions measurements. From the initial 196,468 stars of \citet{huber14}, we cross-matched the three aforementioned catalogs and compiled our base catalog. This contains 182,821 sources for which we had, in principle, enough information to perform common-proper-motion searches.

Figure \ref{fig:spatial_density} shows the density map in Galactic coordinates of our base catalog. The position of the star clusters located in the {\kepler} field (NGC 6791, NGC 6811, NGC 6819 and NGC 6866) are shown as the black squares, with their corresponding exclusion area (see \S \ref{subsec:search_algorithm}) shown as black circumferences around them. It can be seen that the number density of stars is a strong function of latitude, increasing towards the Galactic plane.
\subsection{Catalog Supplementation}
\label{subsec:catalog_supplementation}

Ideally, a search for gravitationally bound wide binaries would use a data set containing the six parameters of position ($\alpha$, $\delta$, parallax) and velocity space (proper motion in $\alpha$ and $\delta$, RV). At this point, our base catalog is, however, missing two of these parameters: parallaxes and RVs. 

Obtaining these parameters usually requires greater observing efforts, but as a catalog supplementation we looked for parallax and RV information in the {\gaia} and Large Sky Area Multi-Object Fibre Spectroscopic Telescope ({\lamost}) data, respectively.

\subsubsection{TGAS Data}
\label{subsubsec:data_tgas}

Although no systematic efforts for obtaining parallaxes for the complete sample of {\kepler} stars have been performed, some of our stars are flagged in the KIC data as also belonging to the {\it Hipparcos} or {\it Tycho-2} catalog. These stars are therefore likely to have parallaxes in the {\it Tycho-Gaia} Astrometric Solution (TGAS; \citealt{gaia16a}) catalog, a subset of the {\gaia} Data Release 1. Using their IDs and VizieR, we looked for parallax information of these stars in the TGAS catalog. 

Considering the very different magnitude distributions of the TGAS catalog \citep{gaia16a} and the stars observed by {\kepler}, we did not expect a considerable overlap between both data sets (though this will change as further {\gaia} releases are published). We find 12,201 ($\approx$ 7\%) of our stars  in TGAS. As explained in \S \ref{subsubsec:distances_photometric}, the TGAS stars tend to be nearby stars, in comparison with most of the {\kepler} stars. 

Given the non-trivial analysis involved in calculating distances from parallaxes, we simply compiled the distance values reported by \citet{astraatmadja16}. Using different priors in the distance distribution, besides assumptions on the Galactic distribution of stars, \citet{astraatmadja16} derived distances from parallaxes for the whole TGAS sample. Following \citet{andrews17}, we chose the distance values calculated using the exponential-disk prior. 
\subsubsection{{\lamost} Data}
\label{subsubsec:data_lamost}

We cross-matched our base catalog with the {\lamost} DR3 catalog and the \citet{frasca16} catalog (which is based on the {\lamost}-{\kepler} survey; \citealt{decat15b}). These catalogs report atmospheric parameters ({\Teff}, {\logg}, and {\FeH}) and RVs. We find 34,368 ($\approx$ 19\%) of our base catalog in {\lamost}. 25,975 of these stars were found in both catalogs, but given the more precise DR3 measurements, we prioritized those values over the \citet{frasca16} ones for these cases.

A fraction of the {\lamost} stars have multiple spectroscopic observations (hence multiple RV and atmospheric parameters measurements). In these cases, for a given parameter, we took the actual value as the weighted average of the individual measurements, using the reported errors as weights. 
\subsection{Base Catalog Characterization}
\label{subsec:catalog_characterization}
\subsubsection{Spatial Footprint and Chip Gaps}
\label{subsubsec:catalog_gaps}

The spatial footprint of the {\kepler} field is shown in Figure \ref{fig:spatial_density}. This pattern is produced by the 21 different CCDs, one per chip, in addition the 90{\degr} rotation on the spacecraft orientation every 3 months \citep{borucki08}. Separating the different chips there are gaps with typical separations of $\sim 2000\arcsec$. 

These gaps introduce a source of incompleteness in our catalog, as we are not capable of finding wide binaries for which one component is within a chip (near the edge) and was observed by {\kepler}, and its companion is in an adjacent gap (and therefore not observed by {\kepler} and missing in our catalog). Since our goal is not to perform an statistically complete binary search, but rather to identify pairs where both stars were observed by {\kepler}, we do not attempt to correct for the edge effects introduced by these gaps.

Moreover,  given the typical distances of the stars in the field (see \S \ref{subsubsec:distances_photometric}), a real wide binary cannot have its two components on different chips, as the gaps width translates to unphysically wide systems (2000$\arcsec$ at $\sim$1 kpc yields a projected physical separation of $\sim 10$ pc). Based on this, we decided to perform an independent binary search in each one of the 21 chips.

In addition to this, in the middle each chip there is a smaller gap with a typical width of $\sim 400\arcsec$. All the chips but the central one show this smaller gap, where the 90{\degr} spacecraft rotation removed it. Similarly as with the gaps in between chips, the smaller gaps within chips also introduce a source of incompleteness, which we do not try to correct for. Since, for example, systems closer than $\sim$ 400 pc could have projected angular separations as wide as $\sim 500\arcsec$ (assuming a physical separation limit of $\sim$ 1 pc), we do not attempt to use the smaller gaps to separate chips in sub-chips.
\subsubsection{Proper Motions}
\label{subsubsec:catalog_pm}

\begin{figure}
\centering
\subfloat{{\includegraphics[width=1.0\linewidth]{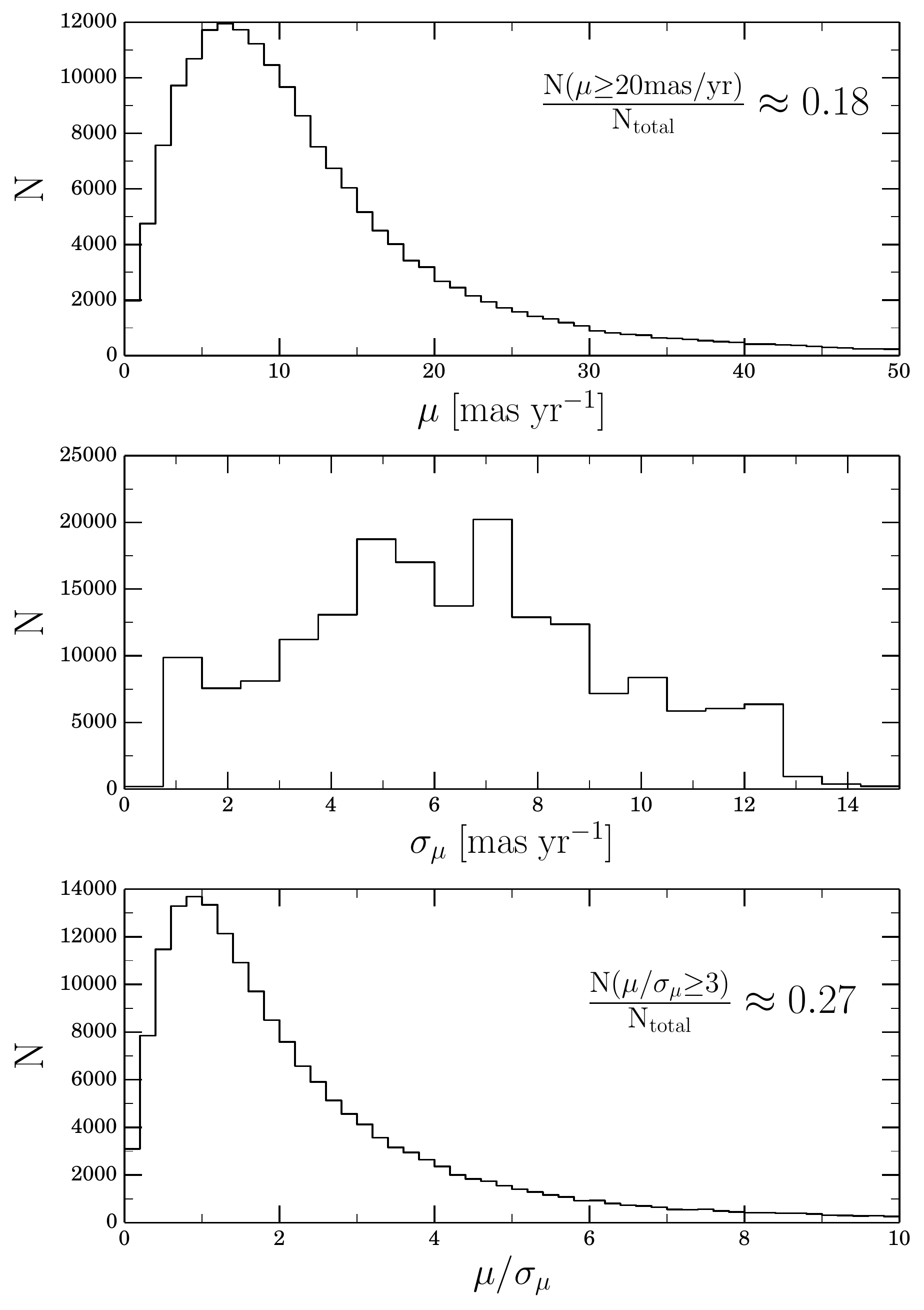}}}
\caption{Distributions of the UCAC4 total proper motion $\mu$ (top), the total proper motion error $\sigma_{\mu}$ (middle), and the ratio $\mu/\sigma_{\mu}$ (bottom) for the entire base catalog.}
\label{fig:histogram_PM}
\end{figure}

In order to understand our base catalog, it is important to characterize the proper motions and the associated errors. For a given star, in order to quantify the quality of its proper motion, we have defined the (dimensionless) parameter
\begin{equation}
\frac{\mu}{\sigma_{\mu}}= \frac{|\vec{\mu}|}{|\vec{\sigma_{\mu}}|} = \frac{\sqrt{\mu_{\alpha}^2+\mu_{\delta}^2}}{\sqrt{\sigma_{\mu_{\alpha}}^2+\sigma_{\mu_{\delta}}^2}},
\label{eqn:mu_sigma_mu_definition}
\end{equation}

where $\mu_{\alpha}$, the proper motion in $\alpha$ from UCAC4, already accounts for the multiplicative $\cos(\delta)$ factor \citep{zacharias13}, $\mu$ is the total proper motion, and $\sigma_{\mu}$ is the total proper motion error.

Figure \ref{fig:histogram_PM} shows, for the entire base catalog, the distribution of the total proper motion $\mu$, the total proper motion error $\sigma_{\mu}$, and the $\mu / \sigma_{\mu}$ parameter. The median values of $\mu$ and $\sigma_{\mu}$ are $\simeq$ 10 mas yr$^{-1}$ and $\simeq$ 6 mas yr$^{-1}$, respectively, with only $\approx$ 18\% of the stars having $\mu \geq 20$ mas yr$^{-1}$. More importantly, only a $\approx$ 27\% of the stars have $\mu/\sigma_{\mu} \geq 3$, which we define as the subset of stars with {\it well measured} proper motions. This limit is later adopted as a requirement in our binary search (see \S \ref{subsec:general_criteria}). We deem $\mu/\sigma_{\mu}=3$ as the limit, as lower values would not allow a reliable proper motion analysis, and higher values would greatly decrease the fraction of the base catalog that can be used in our search.

We note that the same $\approx$ 7\% of stars with TGAS parallaxes also have TGAS proper motions. When comparing these with the UCAC4 values we found good agreement between both data sets, and we adopted the UCAC4 proper motions as the nominal values for our searches. Later on, for completeness, we re-ran our binary searches using the TGAS proper motions instead, to check if we could gain new potential candidates missed otherwise (see \S \ref{subsec:subs1_search_3}).
\subsubsection{Angular Separation Distribution}
\label{subsubsec:catalog_angular_separation}

An important property of the source catalog of a wide binary search is its completeness as a function of angular separation. This depends on the selection function of the targets, and in our case, we inherit all the selection effects of the {\kepler} target selection.

In its observations, {\kepler} attempted to avoid blending and overlapping of targets in the same pixel. Additionally, {\kepler} has an image scale of 3.98$\arcsec$ per pixel \citep{borucki10}, and the mean FWHM of the KIC photometry is 2.4$\arcsec$ \citep{pinsonneault12}. All of these factors affect the completeness of our base catalog as a function of angular separation, making it hard for pairs of stars with small $\Delta \theta$ values to exist in our catalog.

\begin{figure}
\centering
\subfloat{{\includegraphics[width=1.0\linewidth]{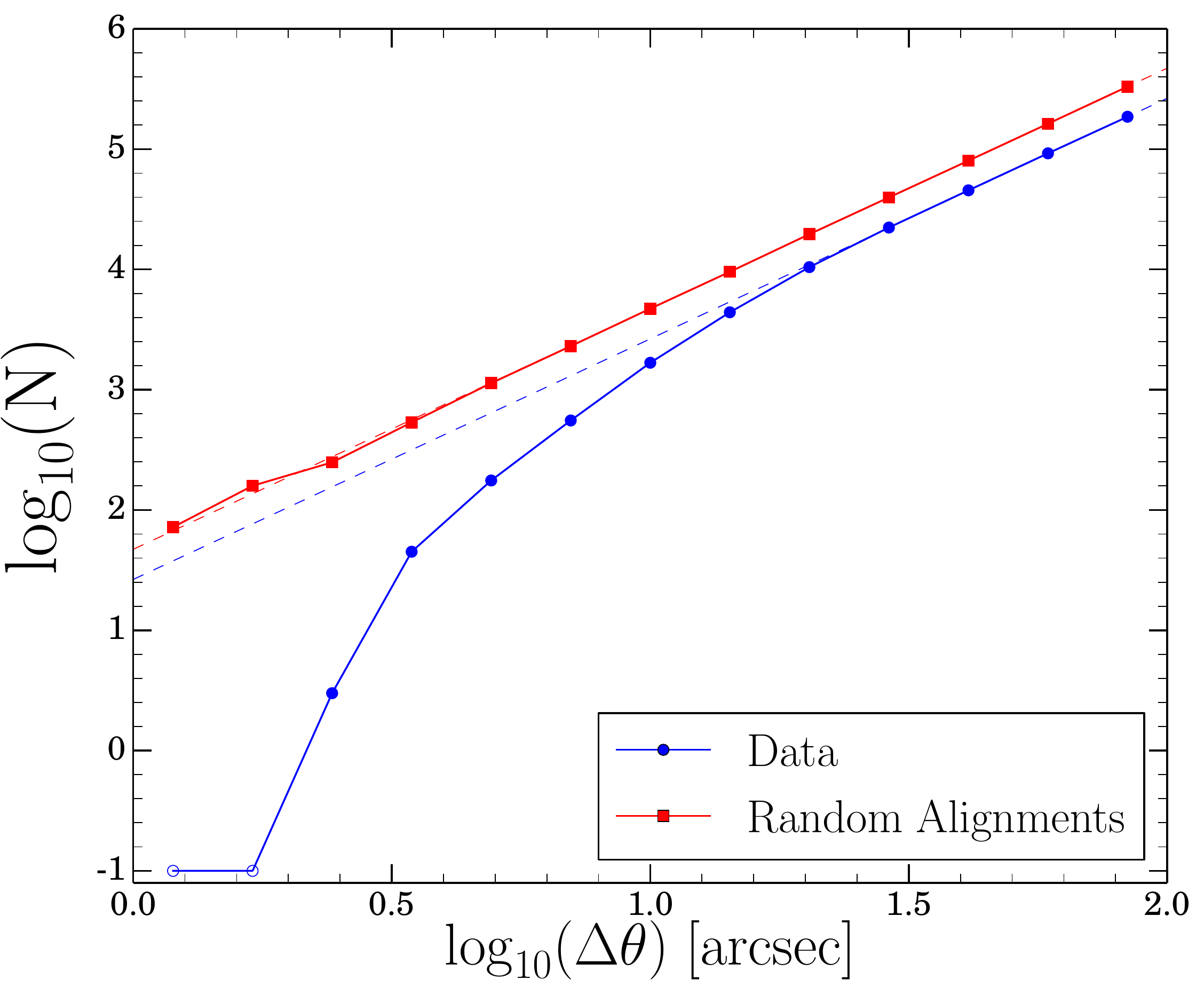}}}
\caption{Angular separation distribution for the base catalog. The distribution of the data (random alignments) sample is shown in the blue circles (red squares). The correspondingly colored dashed line shows the expected behavior for a random population of single stars ignoring angular resolution effects (see text). For plotting purposes only, empty bins (i.e., $\text{N}=0$) have been set to have y-coordinate$=-1$, and are shown with empty symbols. The fact that the blue circles drop below the blue dashed line for $\Delta \theta \lesssim 20\arcsec$ reveals that we are intrinsically biased against the detection of close separation pairs due to the inherited {\kepler} selection function.}
\label{fig:theta_distribution}
\end{figure}

Figure \ref{fig:theta_distribution} shows the angular separation distribution (number of pairs per unit separation, in log-log scale) for the entirety of our base catalog. We show this distribution for both our data sample (blue circles), and for its random alignments counterpart (red squares). For now, we focus on the distribution of the data sample. We explain the generation of the random alignments samples in \S \ref{subsec:random_alignments}, and further describe their angular separation distribution in \S \ref{subsubsec:random_alignments_angular_separation}.

As a comparison, we also show the expected behavior of the distribution for a random population of single stars (a population without binaries in it) ignoring angular separation resolution effects. For each set of points (blue and red) we show this as the correspondingly colored dashed line. Each of these are simply a line with a slope of 2, normalized to fit the widest separation bin (see also \citealt{sesar08,quinn09b}). For plotting purposes only, when a given bin of angular separation has no pairs in it (i.e., $\text{N}=0$), we have set the y-coordinate to be $=-1$, and we show it with an empty symbol.

As expected, the angular separation distribution of the data (blue) is incomplete for small separations, falling below the expected dashed line for $\Delta \theta \lesssim 20 \arcsec$. This result highlights that a search for wide binaries in our base catalog is challenging, as we are biased against the detection of close separation pairs, precisely where most of the binaries would be expected \citep{chaname04,sesar08,quinn09b,andrews17}. 

Given the selection effects present in our base catalog, we do not attempt any degree of completeness in our search. Instead, we focus on finding promising wide binary candidates where both stars were observed by {\kepler}.

\subsubsection{KIC Photometry}
\label{subsubsec:kic_photometry}

The multi-band photometry available from the KIC can be used in our search to derive photometric distances. Using this photometry, however, is not straightforward, as the photometric bands used for the KIC observations are not identical to the SDSS ones (though similar). In order to convert from the KIC system to the SDSS one, we use the coefficients derived by \citet{pinsonneault12}. 

This correction in the KIC photometry is not the only one needed, as these magnitudes are also affected by Galactic extinction. While the KIC does report extinction estimates, these values are based on a simple model of the dust distribution in the Galaxy. We instead use the $A_{V}$ estimates from \citet{schlafly11}, obtained from the IPAC {\it Galactic Dust Reddening and Extinction} web page\footnote{http://irsa.ipac.caltech.edu/applications/DUST/} using the stars' $\alpha$ and $\delta$ coordinates as input. We then used the coefficients reported by \citet{an09} to derive extinctions in the SDSS $griz$ bands from $A_{V}$, and corrected our KIC magnitudes (already in the SDSS system at this point) for this effect.
\subsubsection{Photometric Distances and Trigonometric Parallaxes}
\label{subsubsec:distances_photometric}

\begin{figure}
\centering
\subfloat{{\includegraphics[width=1.0\linewidth]{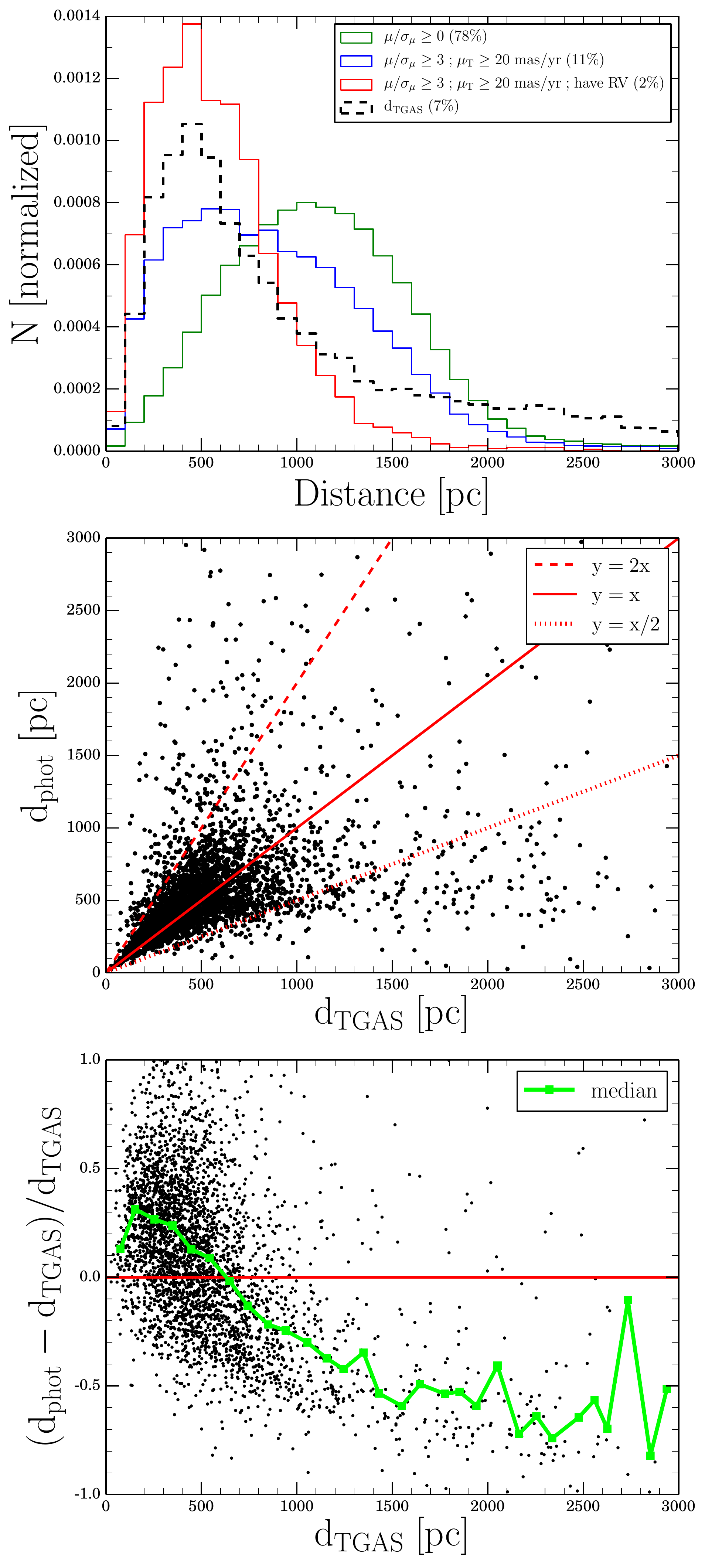}}}
\caption{Top Panel: Distribution of {\it photometric} distances (solid histograms) for the subset of dwarfs ({\logg} $\geq$ 4) that satisfy the criteria: $\mu/\sigma_{\mu} \geq 0$ (green), $\mu/\sigma_{\mu} \geq 3$ and $\mu \geq 20$ mas/yr (blue), and the latter criteria in addition to having RV measurements (red). The distribution of {\it trigonometric} distances for the stars with TGAS parallaxes is also shown (black dashed histogram). The parentheses in the labels show what percentage of the base catalog each subset represents. Middle Panel: Comparison between trigonometric and photometric distances, for the subset of stars with both estimates available. The solid red line shows the 1:1 relation, while the red dashed (dotted) line shows the $y=2x$ ($y=x/2$) relation. Bottom Panel: Fractional error in the photometric distance as a function of trigonometric distance. The red line represents the $\text{d}_{\text{TGAS}}=\text{d}_{\text{phot}}$ relation, and the green line shows the median of the data, binned in 100 pc steps.}
\label{fig:histogram_distance_phot_trig}
\end{figure}

In a search for wide binaries like ours, distance or parallax estimates, if available, can play an important role when distinguishing promising candidates from spurious pairs. Since trigonometric parallaxes are only available for $\approx$ 7\% of the base catalog, we have calculated photometric distances as a supplement.

We use the photometric parallax relations of \citet{dhital10} and \citet{dhital15} (based on SDSS-bands colors), which were obtained by comparing with samples of stars with measured trigonometric parallaxes. The quoted error on these relations is 0.3 mag, which corresponds to a $1\sigma$ error of $\sim$ 14\% of the distance estimate. This error accounts for the intrinsic width of the MS, unresolved binarity, and metallicity effects \citep{dhital10}.

An important consideration when using these photometric parallax relations, is that they only apply to MS dwarfs. According to the gravities of \citet{huber14} and {\lamost} (prioritized over the \citealt{huber14} values when available), $\approx$ 81\% of our base catalog stars have {\logg} $\geq$ 4, and we classify them as dwarfs. Moreover, the relations can only be applied to stars in the appropriate color range ($0.72<r-z<4.53$). We further add the constraint of only calculating photometric distances for the stars with measured magnitudes in all four $griz$-bands. 

After accounting for all of these restrictions, we are able to calculate photometric distances for $\approx$ 78\% of the base catalog. The top panel of Figure \ref{fig:histogram_distance_phot_trig} shows the photometric distance distribution for that subset (green), as well as for subsets obtained by adding the constraints of having $\mu/\sigma_{\mu} \geq 3$ and $\mu \geq 20$ mas/yr (blue), and further adding the constraint of having a measured RV (red). The top panel also shows the trigonometric distance distribution for the $\approx$ 7\% of our stars found in TGAS (black). The middle panel of Figure \ref{fig:histogram_distance_phot_trig}  shows, for the subset of stars with both trigonometric and photometric distances, a comparison of these two estimates. The bottom panel shows the fractional error in the photometric distance as a function of trigonometric distance.

As it can be expected, the top panel of Figure \ref{fig:histogram_distance_phot_trig} shows that by adding the criteria of fast and well measured proper motions, closer by stars are preferentially selected (blue versus green histogram). A similar effect is seen when RVs are required (red versus blue histogram), as higher signal-to-noise spectroscopy is obtained for brighter targets relative to fainter, more distant targets. This subset (red histogram) and the TGAS one (black dashed histogram) peak at a similar distance, although the latter extends to larger values.

The middle panel of Figure \ref{fig:histogram_distance_phot_trig} shows that, although as a population the stars tend to follow the 1:1 relation, there is large scatter at practically all trigonometric distances. Moreover, as shown in the bottom panel of Figure \ref{fig:histogram_distance_phot_trig}, the median fractional error on the photometric distances varies strongly with trigonometric distance, changing for up to $+30\%$ for $\text{d}_{\text{TGAS}} < 500$ pc down to $-30 \%$ for $500$ pc $< \text{d}_{\text{TGAS}} < 1000$ pc. For larger $\text{d}_{\text{TGAS}}$ the median fractional error keeps increasing (toward negative values), although fewer stars are found in this regime. Given all of these, we decided to only use the photometric distances to identify candidate pairs as a last resource (see \S \ref{subsec:subs4_search_2}).
\subsubsection{Parallax, Radial Velocity, and Metallicity Errors}
\label{subsubsec:parallax_rv_feh_errors}

\begin{figure}
\centering
\subfloat{{\includegraphics[width=1.0\linewidth]{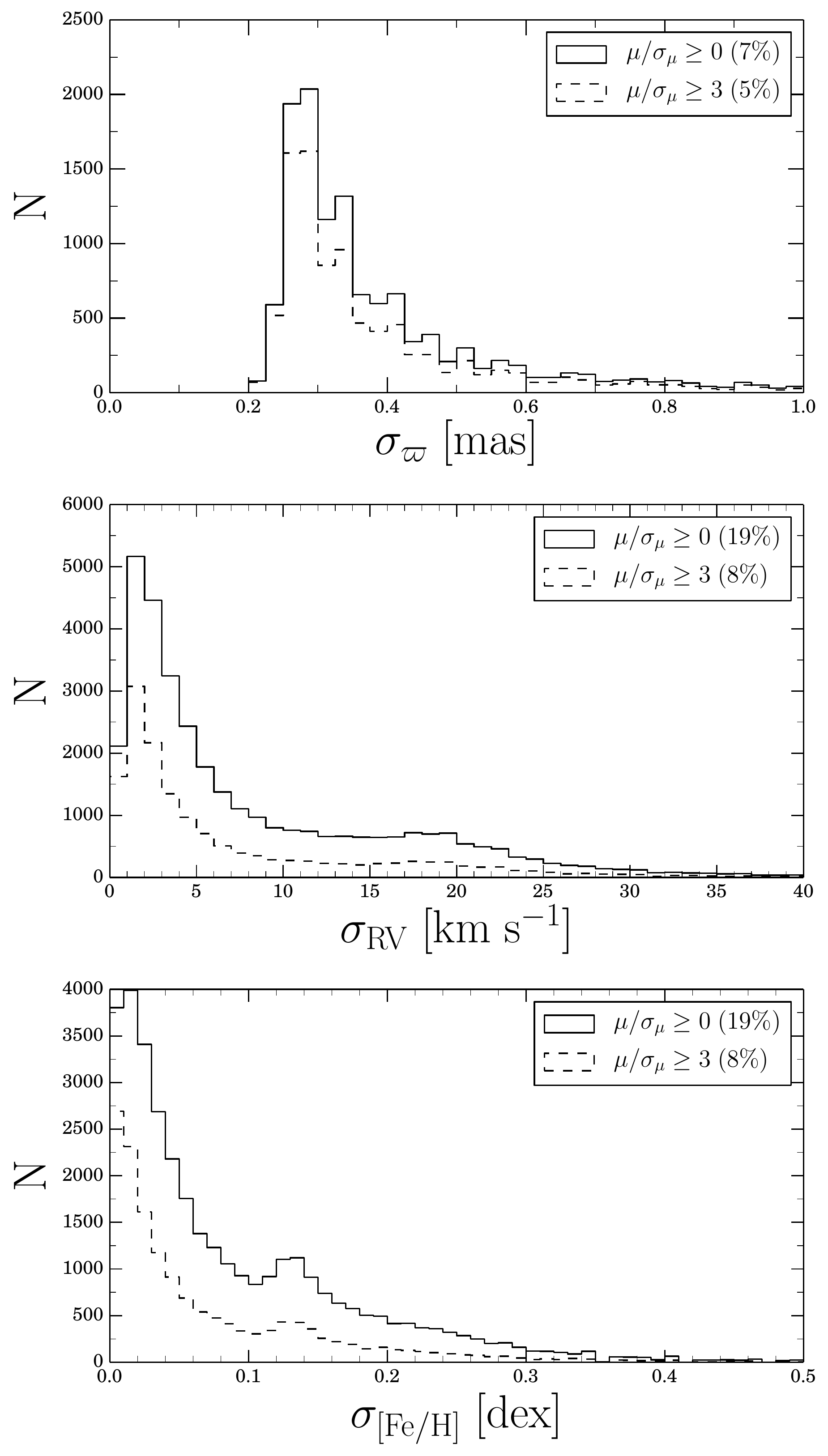}}}
\caption{Top Panel: Distribution of the standard errors of the trigonometric parallaxes for all the stars found in TGAS (no proper motion requirements, i.e., $\mu/\sigma_{\mu} \geq 0$; solid histogram), and the subset with $\mu/\sigma_{\mu} \geq 3$ (dashed histogram). Middle Panel: Distribution of the 1$\sigma$ statistical errors of the RVs for all the stars found in {\lamost} ($\mu/\sigma_{\mu} \geq 0$; solid), and the subset with $\mu/\sigma_{\mu} \geq 3$ (dashed). Bottom Panel: Same as the middle panel, for the {\lamost} metallicities. The parentheses in the labels show what percentage of the base catalog each subset represents.}
\label{fig:histogram_sigma_plx_RV_FeH}
\end{figure}

As a further characterization of our data, the top panel of Figure \ref{fig:histogram_sigma_plx_RV_FeH} shows the distribution of the standard errors of the TGAS parallaxes, and the middle and bottom panels show the distribution of the statistical errors of the {\lamost} RVs and metallicities, respectively.

The TGAS parallax errors have a floor of 0.2 mas, and a roof of 1.0 mas, with the distribution peaking at $\sigma_{\varpi} \simeq$ 0.25 $\textendash$ 0.30 mas. Since for our searches we {\it always} required $\mu/\sigma_{\mu} \geq 3$ (see \S \ref{sec:define_subsamples}), we also show the distribution for that subset of stars (dashed histogram).

For the {\lamost} RV errors we also show the distributions for the subsets of stars with $\mu/\sigma_{\mu} \geq 0$ and $\geq 3$, with both distribution being similar. The errors peak at values of $\sigma_{\text{RV}} < 5$ km s$^{-1}$, with a secondary peak around $\sim$ 17 $\textendash$ 20 km s$^{-1}$. These distributions are dominated by the {\lamost} DR3 values, and the secondary peak arises from the less precise {\lamost}-{\kepler} measurements \citep{frasca16}.

A similar behavior is shown by the {\lamost} metallicity errors, with its main peak at values of $\sigma_{\text{[Fe/H]}} <0.04$ dex (coming from the {\lamost} DR3 values), and a secondary peak around $\sim$ 0.13 dex (arising from the {\lamost}-{\kepler} data).


\section{Subsamples and General Parameters}
\label{sec:define_subsamples}
\subsection{Subsamples of the Base Catalog}
\label{subsec:subsamples_ssc}

Given the availability of the data supplemented for our base catalog, we have performed four independent searches for wide binaries, each associated with a different subsample. We have defined the following subsamples:

\begin{itemize}
\item {\it Subsample 1}: ``PM-$\varpi$-RV''. All six parameters of position and velocity space are required: positions, proper motions, TGAS parallaxes, and {\lamost} RVs.
\item {\it Subsample 2}: ``PM-$\varpi$''. Positions and proper motions are supplemented with TGAS parallaxes. (No RVs required)
\item {\it Subsample 3}: ``PM-RV-Metallicity''. Positions and proper motions are supplemented with {\lamost} RVs and metallicities. (We argue why metallicities can be used as a constraint in \S \ref{subsec:subs1_conclusions}. No trigonometric parallaxes required)
\item {\it Subsample 4}: ``PM-only'' ($+$ photometric distances). Only proper motions were to be used in principle, and a final exercise was carried out using photometric distances as a supplement ({\it Branch A} and {\it Branch B}, respectively; neither RVs nor trigonometric parallaxes required).
\end{itemize}

We note that, although the searches performed in a given subsample are independent from the others, there is some overlap among some of the subsamples. For instance, candidates found in {\it Subsample 1} could also be found in {\it Subsamples 2} or {\it 3} (if they pass the selection criteria). Similarly, there is overlap between {\it Subsamples 4 - Branch A} and {\it Branch B}. If a given candidate pair was already found in a previous subsample, we note this explicitly in both the text and corresponding table.
\subsection{General Parameters and Criteria}
\label{subsec:general_criteria}

Here we describe the parameters that we used as criteria in our wide binary searches. We separate these in parameters defined for individual stars, and those defined for pairs of stars.
\subsubsection{Parameters for Individual Stars}

The common element among the binary searches in the four different subsamples is that they all use proper motions as constraints. Therefore, we limit ourselves to only use the stars with {\it well measured} proper motions. Accordingly, we {\it always} require the criterion:
\begin{equation}
\mu/\sigma_{\mu}\geq 3.
\end{equation}

This requirement limits us to only use a $\approx$ 27\% of the base catalog (see Equation \ref{eqn:mu_sigma_mu_definition} and Figure \ref{fig:histogram_PM}).

As the contribution from chance alignments decreases for faster proper motions, we have also defined a total proper motion parameter:
\begin{equation}
\mu = \sqrt{ \mu_{\alpha}^2 + \mu_{\delta}^2}.
\end{equation}
For a given star, $\mu$ is later compared with the minimum total proper motion allowed for a given search, $\mu_{\text{min}}$, and we only look for companions around that star if:
\begin{equation}
\mu \geq \mu_{\text{min}}.
\end{equation}
\subsubsection{Parameters for Pairs of Stars}
\label{subsubsec:parameters_pairs}

Given two stars, A and B, located nearby in the sky, the angular separation between them is defined as:
\begin{equation}
\Delta \theta (\text{A},\text{B}) \simeq \sqrt{ (\alpha_{\text{A}} - \alpha_{\text{B}})^2 \cos{\delta_{\text{A}}} \cos{\delta_{\text{B}}} + (\delta_{\text{A}} - \delta_{\text{B}})^2}.
\end{equation}

This parameter is later compared with the maximum angular separation allowed for a given search, $\Delta \theta_{\text{max}}$, and the pair AB is only accepted if
\begin{equation}
\Delta \theta \text{(A,B)} \leq \Delta \theta_{\text{max}}.
\end{equation}

As a common proper motion indicator, for two stars A and B with proper motion vectors $\vec{\mu}_{\text{A}}$ and  $\vec{\mu}_{\text{B}}$, we calculate the following parameter:
\begin{equation}
\frac{\mu}{\Delta \mu} (\text{A},\text{B})= \frac{ \text{min}(|\vec{\mu}_{\text{A}}|,|\vec{\mu}_{\text{B}}|)}{|\vec{\mu}_{\text{A}} - \vec{\mu}_{\text{B}}|}.
\label{eqn:mu_delta_mu_definition}
\end{equation}

We have designed this parameter to have a higher value for more promising wide binary candidates. It increases with the proper motion of the pair (numerator of Equation \ref{eqn:mu_delta_mu_definition}, defined here as the minimum of the total proper motion of both stars), as for a faster proper motion the probability of the pair being a chance alignments is lower. Additionally, the more {\it comoving} the stars A and B are, the more alike their proper motions vectors are (hence smaller $\Delta \mu$), and the {\mudeltamu} parameter will also increase. Therefore, the higher the {\mudeltamu} value for a given pair, the more likely it is, at least a priori, to be a real wide binary.

This parameter is then compared with the minimum allowed value for the search, $\mu/\Delta \mu_{\text{min}}$, and the pair AB is only accepted if
\begin{equation}
\mu/\Delta \mu \text{(A,B)} \geq \mu/\Delta \mu_{\text{min}}.
 \end{equation}
 
We note that, for a true wide binary, the $\mu/\Delta \mu$ parameter has the advantage of being a distance-independent quantity.

Further criteria are defined later for the specific searches in each of the four subsamples, according to the data available for them (see \S \ref{subsec:description_final_criteria}).

\section{Binary Search Algorithm and Random Alignments Samples} 
\label{sec:search_algorithm}

Given the inherited selection function of our base catalog, our searches are not meant to be complete or free of selection effects. Instead, our goal is to find candidates that, to the limits of the data, are more likely to be real wide binaries.

\subsection{Search Algorithm}
\label{subsec:search_algorithm}

Our procedure to search for wide binary candidates works as follows. For a given search, we separate the input criteria in individual (e.g., $\mu \geq \mu_{\text{min}}$) and pair criteria (e.g., $\mu/\Delta \mu \geq \mu /\Delta \mu_{\text{min}}$). The individual criteria (which apply to individual stars), are applied first to expedite the calculations and avoid unnecessary pair matching. Then, pairs are constructed by cross-matching the list of stars that passed the individual criteria with itself. Finally, the pair criteria are applied to these pairs, and we only retain those that pass all the requirements and discard the rest.

We perform a search for wide binary candidates in each of the 21 chips independently (see \S \ref{subsubsec:catalog_gaps}), and then combine the output candidates in a single list. The result is, for a given set of input criteria, a list of pairs that fulfill all the requirements and can be later analyzed and compared to the results of other searches.

We reject the pairs that have positions consistent with the clusters, as they might be members of them. For each cluster we define an exclusion area, a circumference centered in the cluster with a radius $r_{\text{excl}}$, and discard the pairs located within it. We estimate the exclusion radii ``by eye'' using the overdensities seen in Figure \ref{fig:spatial_density} as guidance, and find $r_{\text{excl}} \simeq 11\arcmin$ for NGC 6791, $\simeq 32\arcmin$ for NGC 6811, $\simeq 12\arcmin$ for NGC 6819, and $\simeq 27\arcmin$ for NGC 6866.
\subsection{Random Alignments Samples}
\label{subsec:random_alignments}

A critical step when searching for wide binary candidates, is to characterize the contamination from random alignments and spurious pairs. We do this empirically by following the procedure of \citet{lepine07} (see also \citealt{andrews17}). For a given set of (individual and pair) criteria, this method generates a simulated reciprocal sample of pairs {\it completely} made of random alignments, where no genuine binaries are present.

The procedure to generate a random alignments sample is almost identical to the one described in \S \ref{subsec:search_algorithm}, with a crucial difference. After the individual stars that passed the individual criteria are selected, this list is cross-matched with a {\it spatially displaced} version of itself. For a pair formed by stars A and B, this displacement is done by adding a constant $\Theta$ to the $\delta$ coordinate of the star B: $\delta_{\text{B}'} = \delta_{\text{B}} + \Theta$ (see also Equation 3 of \citealt{lepine07}). 

By displacing all the stars in the list by a large enough angle $\Theta$, we ensure that all the original pairs are artificially destroyed (they now have $\Delta \theta \text{(A,B)} > \Delta \theta_{\text{max}}$), and we now have a sample made of pure random alignments. With this, for a given search with its set of criteria, we generate both a list of pairs selected from the data, and its corresponding random alignments counterpart. For the reminder of this paper, we name these list of pairs {\it data} and {\it random alignments} samples, respectively.

For a given search with its $\Delta \theta_{\text{max}}$ criterion, in order to calculate its corresponding random alignments sample counterpart, we set the displacement $\Theta$ to be:
\begin{equation}
\Theta \equiv 2.5 \times \Delta \theta_{\text{max}}.
\end{equation}
With this, we are guaranteed to break all the pairs of the data sample, and at the same time to change the spatial density of stars only by a small amount.

The random alignments samples were decisive to assess the presence and behavior of the contamination in our searches, as well as to identify the appropriate criteria in order to select a sample of promising candidates.
\subsubsection{Random Alignments Angular Separation Distribution}
\label{subsubsec:random_alignments_angular_separation}

Figure \ref{fig:theta_distribution} has shown that, given the selection function inherited from {\kepler} in our base catalog, the angular separation distribution of our data sample is incomplete for $\Delta \theta \lesssim 20 \arcsec$ (blue circles). We now compare the angular separation distribution of the random alignments sample (red squares) with that of the data sample. 

Two main differences arise in this comparison. First, a normalization difference is seen between both distributions. This effect is produced by the algorithm that generates the random alignments samples. When shifting the positions of the stars of the pair AB (say, first for star A keeping star B fixed, and second for star B keeping star A fixed), the resulting modified pairs (A'B and AB', respectively) are actually different. Consequently, the algorithm yields a greater number of random alignments pairs than of data pairs.

Second, and contrary to the behavior of the data, the random alignments distribution tightly follows the expected dashed line. This is not surprising, as by displacing the stars by an angle $\Theta$, we are artificially removing the angular separations selection effects of the underlying catalog. For instance, in the random alignments sample, two stars can easily be closer than the {\kepler} image scale per pixel. 

All of these translates into a dearth of data pairs relative to the random alignments pairs, when compared with their respective dashed lines, for small separations ($\Delta \theta \lesssim 20 \arcsec$), even after accounting for the different normalizations. In other words, only the data sample falls below the expected dashed line, while the random alignments sample follows the expectation for the entire angular separation regime.

\subsection{Expectations from the $\mu/\Delta \mu$ versus $\Delta \theta$ diagram}
\label{subsec:description_mu_delta_mu_versus_delta_theta}

\begin{table*}
\begin{minipage}{\textwidth}
\centering
\caption{Summary of the selection criteria used in all subsamples. The 1st and 2nd column list the subsample \No$\text{ }$and the parameters used in it, respectively. Given that for {\it Subsamples 1} and {\it 2} we used both UCAC4 and TGAS proper motions to select our candidates, we have separated them in different branches and the 3rd column specifies which proper motion source was used. The 4th column lists the final criteria used in selection of our candidates in each subsamples. These criteria are the relaxed version of our searches (e.g., see \S \ref{subsec:subs1_search_2}). For all subsample the $\mu/\Delta \mu_{\text{min}} (\Delta \theta)$ criterion is listed as an array of $(x,y)$ data points, corresponding to $(\Delta \theta,\mu/\Delta \mu_{\text{min}})$ values (where $\Delta \theta$ is in arcsec). The 5th column lists the number of candidates obtained in each subsample.}
\renewcommand{\arraystretch}{1.0}
\begin{tabular}{|c|c|c|c|c|}
\hline
{\it Subsample} \No & Parameters &  \begin{tabular}{@{}c@{}} Proper Motion \\ Source \end{tabular} & Selection Criteria & \begin{tabular}{@{}c@{}} \No $\text{ }$of \\ Candidates \end{tabular} \\
\hline
\hline
1 & PM, $\varpi$, RV & UCAC4 &  \begin{tabular}{@{}c@{}}$\mu/\sigma_{\mu} \geq 3$ ; $\Delta \theta \leq 500 \arcsec$ ; $\Delta \text{RV}\leq 3\sigma$ ;\\ $\Delta \varpi \leq 3\sigma$ ; $\text{s} \leq 1$ pc ;\\ $(\Delta \theta ,\mu/\Delta \mu_{\text{min}})= \{(0,3),(70,3),(500,6)\}$ \end{tabular} & 10  \\ \cline{3-5}
 &  & TGAS & same & 3  \\ 
\hline
\hline
2 & PM, $\varpi$ & UCAC4 & \begin{tabular}{@{}c@{}}$\mu/\sigma_{\mu} \geq 3$ ; $\Delta \theta \leq 500 \arcsec$ ; $\varpi /\sigma_{\varpi}\geq 3$ ;\\ $\Delta \varpi \leq 3\sigma$ ; $\text{s} \leq 1$ pc ;\\ $(\Delta \theta ,\mu/\Delta \mu_{\text{min}})= \{(0,3),(6,3),(100,12),(300,20),(>300,\infty)\}$ \end{tabular} & 14  \\  \cline{3-5}
  & & TGAS & \begin{tabular}{@{}c@{}} same except for: \\ $(\Delta \theta ,\mu/\Delta \mu_{\text{min}})= \{(0,3),(40,3),(500,24)\}$ \end{tabular} & 7  \\  \cline{3-5}
\hline
\hline
3 & \begin{tabular}{@{}c@{}} PM, RV, \\ Metallicity \end{tabular} & UCAC4 & \begin{tabular}{@{}c@{}}$\mu/\sigma_{\mu} \geq 3$ ; $\Delta \theta \leq 200 \arcsec$ ; $\sigma_{\text{RV}}=15$ km s$^{-1}$ ;\\ $\sigma_{\text{[Fe/H]}} = 0.15$ dex ; $\Delta \text{RV} \leq 2 \sigma$ ; $\Delta \text{[Fe/H]} \leq 2 \sigma$ ;\\ $(\Delta \theta ,\mu/\Delta \mu_{\text{min}})= \{(0,3),(30,3),(200,15)\}$ \end{tabular} & 8  \\ 
\hline
\hline
4 -{\it A} & PM & UCAC4 & \begin{tabular}{@{}c@{}}$\mu/\sigma_{\mu} \geq 3$ ; $\Delta \theta \leq 200 \arcsec$ ; $\mu_{\text{min}}= 20$ mas yr$^{-1}$ ;\\ $(\Delta \theta ,\mu/\Delta \mu_{\text{min}})= \{(0,5),(6,5),(55,35),(>55,\infty)\}$ \end{tabular} & 10  \\ 
\hline
\hline
4 -{\it B} & PM, d$_{\text{phot}}$ & UCAC4 & \begin{tabular}{@{}c@{}}$\mu/\sigma_{\mu} \geq 3$ ; $\Delta \theta \leq 200 \arcsec$ ; $\mu_{\text{min}}= 10$ mas yr$^{-1}$ ;\\ $\text{d}_{\text{phot}}$ ; $\Delta \text{d}_{\text{phot}} \leq 1 \sigma$ ; $\text{s}_{\text{phot}} \leq 0.1$ pc ;\\ $(\Delta \theta ,\mu/\Delta \mu_{\text{min}})= \{(0,10),(10,10),(100,25),(>100,\infty)\}$\end{tabular} & 3  \\ 
\hline
\hline
\end{tabular}
\label{tab:summary_criteria_subsamples}
\end{minipage}
\end{table*}

Since positions and proper motions are the only parameters that are available for all of our subsamples, the $\mu/\Delta \mu$ versus $\Delta \theta$ plot plays an important role in all of our analysis. Here we describe the qualitative expectations for such a diagram.

Even though we are biased against the detection of close separation pairs, we expect {\it most} of the wide binary candidates to have rather {\it small} angular separations, and at the same time to have {\it similar} proper motions (therefore high $\mu/\Delta \mu$ values; see \S \ref{subsubsec:parameters_pairs}). Here the terms {\it small} and {\it similar} are used just qualitatively, as their absolute values will depend on the subsample under study.

Conversely, we expect chance alignments to have rather {\it wide} $\Delta \theta$ and {\it low} $\mu/\Delta \mu$ values, as the nature of these pairs is to dominate at wide angular separations (the density of potential binary companions grows with $\Delta \theta$), and to not have similar proper motions. Again, here we use {\it wide} and {\it low} just qualitatively.

Although we always expect some overlapping, in the analysis that follows we try to identify regions in the  $\mu/\Delta \mu$ versus $\Delta \theta$ diagram where we expect to find promising wide binary candidates, and at the same time to minimize the presence of random alignments.
\subsection{The reasoning behind the selection criteria}
\label{subsec:description_final_criteria}

In each subsample, when possible, the analysis on the $\mu/\Delta \mu$ vs $\Delta \theta$ diagram is supplemented by requiring further criteria (e.g., parallax constraints, RV constraints), depending on the available data. This allows us to select promising candidates that have matching parameters in a multi-dimensional phase space.

In all subsamples we combine the available data in a set of multi-dimensional selection criteria to produce a {\it nominal} sample of candidates. Then, in particular in the {\it Subsamples 1} to {\it 3}, as some of the constraints used in our nominal search might be too stringent for some promising candidates, we systematically relax these selection criteria. We continue relaxing them until we enter the `contamination-dominated' region, where chance alignments begin to dominate, and where the quality of our data does not allow us to reliably distinguish candidates from chance alignments.

Given that in each subsample we use a number of constraints, sometimes changing from one subsample to another, for clarity we list the final selection criteria used in Table \ref{tab:summary_criteria_subsamples}. (The listed criteria correspond to the `relaxed' versions.)

\section{Binary Search in the {\it Subsample 1}: ``PM-$\varpi$-RV''}
\label{sec:search_subs1}
\subsection{Finding Promising Candidates}
\label{subsec:subs1_search_1}

\begin{figure*}
\begin{minipage}{\textwidth}
\centering
\subfloat{{\includegraphics[width=0.70\linewidth]{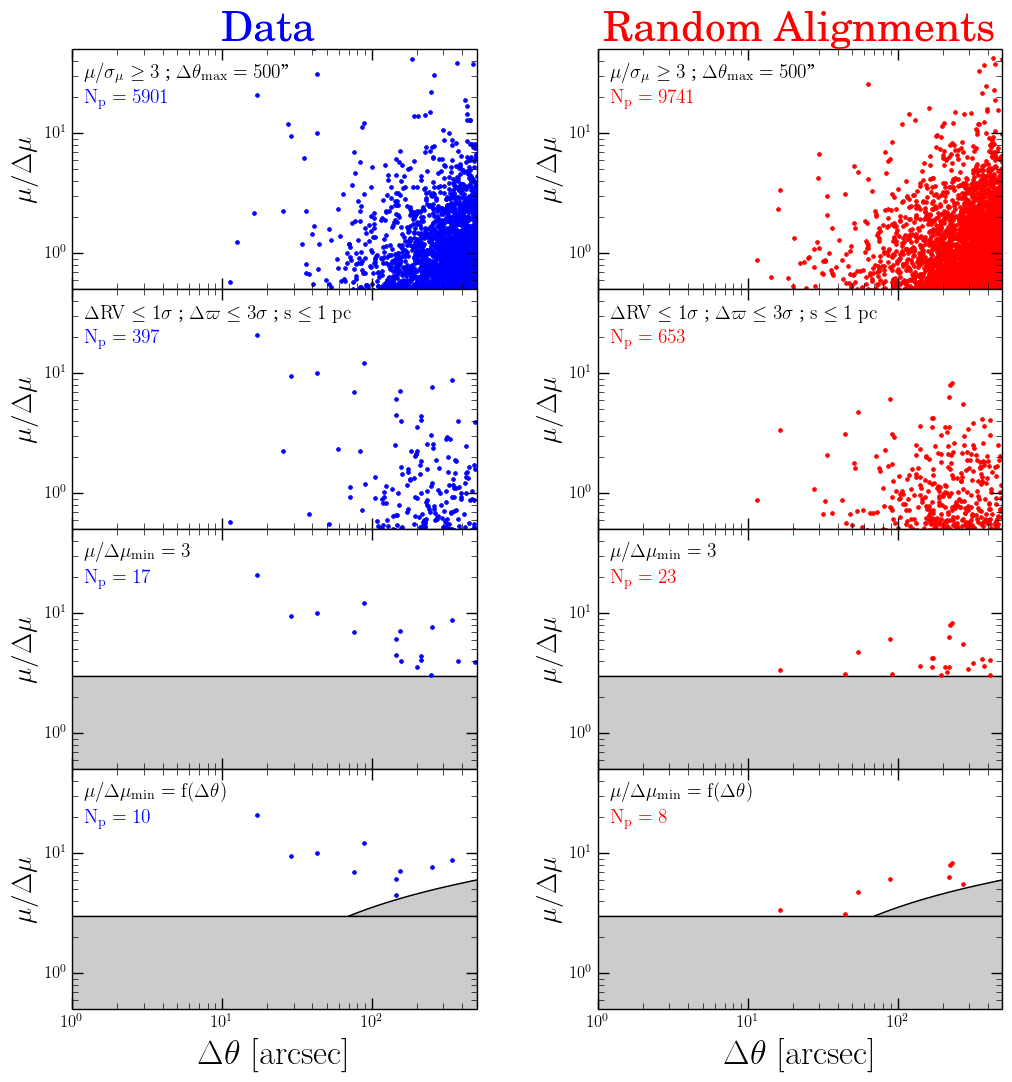}}}
\caption{{\it Subsample 1}: ``PM-$\varpi$-RV''. $\mu/\Delta \mu$ versus $\Delta \theta$ for the pairs that pass the criteria indicated in the top-left corner of each panel. The left (right) column shows in blue (red) the pairs of the data (random alignments) sample. Each row represents a different search, going downwards as more criteria are included. Each row requires the criteria indicated in that panel, in addition to all the criteria used in the panels above it. The number of pairs of each panel is shown at the top left in blue or red, respectively. In the 1st row we only include stars with well measured proper motions ($\mu/\sigma_{\mu} \geq 3$), and search for pairs up to angular separations of $\Delta \theta_{\text{max}}=500 \arcsec$. In the 2nd row, we only keep the pairs whose stars have RVs consistent within 1$\sigma$, parallaxes consistent within 3$\sigma$, and projected physical separations $\text{s} \leq 1$ pc. In the 3rd row we add the common proper motion criterion $\mu/\Delta \mu_{\text{min}} = 3$, discarding everything below that line, shown as the shaded area. Finally, in the 4th row we modify this criterion by adding an angular separation dependence to it, $\mu/\Delta \mu_{\text{min}}=\text{f}(\Delta \theta)$, shown as the extra shaded area in comparison with the 3rd row.}
\label{fig:subs1_mu_theta}
\end{minipage}
\end{figure*}

We begin our wide binary search with the subsample of stars for which all six parameters of position and velocity space are available. This subsample is chosen as the starting point, as more constraints (e.g., RV consistency, parallax consistency) can be applied. This provides the advantage of allowing us to constrain different parameters simultaneously, without being overly stringent in any one of them in particular.

Our initial pool consists of 4,642 stars with $\mu/\sigma_{\mu} \geq 3$ for which RVs and trigonometric parallaxes are available. Given the peak of the trigonometric distance distribution of $\sim$ 400\textendash500 pc (see Figure \ref{fig:histogram_distance_phot_trig}), we set the maximum angular separation of pairs to be $\Delta \theta_{\text{max}} = 500 \arcsec$, which would allow us to find wide binaries with projected physical separations up to $\sim$ 1 pc (if they are present). 

For now, we do not require any common proper motion constraint ($\mu/\Delta \mu_{\text{min}}=0$). Given these initial criteria, we run our search algorithm and find 5,901 pairs in the data sample, and 9,741 pairs in its random alignments counterpart. The top row of Figure \ref{fig:subs1_mu_theta} shows the distribution of these pairs in the $\mu/\Delta \mu$ versus $\Delta \theta$ space for both the data (left column, blue) and the random alignments samples (right column, red). The following rows of Figure \ref{fig:subs1_mu_theta} show the results of subsequent searches when further criteria are added.

In this initial search, the top row of Figure \ref{fig:subs1_mu_theta} shows that for both the data and random alignments samples, most of the pairs have $\mu/\Delta \mu <5$. Moreover, most of the pairs are piled up at low $\mu/\Delta \mu$ and large $\Delta \theta$ values, as expected from samples dominated by chance alignments (see \S \ref{subsec:description_mu_delta_mu_versus_delta_theta}). While both panels look similar at first glance, there are some important differences. First, as expected, the random alignments sample shows more pairs with $\Delta \theta < 30 \arcsec$ than the data sample (particularly at $\mu/\Delta \mu <2$). As discussed in \S \ref{subsubsec:random_alignments_angular_separation}, this is an artifact of the algorithm that generates the random alignments samples. Another difference is that for $\Delta \theta \lesssim 50\arcsec$, there are a some pairs in the data sample that stand out for their high $\mu/\Delta \mu$ values ($\gtrsim 8$), in comparison with the random alignments counterpart.

We now wish to add further constraints to our search. For real wide binaries, one would expect them to have consistent parallax and RV values. We favor a parallax constraint instead of a distance one, because this is the actual measured quantity by {\gaia}, while the distance estimates of \citet{astraatmadja16} take into account a pre-established model of the Galaxy. Nevertheless, we do use their distances in order to estimate projected physical separations.

For a given pair, we define our criteria as a requirement of: consistency of the RV values of the two stars within their errorbars (1$\sigma$), and consistency of their parallax values within 3 times their errorbars (3$\sigma$). While this may not seem as a very stringent parallax criterion given the size of the associated errorbars, later on we assign a qualitative flag to our candidate pairs depending on how consistent (or inconsistent) they look when taking all the parameters into account.

Using the distance estimates of \citet{astraatmadja16} we can estimate projected physical separations for the pairs. We calculate this as the average trigonometric distance of a given pair, times its angular separation:
\begin{equation}
\text{s} =  \left( \frac{\text{d}_1 + \text{d}_2}{2}\right) \times \Delta \theta,
\label{eqn:physical_separation_calculation}
\end{equation}
where $\text{d}_1$ and $\text{d}_2$ are the distance estimates of the stars in the pair. We require the projected physical separation to be of order of the tidal limit for wide binaries: s $\leq 1$ pc \citep{chaname04,jiang10,yoo04}. 

We note that recent works using TGAS claim to identify large numbers of comoving pairs with projected physical separations exceeding the s $\sim$ 1 pc limit, even with separations of up to $\sim 10$ pc \citep{oh17a,oelkers17}. At these scales, these systems are not expected to be gravitationally bound. Most importantly, \citet{andrews17,andrews18} demonstrate that those pairs at very large separations are mostly composed of chance alignments, and TGAS data are not enough to distinguish these from the population of ionized former wide binaries predicted by \citet{jiang10}. While the search for this population is interesting on its own, our goal is to select pairs that are likely to be gravitationally bound wide binaries. Therefore, we restrict our search to pairs with separations smaller than 1 pc.

The remaining pairs after adding these three criteria ($\Delta \text{RV}\leq1\sigma$; $\Delta \varpi\leq3\sigma$; s $\leq$1 pc) are shown in the 2nd row of Figure \ref{fig:subs1_mu_theta}. Now, the number of pairs in both samples has decreased dramatically, implying most of the pairs of the 1st row were chance alignments. Notably, most of the pairs with $\Delta \theta >100 \arcsec$ have been discarded. Although not shown explicitly, we note that this mainly comes mainly from the projected physical separation criterion, highlighting the importance of having distance estimates in our searches.

So far no common proper motion criterion has been required (i.e., $\mu /\Delta \mu_{\min}= 0$), so we now add the constraint of $\mu /\Delta \mu_{\min}= 3$ and discard all the pairs below that line. This is shown in the 3rd row of Figure \ref{fig:subs1_mu_theta}, with the shaded area showing the discarded portion of the diagram. This value was chosen as the random alignment sample at this point is dominated by pairs that do not pass this requirement. The number of pairs in both samples drops dramatically again, but now clear differences between them can be seen. 

In the 3rd row of Figure \ref{fig:subs1_mu_theta} most of the random alignments pairs are still concentrated at low $\mu/\Delta\mu$ ($<5$) and wide $\Delta \theta$ ($>100\arcsec$) values. While using a higher $\mu/\Delta \mu_{\text{min}}$ value would discard most of the random alignments pairs, it would also discard most of the data pairs. We prefer not to do this, to avoid getting rid of potential promising candidates, and we instead use a $\mu/\Delta \mu_{\text{min}}$ criterion with an angular separation dependence: $\mu /\Delta \mu_{\text{min}} = \text{f}(\Delta \theta)$. This is shown as the new shaded area in the 4th row of Figure \ref{fig:subs1_mu_theta}. 

The function $\text{f}(\Delta \theta)$ has been defined qualitatively, as at this point we are trying to delineate a region in the $\mu/\Delta \mu$ versus $\Delta \theta$ phase space where the random alignments sample pairs are greatly diminished, and where simultaneously the data sample pairs stand out as promising candidates.

After applying this criterion, we are left with 10 pairs in the data sample and 8 pairs in the random alignments one. This constitutes the nominal search of the {\it Subsample 1}, although we expand in the following subsections. 

We note, however, that the ratio of the number of random alignments pairs to the number of data sample pairs, 8/10, does not mean that the contamination rate in our candidate pairs is 80\%. We expect the actual contamination rate to be lower for two reasons. 

First, because the algorithm that generates the random alignments samples produces many more of such pairs than data pairs for a given set of criteria. To illustrate this, compare the normalizations for the blue and red dashed lines in Figure \ref{fig:theta_distribution}, and also note the number of pairs quoted in both panels of the 1st row of Figure \ref{fig:subs1_mu_theta}. Second, because even after accounting for the different normalizations, the angular separation distribution of the random alignments is different from that of the data sample pairs (see \S \ref{subsubsec:random_alignments_angular_separation}). This translates to the former distribution having an over-abundance of small $\Delta \theta$ pairs compared to the latter. 

\begin{figure*}
\begin{minipage}{\textwidth}
\centering
\subfloat{{\includegraphics[width=0.65\linewidth]{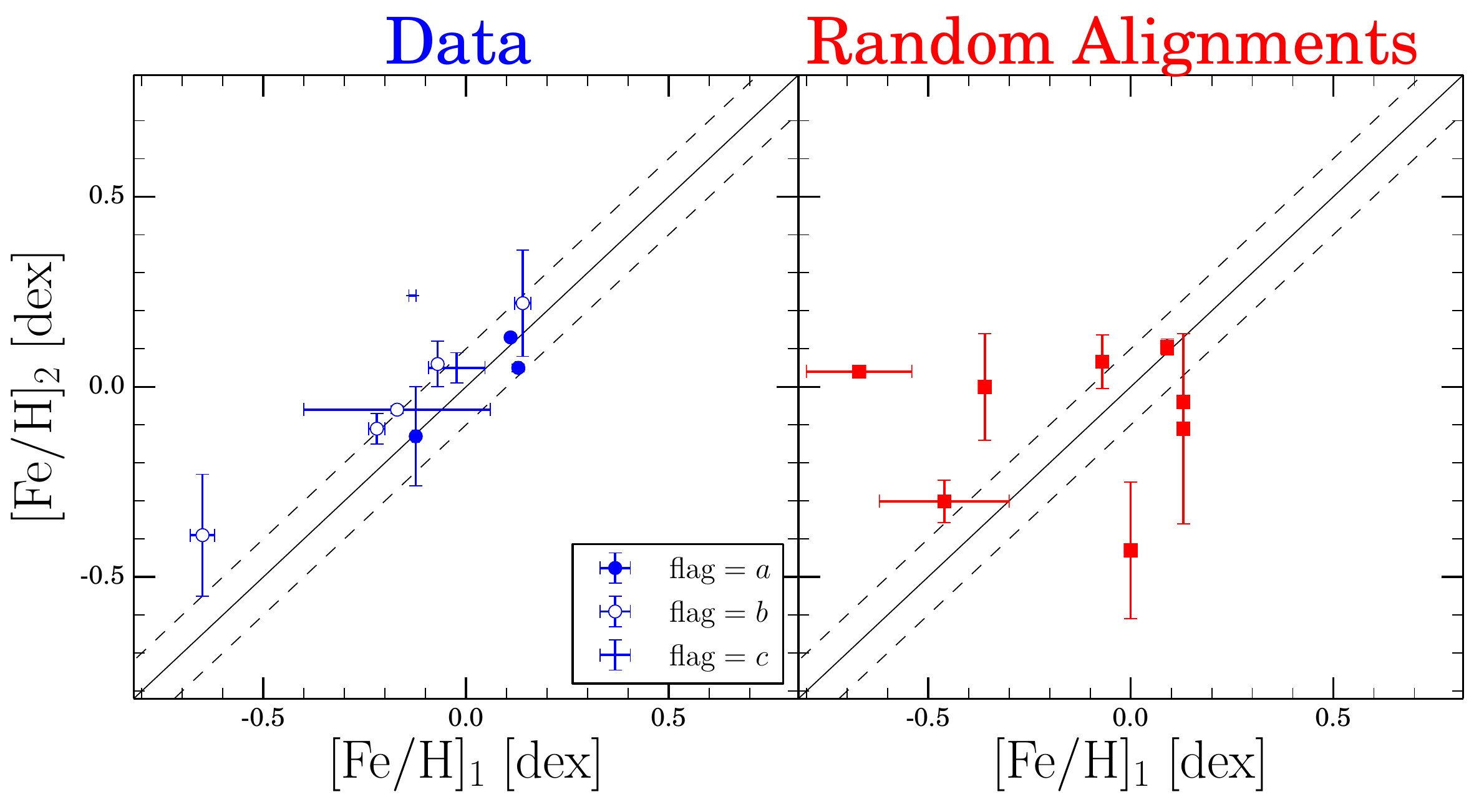}}}
\caption{{\it Subsample 1}: ``PM-$\varpi$-RV''. Comparison of the {\lamost} metallicities of the two components of each pair, for the data sample (left column, blue) and random alignments sample (right column, red). The pairs shown are the ones obtained in the 4th row of Figure \ref{fig:subs1_mu_theta}, after applying the constraints of RV, parallax, projected physical separation, and proper motion consistency. The black solid line is the 1:1 relation, and for guidance we show a $\pm 0.1$ dex displacement from it in the black dashed lines. For the data sample pairs we have assigned three type of flags. For a given pair these flags mean: ``{\it a}'': it is consistent in all parameters (filled circles); ``{\it b}'': it is consistent in some parameters but not in others (empty circles); ``{\it c}'': it shows clear inconsistencies (crosses). In the data sample shown there are 3 ``{\it a}'', 5 ``{\it b}'', and 2 ``{\it c}'' pairs. The pairs assigned with flag$=$``{\it c}'' have been discarded from our candidate list.}
\label{fig:subs1_FeH}
\end{minipage}
\end{figure*}

At this point we would like to assess the reliability of the pairs found in the data sample. One way of doing this, is to compare them with the random alignments sample pairs in a region of phase space that has not been involved in the search. For this purpose we use the {\lamost} metallicities, and compare the [Fe/H] values of the two components of each pair in Figure \ref{fig:subs1_FeH}.

Given our current understanding on the formation of wide binaries, when comparing the abundances of their components, we would expect both stars to have similar metallicities, as both stars come from the same parental material. Empirical evidence for this expectation has been reported in the works of \citet{andrews18}, \citet{desidera04} and \citet{desidera06}, reinforcing the idea that wide binaries can be thought of as the smallest versions of open clusters (cochemical and coeval stars). 

Figure \ref{fig:subs1_FeH} shows the metallicity comparison of the data sample pairs on the left (blue), and of the random alignments pairs on the right (red). The 1:1 relation, and a $\pm 0.1$ dex displacement from it, are shown as the solid and dashed lines, respectively. It can be seen that the data pairs tend to follow the 1:1 relation, while the random alignments pairs do not. 

We take this result as a confirmation that our multiple-parameter criteria used to select the data sample pairs is indeed recovering promising wide binary candidates, as for most of the components of the pairs their metallicities tend to be similar (even though the metallicity was not used as a criterion). This result is in agreement with the expectations.

In spite of these promising results, some contamination might still be present in our nominal data sample. For instance, one pair in Figure \ref{fig:subs1_FeH} is located well off the 1:1 relation, with [Fe/H]$_{1}=-0.132$ dex and [Fe/H]$_{2}=+0.24$ dex. Because of this, we decide to assign qualitative flags to our candidates depending on how consistent or inconsistent they appear to be when taking all the parameters into account. 

For any given pair, we assign it one of three types of flags: ``{\it a}'', when it is consistent in all the parameters and most likely is a real binary; ``{\it b}'', when it appears to be consistent in some regards but not in others; and ``{\it c}'', when even though it passed all the required criteria of the search, it shows clear inconsistencies under a more detailed inspection, and most likely correspond to a chance alignment. 

Out of the 10 pairs obtained in the 4th row of Figure \ref{fig:subs1_mu_theta}, we classify 3 of them as ``{\it a}'', 5 of them as ``{\it b}'', and 2 of them as ``{\it c}''. Figure \ref{fig:subs1_FeH} also shows the data sample pairs labeled according to this classification. We report the ``{\it a}'' and ``{\it b}'' pairs in the first block of Table \ref{tab:subs1_table}, where we also provide the parameters associated with the pairs and their components stars. The pairs assigned with flag$=$``{\it c}'' are discarded from our candidate list and not further analyzed.

\subsection{Relaxing the $\Delta$RV Criterion}
\label{subsec:subs1_search_2}

Some of the candidates found in \S \ref{subsec:subs1_search_1} have very precise RV (and metallicity) measurements (see Table \ref{tab:subs1_table}). While this is a convenient aspect of the {\lamost} DR3 data, in the event that some of the RV errors are underestimated, we might be discarding promising pairs because the RV criterion we have used ($\Delta \text{RV} \leq 1\sigma$) is too stringent. In other words, there might be promising pairs that are only consistent within 2$\sigma$ or 3$\sigma$ in $\Delta$RV, that are being rejected by our algorithm. 

Consequently, we run our search algorithm modifying the RV criteria used in the final search of \S \ref{subsec:subs1_search_1} to $\Delta \text{RV} \leq 2\sigma$ (with the rest of the criteria unchanged). By doing this we find 4 new pairs, out of which we classify 1 with flag$=$``{\it a}'', 1 with flag$=$``{\it b}'', and 2 with flag$=$``{\it c}''. We report the pairs classified as ``{\it a}'' and ``{\it b}'' in the second block of Table \ref{tab:subs1_table} and discard the ones classified as `{\it c}''. 

We also run our algorithm using $\Delta \text{RV} \leq 3\sigma$, but we only recovered 2 new pairs, both being classified with flag$=$``{\it c}'', with no new promising pairs being gained. This implies that we are probably entering the domain where chance alignments dominate, and we therefore decide to not continue relaxing the criteria and terminate the binary search of this subsample (using the UCAC4 proper motions) here.

\begin{table*}
\begin{minipage}{\textwidth}
\caption{List of our wide binary candidates in the {\it Subsample 1}: ``PM-$\varpi$-RV''. We only report the pairs qualitatively flagged as ``{\it a}'' and ``{\it b}'', as the pairs flagged as ``{\it c}'' have been discarded from our candidate list. The table is separated in three blocks, each one of them containing the candidates found in \S \ref{subsec:subs1_search_1},  \S \ref{subsec:subs1_search_2}, and  \S \ref{subsec:subs1_search_3}, respectively. In any given block of the table, the pairs are separated by horizontal lines. For each star we report KIC ID, proper motion, parallax, RV, and metallicity. For each pair we report $\mu /\Delta \mu$, angular separation, projected physical separation, and the assigned qualitative flag. The 12th column, RV-flag, indicates whether or not a given star has multiple {\lamost} measurements. For stars with only one measurement, the column is left blank. For stars with multiple measurements, RV-flag$=$*, and the reported RV and metallicity are the weighted average of the individual measurements (see \S \ref{subsubsec:data_lamost}).}
\renewcommand{\arraystretch}{1.0}
\begin{tabular}{ccccccccccccccc}
\hline
Pair \No & KIC ID & $\mu_{\alpha}$ & $\mu_{\delta}$ & $\mu/\Delta \mu$ & $\Delta \theta$ & $\varpi$ & $\sigma_{\varpi}$ & s & RV  & $\sigma_{\text{RV}}$ & RV-flag & [Fe/H] & $\sigma_{\text{[Fe/H]}}$& flag \\
  - & - & \multicolumn{2}{c}{[mas yr$^{-1}$] }& - & [$\arcsec$] & \multicolumn{2}{c}{[mas]} & [pc] & \multicolumn{2}{c}{[km s$^{-1}$]} & - & \multicolumn{2}{c}{[dex]} & -  \\
\hline
\hline
\multicolumn{15}{c}{List of pairs obtained in \S \ref{subsec:subs1_search_1}} \\
\hline
1 & 8689121 & -3.9 & -7.4 & 4.5 & 144.1 & 1.48 & 0.38 & 0.699 & -21.8 & 2.2 &  & 0.14 & 0.02 & {\it b} \\ 
 & 8753234 & -2.8 & -8.9 &  &  & 0.89 & 0.26 &  & -26.0 & 26.1 &  & 0.22 & 0.14 &  \\ 
\hline
2 & 8818205 & -3.2 & -9.4 & 12.2 & 88.5 & 1.46 & 0.34 & 0.848 & 4.5 & 2.0 &  & -0.22 & 0.02 & {\it b} \\ 
 & 8818252 & -2.4 & -9.5 &  &  & 0.07 & 0.45 &  & -0.8 & 3.5 &  & -0.11 & 0.04 &  \\ 
\hline
3 & 10164839 & 11.1 & 9.6 & 20.8 & 17.0 & 2.22 & 0.3 & 0.041 & -43.0 & 0.8 &  & 0.11 & 0.0 & {\it a} \\ 
 & 10164867 & 11.6 & 10.1 &  &  & 1.98 & 0.33 &  & -44.3 & 0.6 &  & 0.13 & 0.0 &  \\ 
\hline
4 & 5183581 & 13.1 & 29.0 & 8.8 & 343.0 & 3.97 & 0.3 & 0.447 & -26.2 & 0.7 & * & -0.07 & 0.0 & {\it b} \\ 
 & 5271947 & 16.7 & 29.5 &  &  & 3.61 & 0.34 &  & -30.1 & 5.2 &  & 0.06 & 0.06 &  \\ 
\hline
5 & 5523975 & -5.4 & -9.9 & 7.0 & 75.4 & 2.23 & 0.26 & 0.178 & -9.4 & 2.8 &  & -0.65 & 0.03 & {\it b} \\ 
 & 5524045 & -4.3 & -9.0 &  &  & 2.0 & 0.29 &  & 7.8 & 36.2 &  & -0.39 & 0.16 &  \\ 
\hline
6 & 7778058 & -5.0 & -10.8 & 10.0 & 42.8 & 1.45 & 0.54 & 0.187 & -0.2 & 1.0 & * & -0.124 & 0.009 & {\it a} \\ 
 & 7778114 & -4.4 & -9.9 &  &  & 1.42 & 0.5 &  & -4.2 & 18.6 &  & -0.13 & 0.13 &  \\ 
\hline
7 & 7914562 & -4.2 & -5.1 & 7.2 & 154.7 & 1.81 & 0.3 & 0.898 & -37.6 & 24.0 &  & -0.17 & 0.23 & {\it b} \\ 
 & 7983221 & -3.5 & -5.7 &  &  & 0.65 & 0.29 &  & -22.4 & 1.2 &  & -0.06 & 0.01 &  \\ 
\hline
8 & 4386086 & 27.1 & -6.2 & 9.4 & 28.7 & 10.67 & 0.38 & 0.013 & -17.1 & 1.4 & * & 0.129 & 0.007 & {\it a} \\ 
 & 4484238 & 27.6 & -3.3 &  &  & 10.6 & 0.28 &  & -17.5 & 1.5 &  & 0.05 & 0.01 &  \\ 
\hline
\multicolumn{15}{c}{List of pairs obtained in \S \ref{subsec:subs1_search_2}} \\
\hline
9 & 11071635 & 0.8 & 9.2 & 4.7 & 77.6 & 3.07 & 0.31 & 0.146 & -21.2 & 0.5 & * & -0.17 & 0.0 & {\it b} \\ 
 & 11124904 & 2.2 & 8.1 &  &  & 2.35 & 0.38 &  & -18.7 & 1.2 & * & 0.14 & 0.0 &  \\ 
\hline
10 & 5790787 & 4.5 & 10.6 & 12.1 & 27.7 & 2.64 & 0.62 & 0.055 & -29.9 & 0.8 & * & 0.09 & 0.0 & {\it a} \\ 
 & 5790807 & 5.1 & 9.9 &  &  & 2.57 & 0.3 &  & -33.0 & 0.9 &  & 0.1 & 0.0 &  \\ 
\hline
\multicolumn{15}{c}{List of pairs obtained in \S \ref{subsec:subs1_search_3}} \\
\hline
11 & 11874623 & -4.657 & -18.23 & 4.3 & 123.2 & 1.69 & 0.41 & 0.352 & -19.7 & 0.8 &  & -0.12 & 0.01 & {\it b} \\ 
 & 11874676 & -8.336 & -16.104 &  &  & 1.99 & 0.3 &  & -16.7 & 0.4 &  & -0.28 & 0.0 &  \\ 
\hline
12 & 10033625 & 10.111 & 20.748 & 6.0 & 382.3 & 4.23 & 0.26 & 0.461 & -22.3 & 0.7 &  & 0.02 & 0.0 & {\it b} \\ 
 & 10097397 & 6.877 & 19.526 &  &  & 3.88 & 0.27 &  & -20.8 & 0.6 &  & -0.13 & 0.0 &  \\ 
\hline
13 & 5295387 & -4.656 & -14.843 & 5.4 & 157.2 & 1.63 & 0.46 & 0.473 & 5.5 & 22.1 &  & -0.15 & 0.14 & {\it b} \\ 
 & 5295670 & -7.213 & -16.152 &  &  & 1.98 & 0.34 &  & 4.4 & 0.7 & * & -0.32 & 0.0 &  \\ 
\hline
\end{tabular}
\label{tab:subs1_table}
\end{minipage}
\end{table*}
\subsection{Using the TGAS Proper Motions}
\label{subsec:subs1_search_3}

The results of \S \ref{subsec:subs1_search_1} and \S \ref{subsec:subs1_search_2} were obtained using the UCAC4 proper motions. As a final exercise, in order to find potential candidates missed otherwise, we re-run our search algorithm using the TGAS proper motions instead, and check if new candidate pairs pass the criteria.

Similarly, we use the nominal $\Delta \text{RV} \leq 1\sigma$ criterion, but also relax it to 2$\sigma$ and 3$\sigma$. The rest of the criteria are left unchanged. When $\Delta \text{RV} \leq 1\sigma$, we gain 3 new pairs, 1 of them being classified as flag$=$``{\it b}'' and 2 as flag$=$``{\it c}''. For $\Delta \text{RV} \leq 2\sigma$, 2 new pairs are gained, 1 classified as flag$=$``{\it b}'' and 1 as flag$=$``{\it c}''. Finally, for the $\Delta \text{RV} \leq 3\sigma$ case, 2 new pairs are gained as well, 1 being classified as flag$=$``{\it b}'' and 1 as flag$=$``{\it c}''. The 3 promising pairs gained when using the TGAS proper motions (all classified as flag$=$``{\it b}'') are reported in the third block of Table \ref{tab:subs1_table}.
\subsection{Conclusions from the {\it Subsample 1}}
\label{subsec:subs1_conclusions}

By having started performing our wide binary searches in the {\it Subsample 1}, the only subsample where all six parameters of the position and velocity phase space can be constrained simultaneously, we have gained insight on how to better select promising candidates. This insight can now be applied for the remaining subsamples, where not all the parameters are available, and therefore more specific constraints are required.

The approach used in this subsample, where we started requiring criteria first for the individual stars (e.g., $\mu/\sigma_{\mu} \geq 3$, availability of parallax and RV measurements) and then for the pairs (e.g., $\Delta$RV, $\mu/\Delta \mu_{\text{min}}$), is also applied in the other subsamples. In this way we first get a picture of the phase space without any constraints on the pairs (1st row of Figure \ref{fig:subs1_mu_theta}), and then we move onto applying several criteria in order to select our candidates. A summarized version of the selection criteria used in this subsample can be found in Table \ref{tab:summary_criteria_subsamples}.

Additionally, Figure \ref{fig:subs1_FeH} has lead us to conclude that the component of wide binaries {\it tend} to have similar metallicities, in agreement with theoretical expectations and the observational results of \citet{andrews18}, \citet{desidera04}, and \citet{desidera06}. With this result in mind, in an attempt to compensate for the absence of trigonometric parallaxes, in the search performed in the {\it Subsample 3} we {\it require} the metallicities of the components of candidate pairs to be consistent. 

We also point out that, for a given set of criteria, the reader should not interpret the ratio of the number of random alignments pairs to the number of data pairs as the contamination rate of our candidates. This ratio is biased towards higher values (see \S \ref{subsec:subs1_search_1}), and we expect the actual contamination rate to be lower than that.

Finally, we note that the goal of our procedure is to find samples of wide binary candidates where both stars were observed by {\kepler}. This, however, does not imply that our search is complete. Besides of the inherited {\kepler} selection function, the criteria used in our search might be too stringent for some binary candidates. For these potentially promising but nevertheless discarded pairs, the quality of our data does not allow us to reliably distinguish them from chance alignments.

The procedure to select our binary candidates in the remaining subsamples is analogous to the one used here, to the extent of the available data. The reader may wish to skip the details of the selection process and continue in \S \ref{sec:search_all_subs}, after our final sample of candidates has been compiled.


\section{Binary Search in the {\it Subsample 2}: ``PM-$\varpi$''}
\label{sec:search_subs2}
\subsection{Finding Promising Candidates}
\label{subsec:subs2_search_1}

\begin{figure*}
\begin{minipage}{\textwidth}
\centering
\subfloat{{\includegraphics[width=0.70\linewidth]{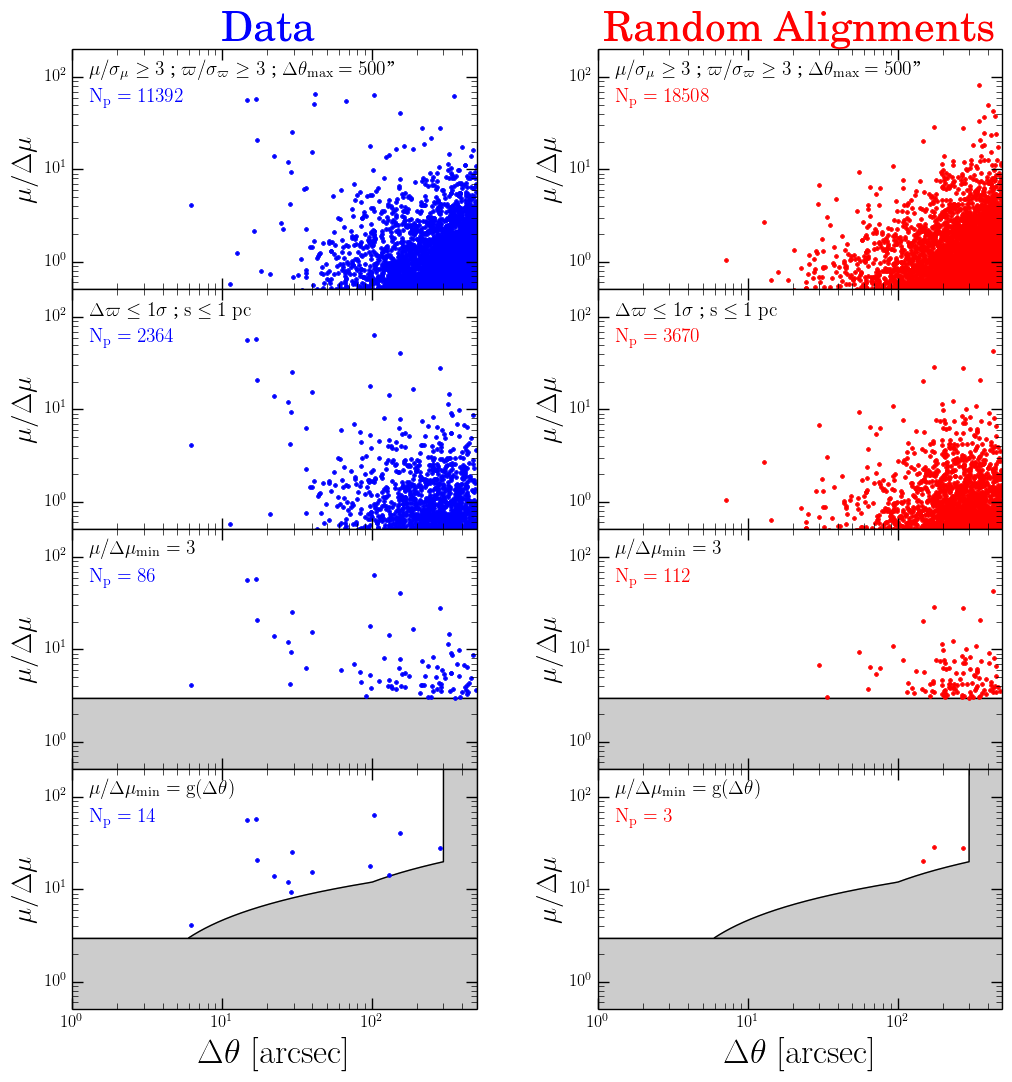}}}
\caption{{\it Subsample 2}: ``PM-$\varpi$''. Analogous to Figure \ref{fig:subs1_mu_theta}, for the subsample of stars with proper motions and trigonometric parallaxes available (no RVs required). The data (random alignments) pairs are shown in the left (right) column in blue (red), with every row representing a different search going downwards as more criteria are included. The criteria used in each row, together with the number of pairs in each panel, are shown at the top left corner. In the 1st row we only include stars with well measured proper motions ($\mu/\sigma_{\mu} \geq 3$) and parallaxes ($\varpi/\sigma_{\varpi} \geq 3$), and search for pairs up to angular separations of $\Delta \theta_{\text{max}} = 500 \arcsec$. In the 2nd row we only keep the pairs whose stars have parallaxes consistent within 1$\sigma$ and projected physical separations s $\leq 1$ pc. In the 3rd row we add the $\mu/\Delta \mu_{\text{min}}=3$ common proper motion constraint, and in the 4th row we add an angular separation dependence to it, $\mu/\Delta \mu_{\text{min}}=\text{g}(\Delta \theta)$.} 
\label{fig:subs2_mu_theta}
\end{minipage}
\end{figure*}

\begin{figure*}
\begin{minipage}{\textwidth}
\centering
\subfloat{{\includegraphics[width=0.65\linewidth]{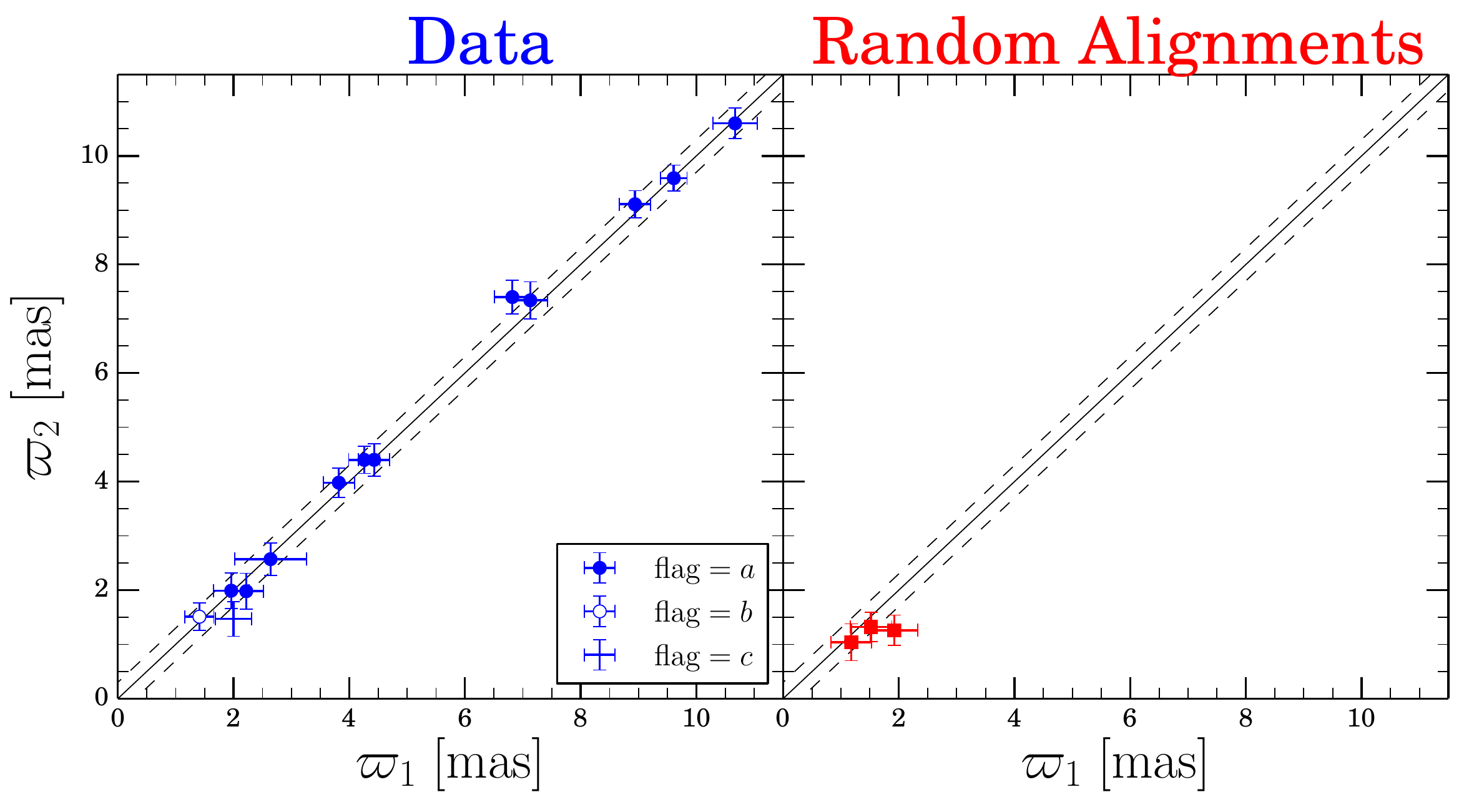}}}
\caption{{\it Subsample 2}: ``PM-$\varpi$''. Comparison of the TGAS parallaxes of the two components of each pair in the data sample (left column, blue) and random alignments sample (right column, red). The pairs shown are the ones obtained in the 4th row of Figure \ref{fig:subs2_mu_theta}, after applying the constraints of parallax consistency, projected physical separation, and proper motion consistency. The black solid line shows the 1:1 relation, and for guidance we show a $\pm$0.3 mas displacement from it in the black dashed lines. Similarly as for {\it Subsample 1}, we have assigned the qualitative flags ``{\it a/b/c}'' based on the overall consistency of a given pair. The symbols used for the flags are the same used in Figure \ref{fig:subs1_FeH}. In the data sample shown there are 11 ``{\it a}'', 1 ``{\it b}'', and 1 ``{\it c}'' pairs. We note that the well known wide binary 16 Cygni (KIC 12069424/KIC 12069449) is recovered by our nominal search, but due to its large parallax value ($\varpi \simeq 47$ mas), it is off the scale of the plot.} 
\label{fig:subs2_plx}
\end{minipage}
\end{figure*}

We now proceed to look for wide binary candidates in the subsample of stars with proper motions and trigonometric parallaxes available (no RVs required). Since this is intrinsically a subsample where less constraints can be applied (no $\Delta$RV), we add a parallax quality criterion. Similarly as with the proper motion quality cut, we only work with stars that satisfy:

\begin{equation}
\varpi/\sigma_{\varpi}\geq 3.
\label{eqn:plx_quality}
\end{equation}
This criterion preferentially retains stars with larger parallax values.

With this, our initial pool now consists of 6,688 stars with $\mu/\sigma_{\mu} \geq 3$ that also satisfy Equation \ref{eqn:plx_quality}. Similarly as with the {\it Subsample 1}, based on the trigonometric distance distribution of Figure \ref{fig:histogram_distance_phot_trig}, we set the angular separation search radius to be $\Delta \theta_{\text{max}}=500 \arcsec$, again allowing us to find binaries with s $\simeq$ 1 pc if present.

For now, we do not require any common proper motion constraint ($\mu/\Delta \mu_{\text{min}}=0$). With these criteria we run our search algorithm and find 11,392 pairs in the data sample and 18,508 pairs in its random alignments counterpart. The $\mu/\Delta \mu$ versus $\Delta \theta$ phase space for this search is shown in the top row of Figure \ref{fig:subs2_mu_theta}, with the result of subsequent searches shown in the following rows.

Similarly as in the {\it Subsample 1}, we find most of the pairs piled up at low $\mu/\Delta \mu$ and large $\Delta \theta$ values, indicating the region where the chance alignments dominate. In this subsample, however, there is an even more clear over-abundance of data sample pairs with high $\mu/\Delta \mu$ values ($\gtrsim 10$) and $\Delta \theta \lesssim 40 \arcsec$.

We now add the parallax constraint $\Delta \varpi \leq 1\sigma$, meaning we only keep the pairs whose stars have parallaxes consistent within their errorbars. We also add the projected physical separation constraint s $\leq 1$ pc. The remaining pairs after applying these two criteria are shown in the 2nd row of Figure \ref{fig:subs2_mu_theta}.

From the 1st to the 2nd row of Figure \ref{fig:subs2_mu_theta} the number of pairs in both panels has dropped dramatically. Once more, most of the pairs with $\Delta \theta > 100 \arcsec$ have been discarded due to the s $\leq 1$ pc criterion.

We now add the common proper motion constraint $\mu /\Delta \mu_{\text{min}}=3$, discarding all the pairs located in the shaded region of the 3rd row of Figure \ref{fig:subs2_mu_theta}. The difference between both panels is now more evident, with the random alignments sample pairs being concentrated at the bottom right corner of the plot, and the data sample pairs covering the same portion of the plot but clearly extending to smaller $\Delta \theta$ and greater $\mu/\Delta \mu$ values. Moreover, only the widest ($\Delta \theta>100 \arcsec$) random alignments pairs have $\mu / \Delta \mu>$10, while there are data pairs with $\mu/\Delta \mu>$10 across the entire 10$\arcsec<\Delta \theta<500\arcsec$ range.

As a last step, we add an angular separation dependence to the common proper motion criterion, $\mu/\Delta \mu_{\text{min}}=\text{g}(\Delta \theta)$. As in {\it Subsample 1}, we define the $\text{g}(\Delta \theta)$ function qualitatively, trying to clip as many random alignments pairs as possible, and at the same time finding a region where the data pairs stand out for their high $\mu /\Delta \mu$ value.

After adding this criterion in the 4th row of Figure \ref{fig:subs2_mu_theta}, we are left with 14 pairs in the data sample and 3 pairs in the random alignments sample. This constitutes the nominal search of the {\it Subsample 2}, although we expand it in the following subsections.

Contrary to the pairs left in the data sample, the 3 pairs left in the random alignment sample all have $\Delta \theta > 100 \arcsec$, showing that we expect the contamination of our wide binary candidates to be not only small, but present only at large $\Delta \theta$ values.

Even though the parallax values were used as constraints in this subsample (therefore, for any given pair, matching by construction), we can still get relevant conclusions from studying their distribution. Figure \ref{fig:subs2_plx} compares the parallax value of the two components of the data pairs (left column, blue) and random alignments pairs (right column, red) with the 1:1 relation (black solid line).

Besides having wide angular separations, Figure \ref{fig:subs2_plx} shows that the 3 random alignments pairs left also have small parallax values when compared with the data sample pairs. The random alignments pairs are concentrated at $\varpi \lesssim 2$ mas, while the data pairs extend across the entire $2 \lesssim \varpi \lesssim 10$ range. Therefore, we expect most of the contamination in our candidates to exist for the pairs with smallest parallax values.

Similarly as for the {\it Subsample 1}, we assign qualitative flags (``{\it a}'', ``{\it b}'', or ``{\it c}'' ) to our wide binary candidates. We note that in this case, when assessing the overall consistency of our pairs, we do not have RVs available for most of them, and therefore our qualitative statement is based only on parallaxes, proper motions, and projected physical separations. 

We note that out of the 14 pairs obtained in the 4th row of Figure \ref{fig:subs2_mu_theta}, 3 of them were also obtained in {\it Subsample 1} (all of them classified as ``{\it a}'', these are the pairs \No 3, \No 8, and \No 10 of Table \ref{tab:subs1_table}). Of the 11 {\it new} pairs obtained in the 4th row of Figure \ref{fig:subs2_mu_theta}, we classify 9 of them as ``{\it a}'', 1 of them as ``{\it b}'', and 1 of them as ``{\it c}''. The data sample pairs shown in Figure \ref{fig:subs2_plx} are labeled according to this classification, and we report the ``{\it a}'' and ``{\it b}'' pairs in the first block of Table \ref{tab:subs2_table}.

\begin{table*}
\begin{minipage}{\textwidth}
\centering
\caption{List of our wide binary candidates in the {\it Subsample 2}: ``PM-$\varpi$''. Analogous to Table \ref{tab:subs1_table}, for the subsample of stars with only proper motions and trigonometric parallaxes available (no RVs required). We only report the pairs qualitatively flagged as ``{\it a}'' and ``{\it b}''. The table is separated in three blocks, each one of them containing the candidates found in \S \ref{subsec:subs2_search_1},  \S \ref{subsec:subs2_search_2}, and  \S \ref{subsec:subs2_search_3}, respectively. The 3 pairs found in \S \ref{subsec:subs2_search_1} that were also found in {\it Subsample 1} (pairs \No3, \No8, and \No10 of Table \ref{tab:subs1_table}) are not reported in this table.}
\renewcommand{\arraystretch}{1.0}
\begin{tabular}{cccccccccc}
\hline
Pair \No & KIC ID & $\mu_{\alpha}$ & $\mu_{\delta}$ & $\mu/\Delta \mu$ & $\Delta \theta$ & $\varpi$ & $\sigma_{\varpi}$ & s & flag \\
  - & - & \multicolumn{2}{c}{[mas yr$^{-1}$] }& - & [$\arcsec$] & \multicolumn{2}{c}{[mas]} & [pc] & -  \\
\hline
\hline
\multicolumn{10}{c}{List of pairs obtained in \S \ref{subsec:subs2_search_1}} \\
\hline
1 & 12156630 & 20.2 & 40.8 & 40.7 & 154.4 & 6.82 & 0.31 & 0.106 & {\it a} \\ 
 & 12156742 & 19.7 & 41.8 &  &  & 7.4 & 0.31 &  &  \\ 
\hline
2 & 9139151 & 22.6 & 58.4 & 25.7 & 29.3 & 9.61 & 0.23 & 0.015 & {\it a} \\ 
 & 9139163 & 25.0 & 58.0 &  &  & 9.59 & 0.24 &  &  \\ 
\hline
3 & 12069424 & -147.8 & -159.0 & 15.6 & 39.6 & 46.98 & 0.25 & 0.004 & {\it a} \\ 
 & 12069449 & -135.1 & -163.8 &  &  & 47.12 & 0.23 &  &  \\ 
\hline
4 & 7013635 & 11.9 & -44.7 & 63.5 & 102.9 & 8.94 & 0.27 & 0.055 & {\it a} \\ 
 & 7013649 & 12.1 & -45.4 &  &  & 9.11 & 0.25 &  &  \\ 
\hline
5 & 10614190 & 2.1 & 2.7 & 14.3 & 129.4 & 1.41 & 0.25 & 0.45 & {\it b} \\ 
 & 10614382 & 2.0 & 2.5 &  &  & 1.51 & 0.25 &  &  \\ 
\hline
6 & 10616124 & -2.7 & -17.5 & 56.0 & 14.6 & 1.96 & 0.31 & 0.037 & {\it a} \\ 
 & 10616138 & -3.0 & -17.6 &  &  & 1.99 & 0.33 &  &  \\ 
\hline
7 & 2696938 & 6.0 & -19.8 & 4.1 & 6.2 & 3.82 & 0.27 & 0.008 & {\it a} \\ 
 & 2696944 & 1.0 & -20.6 &  &  & 3.98 & 0.27 &  &  \\ 
\hline
8 & 8174654 & 12.5 & 18.0 & 18.2 & 97.5 & 4.26 & 0.27 & 0.11 & {\it a} \\ 
 & 8242135 & 13.3 & 18.9 &  &  & 4.4 & 0.25 &  &  \\ 
\hline
9 & 8241071 & 38.2 & -18.7 & 14.2 & 22.2 & 4.43 & 0.27 & 0.025 & {\it a} \\ 
 & 8241074 & 40.8 & -20.2 &  &  & 4.4 & 0.3 &  &  \\ 
\hline
10 & 2992956 & 14.1 & 50.7 & 58.1 & 16.9 & 7.13 & 0.3 & 0.011 & {\it a} \\ 
 & 2992960 & 14.5 & 49.9 &  &  & 7.34 & 0.34 &  &  \\ 
\hline
\multicolumn{10}{c}{List of pairs obtained in \S \ref{subsec:subs2_search_2}} \\
\hline
11 & 9944337 & -29.0 & -28.3 & 51.9 & 41.3 & 4.59 & 0.36 & 0.051 & {\it a} \\ 
 & 9944356 & -29.6 & -28.8 &  &  & 3.57 & 0.48 &  &  \\ 
\hline
12 & 12366681 & -1.9 & 14.8 & 65.8 & 41.8 & 2.68 & 0.3 & 0.09 & {\it b} \\ 
 & 12366719 & -1.8 & 14.6 &  &  & 2.02 & 0.27 &  &  \\ 
\hline
13 & 4043389 & -124.2 & -176.0 & 18.1 & 162.8 & 36.84 & 0.54 & 0.022 & {\it a} \\ 
 & 4142913 & -114.4 & -182.7 &  &  & 35.85 & 0.37 &  &  \\ 
\hline
14 & 6225718 & 105.4 & -174.5 & 54.8 & 66.8 & 19.15 & 0.27 & 0.017 & {\it a} \\ 
 & 6225816 & 108.4 & -176.7 &  &  & 18.46 & 0.34 &  &  \\ 
\hline
\multicolumn{10}{c}{List of pairs obtained in \S \ref{subsec:subs2_search_3}} \\
\hline
15 & 11551404 & 16.864 & 12.296 & 26.5 & 55.7 & 3.04 & 0.28 & 0.088 & {\it a} \\ 
 & 11551430 & 16.466 & 12.977 &  &  & 3.21 & 0.32 &  &  \\ 
\hline
16 & 11603064 & -4.169 & -19.632 & 21.6 & 62.0 & 1.99 & 0.22 & 0.152 & {\it a} \\ 
 & 11603098 & -4.821 & -19.002 &  &  & 2.05 & 0.28 &  &  \\ 
\hline
17 & 6924906 & 5.57 & -16.536 & 8.0 & 72.5 & 1.37 & 0.37 & 0.242 & {\it b} \\ 
 & 6924968 & 5.927 & -14.599 &  &  & 1.88 & 0.28 &  &  \\ 
\hline
18 & 7418359 & 0.788 & 11.436 & 147.4 & 4.4 & 2.45 & 0.36 & 0.009 & {\it a} \\ 
 & 7418367 & 0.799 & 11.513 &  &  & 2.31 & 0.32 &  &  \\ 
\hline
19 & 5967147 & -6.337 & -16.416 & 6.2 & 54.6 & 1.97 & 0.46 & 0.122 & {\it b} \\ 
 & 5967153 & -8.924 & -17.612 &  &  & 2.87 & 0.33 &  &  \\ 
\hline
20 & 7975212 & 7.995 & -1.314 & 4.0 & 36.1 & 2.2 & 0.39 & 0.074 & {\it b} \\ 
 & 7975257 & 6.731 & -2.604 &  &  & 2.84 & 0.36 &  &  \\ 
\hline
21 & 4567525 & 4.548 & 6.37 & 7.9 & 61.0 & 2.65 & 0.28 & 0.114 & {\it b} \\ 
 & 4567531 & 3.83 & 5.848 &  &  & 2.68 & 0.37 &  &  \\ 
\hline
\end{tabular}
\label{tab:subs2_table}
\end{minipage}
\end{table*}

\subsection{Relaxing the $\Delta \varpi$ Criterion}
\label{subsec:subs2_search_2}

Since our parallax constraint of $\Delta \varpi \leq 1\sigma$ might be too stringent for pairs with large parallaxes but underestimated errors, we run our search algorithm relaxing the criteria used. First we use $\Delta \varpi \leq 2\sigma$ and leave the rest of the criteria unchanged. By doing this we find 7 new pairs, out of which we classify 3 as ``{\it a}'', 1 as ``{\it b}'', and 3 as ``{\it c}''. We then further relax our parallax consistency criterion to $\Delta \varpi \leq 3\sigma$ and find only 1 new pair, classifying it as ``{\it c}''. We decide to conclude this binary search (using the UCAC4 proper motions) here, as a more relaxed $\Delta \varpi$ criterion will predominantly select chance alignments. The ``{\it a}'' and ``{\it b}'' pairs gained by this exercise are reported in the second block of Table \ref{tab:subs2_table}.
\subsection{Using the TGAS Proper Motions}
\label{subsec:subs2_search_3}

Finally, we re-run our search algorithm using the TGAS proper motions in order to find new pairs that would missed otherwise. We use a slightly modified angular separation dependence on the common proper motion consistency criterion, namely $\mu/\Delta\mu_{\text{min}}=\text{g}'(\Delta \theta)$. This function is similar to $\text{g}(\Delta \theta)$ shown in the 4th row of Figure \ref{fig:subs2_mu_theta}, but defined using the TGAS proper motions instead. 

We use the nominal $\Delta \varpi \leq 1\sigma$, but also relax it to $2\sigma$ and $3\sigma$, while keeping the rest of the criteria are unchanged. For $\Delta \varpi \leq 1\sigma$ we gain 10 new pairs, 3 of them being classified as ``{\it a}'', 3 as ``{\it b}'', and 4 as ``{\it c}''. For $\Delta \varpi \leq 2\sigma$ we gain 4 new pairs, 1 of them being classified as ``{\it b}'' and 3 as ``{\it c}''. Finally, for $\Delta \varpi \leq 3\sigma$, only 1 ``{\it c}'' pair is gained. The ``{\it a}'' and ``{\it b}'' pairs gained by this exercise are reported in the third block of Table \ref{tab:subs2_table}. A summarized version of the selection criteria used in this subsample can be found in Table \ref{tab:summary_criteria_subsamples}.

\begin{figure*}
\begin{minipage}{\textwidth}
\centering
\subfloat{{\includegraphics[width=0.70\linewidth]{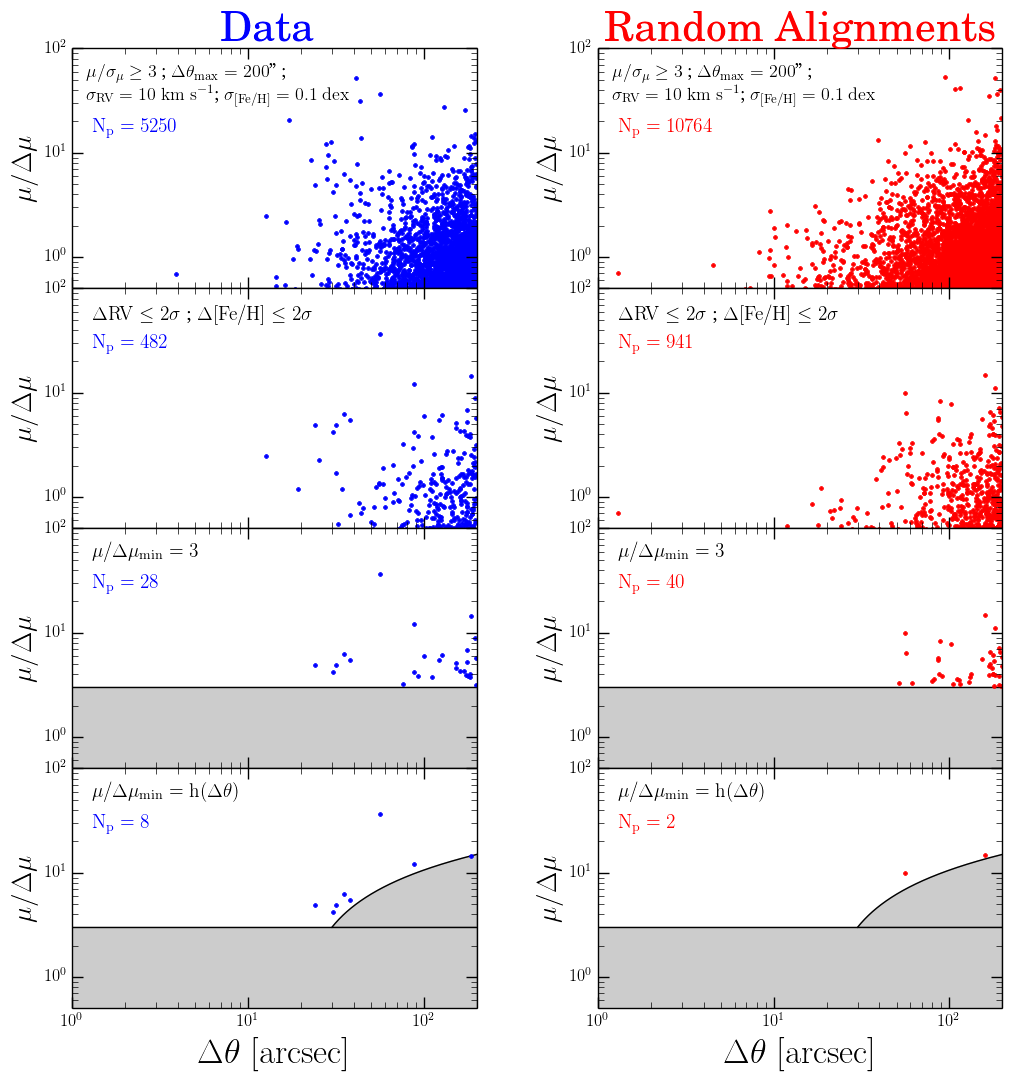}}}
\caption{{\it Subsample 3}: ``PM-RV-Metallicity''. Analogous to Figure \ref{fig:subs1_mu_theta}, for the subsample of stars with proper motions, {\lamost} RVs and metallicities available (no parallaxes required). In the 1st row we only include stars with well measured proper motions ($\mu/\sigma_{\mu} \geq 3$), RVs ($\sigma_{\text{RV}}\leq 10$ km s$^{-1}$), and metallicities ($\sigma_{\text{[Fe/H]}} \leq 0.1$ dex), and search for pairs up to angular separations of $\Delta \theta_{\text{max}}=200\arcsec$. In the 2nd row we only keep the pairs whose stars have RVs and metallicities consistent within 2$\sigma$. In the 3rd row we add the $\mu/\Delta \mu_{\text{min}}=3$ common proper motion constraint, and in the 4th row we add an angular separation dependence to it, $\mu/\Delta \mu_{\text{min}}=\text{h}(\Delta \theta)$.} 
\label{fig:subs3_mu_theta}
\end{minipage}
\end{figure*}


\section{Binary Search in the {\it Subsample 3}: ``PM-RV-Metallicity''}
\label{sec:search_subs3}

\subsection{Finding Promising Candidates}
\label{subsec:subs3_search_1}

\begin{figure*}
\begin{minipage}{\textwidth}
\centering
\subfloat{{\includegraphics[width=0.65\linewidth]{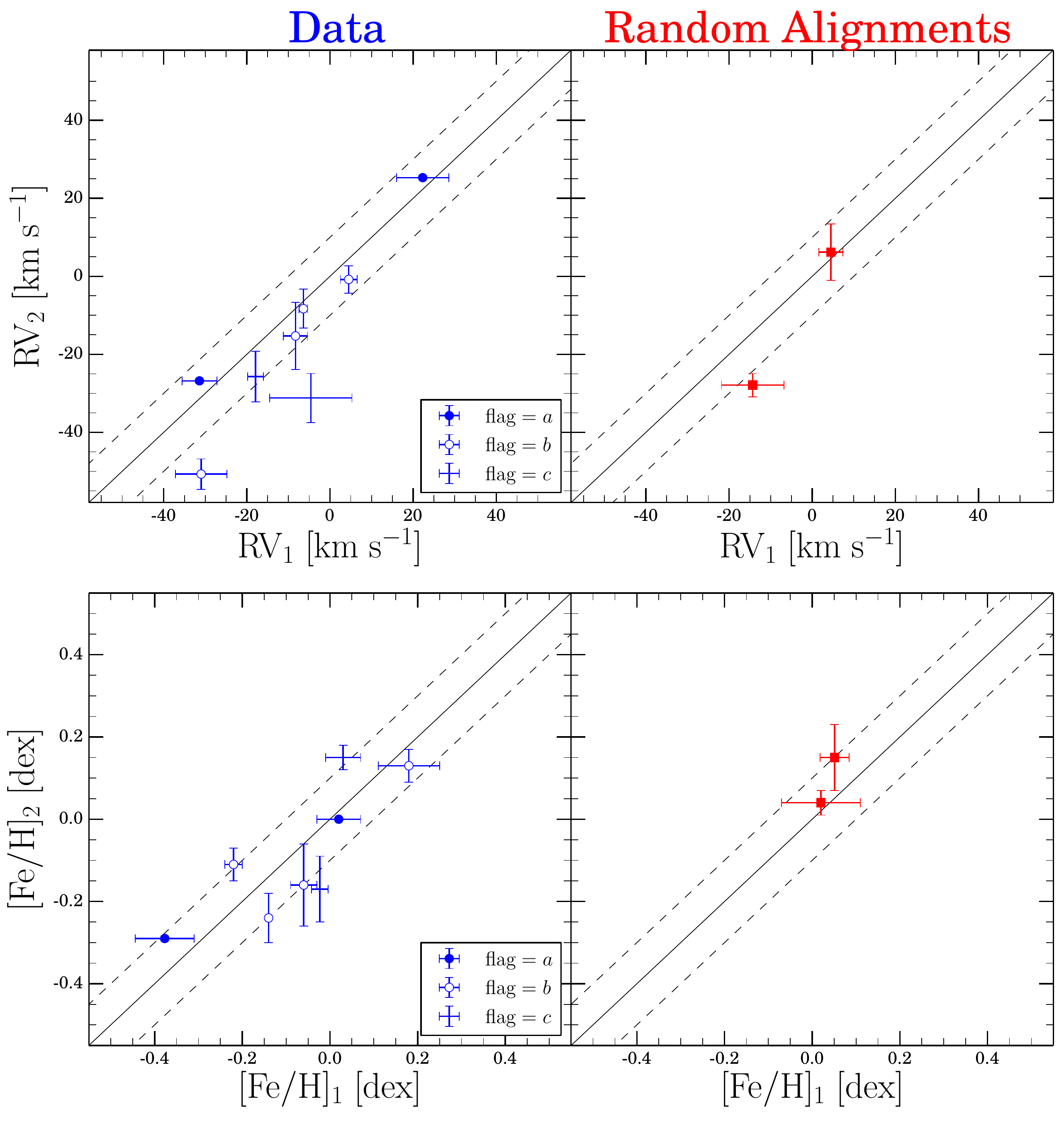}}}
\caption{{\it Subsample 3}: ``PM-RV-Metallicity''. Comparison of the {\lamost} RVs (top panel) and metallicities (bottom panel) of the two components of each pair in the data sample (left column, blue) and random alignments sample (right column, red). The pairs shown are the ones obtained in the 4th row of Figure \ref{fig:subs3_mu_theta}, after applying the constraints of RV and metallicity quality, as well as proper motion, RV, and metallicity consistency. The black solid line in all panels shows the 1:1 relation, and for guidance we show a $\pm$ 10 km s$^{-1}$ (top) and $\pm$0.1 dex (bottom) displacement from it in the black dashed lines. As for the previous subsamples, we have assigned qualitative flags ``{\it a/b/c}'' based on the overall consistency of a given pair. In the data sample shown there are 2 ``{\it a}'', 4 ``{\it b}'', and 2 ``{\it c}'' pairs.} 
\label{fig:subs3_RV_FeH}
\end{minipage}
\end{figure*}

\begin{table*}
\begin{minipage}{\textwidth}
\centering
\caption{List of our wide binary candidates in the {\it Subsample 3}: ``PM-RV-Metallicity''. Analogous to Table \ref{tab:subs1_table}, for the subsample of stars with proper motions and {\lamost} RVs and metallicities available (no parallaxes required). We only report the pairs qualitatively flagged as ``{\it a}'' and ``{\it b}''. The table is separated in two blocks, each one of them containing the candidates found in \S \ref{subsec:subs3_search_1} and \S \ref{subsec:subs3_search_2}, respectively. The pair found in \S \ref{subsec:subs3_search_1} that was also found in {\it Subsample 1} (pair \No2 of Table \ref{tab:subs1_table}) is not reported in this table. The 9th column, RV-flag, indicates whether or not a given star has multiple {\lamost} measurements. For stars with only one measurement, the column is left blank. For stars with multiple measurements, RV-flag$=$*, and the reported RV and metallicity are the weighted average of the individual measurements (see \S \ref{subsubsec:data_lamost}).}
\renewcommand{\arraystretch}{1.0}
\begin{tabular}{cccccccccccc}
\hline
Pair \No & KIC ID & $\mu_{\alpha}$ & $\mu_{\delta}$ & $\mu/\Delta \mu$ & $\Delta \theta$ & RV & $\sigma_{\text{RV}}$ & RV-flag & [Fe/H] &  $\sigma_{\text{[Fe/H]}}$& flag \\
  - & - & \multicolumn{2}{c}{[mas yr$^{-1}$] }& - & [$\arcsec$] & \multicolumn{2}{c}{[km s$^{-1}$]} & - & \multicolumn{2}{c}{[dex]} & -  \\
\hline
\hline
\multicolumn{12}{c}{List of pairs obtained in \S \ref{subsec:subs3_search_1}} \\
\hline
1 & 11069655 & 7.3 & -39.7 & 5.0 & 23.9 & -6.4 & 1.0 & * & -0.14 & 0.0 & {\it b} \\ 
 & 11069662 & 9.6 & -47.5 &  &  & -8.3 & 5.0 &  & -0.24 & 0.06 &  \\ 
\hline
2 & 10663891 & -7.3 & -20.9 & 4.9 & 31.6 & 22.3 & 6.3 & * & -0.377 & 0.067 & {\it a} \\ 
 & 10663892 & -8.4 & -25.3 &  &  & 25.3 & 0.6 & * & -0.29 & 0.006 &  \\ 
\hline
3 & 10736547 & 5.2 & 83.7 & 36.8 & 56.3 & -31.0 & 6.2 &  & 0.18 & 0.07 & {\it b} \\ 
 & 10736623 & 5.8 & 85.9 &  &  & -50.7 & 3.9 &  & 0.13 & 0.04 &  \\ 
\hline
4 & 11467819 & -10.6 & -15.8 & 4.2 & 30.3 & -8.3 & 2.9 &  & -0.06 & 0.03 & {\it b} \\ 
 & 11467837 & -6.5 & -16.2 &  &  & -15.3 & 8.6 &  & -0.16 & 0.1 &  \\ 
\hline
5 & 7748234 & -8.3 & -0.3 & 5.5 & 38.0 & -31.4 & 4.2 &  & 0.02 & 0.05 & {\it a} \\ 
 & 7748238 & -7.6 & 0.9 &  &  & -26.8 & 0.5 &  & 0.0 & 0.0 &  \\ 
\hline
\multicolumn{12}{c}{List of pairs obtained in \S \ref{subsec:subs3_search_2}} \\
\hline
6 & 8293539 & 0.2 & -41.5 & 9.3 & 52.7 & 13.2 & 0.5 &  & -0.25 & 0.0 & {\it b} \\ 
 & 8293571 & -3.8 & -39.9 &  &  & 11.6 & 9.7 & * & -0.123 & 0.116 &  \\ 
\hline
7 & 10489092 & -7.9 & -93.7 & 12.6 & 135.4 & -4.7 & 2.4 &  & -0.39 & 0.02 & {\it a} \\ 
 & 10489173 & -14.5 & -90.6 &  &  & -1.8 & 10.8 &  & -0.27 & 0.13 &  \\ 
\hline
8 & 5907690 & 15.6 & 3.6 & 11.3 & 105.3 & -34.0 & 11.4 & * & 0.064 & 0.146 & {\it b} \\ 
 & 5907881 & 15.8 & 2.2 &  &  & -35.6 & 6.8 &  & 0.26 & 0.08 &  \\ 
\hline
\end{tabular}
\label{tab:subs3_table}
\end{minipage}
\end{table*}

We now search for wide binary candidates in the subsample of stars with {\lamost} RVs and metallicities available (no parallaxes required). In comparison with {\it Subsample 1}, less constraints can be used in this search (no $\Delta \varpi$). In an attempt to compensate for this and further constrain our potential candidates, in addition to using the RVs, we decide to use the metallicities as a criterion. This is based on the conclusion that the components of wide binaries tend to have similar abundances (see \S \ref{subsec:subs1_conclusions}).

First, we perform a quality cut in the parameters that will be used as constraints. We only use stars that satisfy:

\begin{equation}
\sigma_{\text{RV}} \leq 10 \text{ km s}^{-1},
\label{eqn:RV_quality}
\end{equation}
and
\begin{equation}
\sigma_{\text{[Fe/H]}} \leq 0.1 \text{ dex}.
\label{eqn:FeH_quality}
\end{equation}

These quality cuts, in addition to the $\mu/\sigma_{\mu} \geq 3$ criterion, leave us with 10,944 stars in the initial pool for this subsample. In this case we set the angular separation search radius to be $\Delta \theta_{\text{max}} =200 \arcsec$.

For now, we do not require any common proper motion constraint ($\mu/\Delta \mu_{\text{min}}=0$). Using these criteria we run our search algorithm and find 5,250 pairs in the data sample and 10,764 pairs in its random alignment counterpart. The $\mu/\Delta \mu$ versus $\Delta \theta$ phase space for this search is shown in the top row of Figure \ref{fig:subs3_mu_theta}. The distribution of pairs seen is similar as the ones found in the previous subsamples, with most pairs piled up at low $\mu/\Delta \mu$ and large $\Delta \theta$ values. There is, however, a small overabundance of data sample pairs with respect to the random alignments sample pairs at $\Delta \theta \simeq 30\arcsec$ and $\mu/\Delta \mu \simeq 5 \textendash 10$.

We now require the pairs to have RVs  and metallicities consistent within two times their errorbars, i.e., $\Delta$RV $\leq 2\sigma$ and $\Delta$[Fe/H] $\leq 2\sigma$. The same subsequent searches were initially performed with a 1$\sigma$ consistency criterion for RVs and metallicities, however this yielded an extremely small sample size, and we opted to use a 2$\sigma$ consistency instead (see below). 

The pairs remaining after using these criteria are shown in the 2nd row of Figure \ref{fig:subs3_mu_theta}. Although the number of pairs has dropped greatly in both panels, now the overabundance of data sample pairs can be seen more clearly (see $\Delta \theta \simeq 20 \textendash 40 \arcsec$, $\mu/\Delta \mu \simeq 5$). 

As with the previous subsamples, in the 3rd row of Figure \ref{fig:subs3_mu_theta} we add the common proper motion constraint $\mu/\Delta \mu_{\text{min}} = 3$. Now, both panels show clear differences. The random alignments sample pairs are all located at $\Delta \theta \gtrsim 50\arcsec$. While the data sample pairs span the whole $20 \arcsec \leq \Delta \theta \leq 200 \arcsec$ range, the pairs with $\Delta \theta \leq 40\arcsec$ have no counterpart in the random alignments sample, indicating they could be promising candidates.

Finally, we add an angular dependence to the common proper motion criterion, $\mu/\Delta \mu_{\text{min}}= \text{h} (\Delta \theta)$, shown in the 4th row of Figure \ref{fig:subs3_mu_theta}. We have defined $\text{h}(\Delta \theta)$ function qualitatively, trying to discard as many random alignments sample pairs as possible and keeping the promising candidate pairs of the data sample.

With this, we are left with 8 pairs in the data sample and 2 pairs in its random alignments counterpart. This constitutes the nominal search of the {\it Subsample 3}, although we expand it in \S \ref{subsec:subs3_search_2}.

At this point we note that, ideally, the criteria used for this search would have been more stringent regarding the RV and metallicity consistency (e.g., 1$\sigma$ instead of 2$\sigma$). While that was our original search, by doing this (and using the same $\mu/\Delta \mu_{\text{min}}=\text{h}(\Delta \theta)$ criterion) we were left with only 2 pairs in the data sample and 1 pair in its random alignments counterpart. We considered this to be too small of a sample size, and we relaxed the $\Delta \text{RV}$ and $\Delta \text{[Fe/H]}$ criteria to a $2\sigma$ level.

As with the previous subsamples, we assign qualitative flags (``{\it a}'', ``{\it b}'', or ``{\it c}'' ) to our wide binary candidates. The qualitative flags, in this case, do not take into account parallaxes, and they are only based on proper motions, RV and metallicity values. Out of the 8 candidate pairs, we classify 2 of them as ``{\it a}'', 4 as ``{\it b}'', and 2 of them as ``{\it c}''. One of the  ``{\it b}''-flagged pairs was also obtained in {\it Subsample 1} (pair \No2 of Table \ref{tab:subs1_table}). The remaining pairs are reported in the first block of Table \ref{tab:subs3_table}.

Figure \ref{fig:subs3_RV_FeH} shows the comparison of the RV and metallicities of the two components of the data sample (left column, blue) and random alignments sample (right column, red) pairs with the 1:1 relation (black solid line). 

\subsection{Relaxing the $\sigma_{\text{RV}}$ and $\sigma_{\text{[Fe/H]}}$ Criteria}
\label{subsec:subs3_search_2}

In the previous subsamples, when trying to expand our candidate pairs lists, we have relaxed the consistency criteria ($\Delta$RV, $\Delta \varpi$). For this subsample, however, we have already used a 2$\sigma$-consistency level in order to define our candidate list. 

Now, to further look for potential candidates, we slightly relax the quality criterion used on the RVs and metallicities. We set $\sigma_{\text{RV}}=15$ km s$^{-1}$ and $\sigma_{\text{[Fe/H]}}=0.15$ dex. Using these values, and keeping all the other criteria constant, we run our search algorithm and find 4 new pairs in the data sample and 5 new pairs in the random alignments sample. Out of these 4 new pairs, we classify 1 of them as ``{\it a}'', 2 of them as ``{\it b}'', and 1 of them as ``{\it c}'', and report them in the second block of Table \ref{tab:subs3_table}. We notice that, by using these relaxed RV and metallicity quality criteria, the random alignments sample has increased more than the data sample, suggesting we are entering the contamination-dominated region. Accordingly, we terminate this binary search here. A summarized version of the selection criteria used in this subsample can be found in Table \ref{tab:summary_criteria_subsamples}.
\section{Binary Search in the {\it Subsample 4}: ``PM-only''}
\label{sec:search_subs4}

Finally, we proceed to search for wide binary candidates in the subsample of stars with only proper motions.  We design this search to be in a completely different data set as the previous ones. We do this by excluding all the pairs that have TGAS parallaxes and/or {\lamost} RVs and metallicities for both component stars. In this way, every candidate that we could find here, is, by construction, not found before.

We separate this search into two branches. {\it Branch A} uses {\it only} proper motions quality and consistency as criteria. {\it Branch B}, on the other hand, uses proper motions as well as photometric distances (and photometric projected physical separations). 
\subsection{Branch A: ``PM-only''}
\label{subsec:subs4_search_1}

\begin{figure*}
\begin{minipage}{\textwidth}
\centering
\subfloat{{\includegraphics[width=0.70\linewidth]{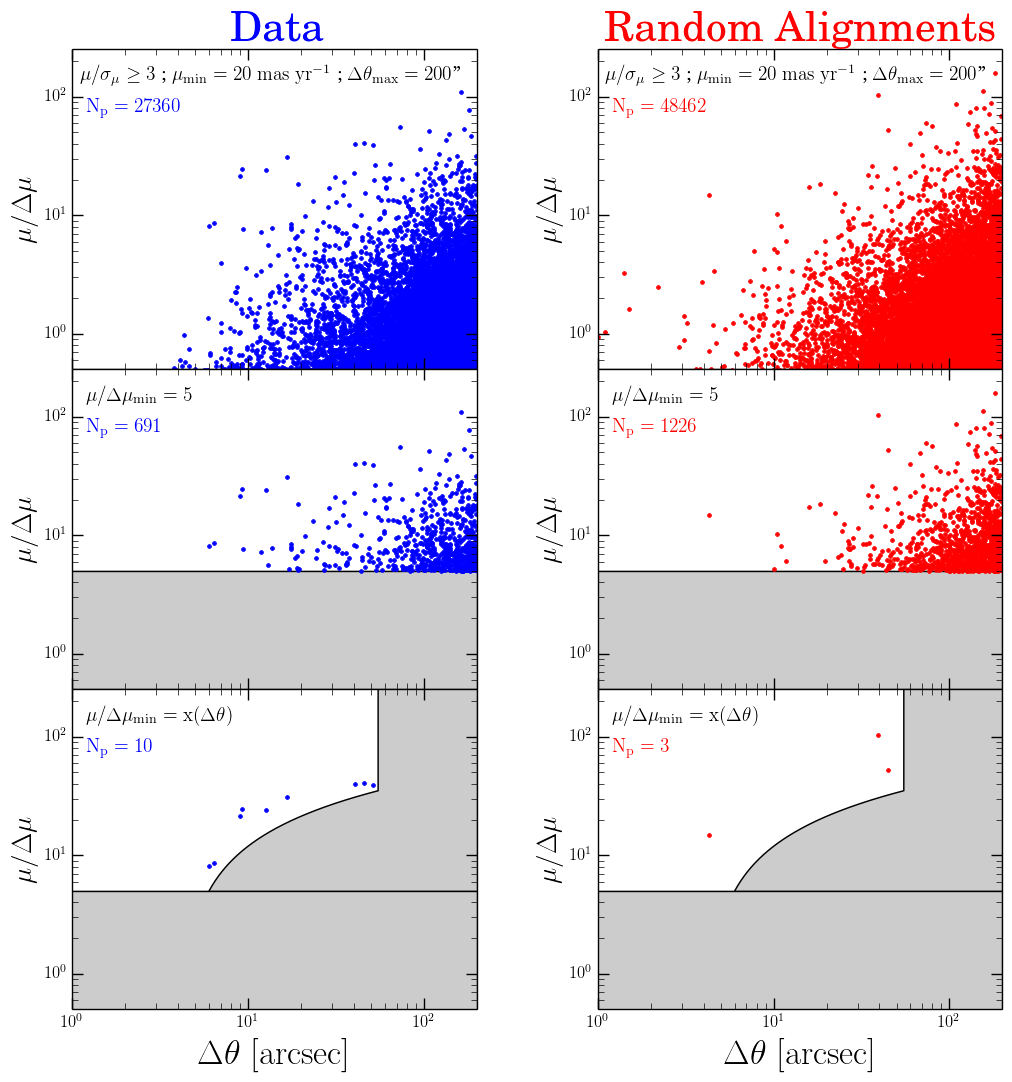}}}
\caption{{\it Subsample 4 - Branch A}: ``PM-only''. Analogous to Figure \ref{fig:subs1_mu_theta} for the subsample of stars with only proper motions (neither TGAS nor {\lamost} data required). In the 1st row we only include stars with {\it fast} and well measured proper motions ($\mu_{\text{min}}=20$ mas yr$^{-1}$; $\mu/\sigma_{\mu} \geq 3$), and search for pairs up to angular separations of $\Delta \theta_{\text{max}}=200 \arcsec$. In the 2nd row we add the $\mu/\Delta \mu_{\text{min}}=5$ common proper motion constraint, and in the 3rd row we add an angular separation dependence to it, $\mu/\Delta \mu_{\text{min}}=\text{x}(\Delta \theta)$.} 
\label{fig:subs4A_mu_theta}
\end{minipage}
\end{figure*}

In all the previous subsamples we have not required a minimum total proper motion criterion (i.e., $\mu_{\text{min}}=0$ mas yr$^{-1}$), as we were simultaneously applying several other constraints (parallaxes and/or RVs and metallicities; common proper motions). In this case, however, fewer constraints can be used, and we therefore expect chance alignments to have a greater impact in our search. To mitigate this, as their contribution increases rapidly with slowly moving stars, we set $\mu_{\text{min}}=20$ mas yr$^{-1}$. 

With this, our initial pool consists of 24,428 stars with {\it fast} and well measured proper motions ($\mu \geq 20$ mas yr$^{-1}$; $\mu/\sigma_{\mu}\geq 3$). For the angular separation search radius we set $\Delta \theta_{\text{max}}= 200 \arcsec$.  

For now, we do not require any common proper motion constraint ($\mu/\Delta \mu_{\text{min}}=0$). With these criteria we run our search algorithm, and find 27,360 pairs in data sample and 48,462 pairs in its random alignments counterpart. The $\mu/\Delta \mu$ versus $\Delta \theta$ phase space for this search is shown in the top row of Figure \ref{fig:subs4A_mu_theta}.

In this case, as many more stars are being used in comparison with the other subsamples, the $\mu/\Delta \mu$ versus $\Delta \theta$ phase space is more populated for both the data and random alignments samples. The distribution of points, however, is similar to the ones seen in previous subsamples, though in this case it naturally extends to smaller $\Delta \theta$ and higher $\mu/\Delta \mu$ values. The selection effects of the TGAS and/or {\lamost} stars are not imprinted in this search.

We now add the common proper motion constraint $\mu/\Delta \mu_{\text{min}}=5$, shown in the 2nd row of Figure \ref{fig:subs4A_mu_theta}. In this case, the $\mu/\Delta \mu_{\text{min}}$ value used is more stringent than in the previous subsamples, as fewer constraints are being used, and proper motions are our only mean to select candidates. 

At this point, although there is still mostly overlap between both samples in the phase space, some data sample pairs with close separations ($\Delta \theta \lesssim 20 \arcsec$) and high $\mu/\Delta \mu$ values do not have analogous counterparts in the random alignments sample. At the same time, for separations wider than $\Delta \theta \gtrsim 60 \arcsec$, the random alignments sample pairs have even higher $\mu/\Delta \mu$ values than the data sample pairs, showing how complicated it would be to look for promising candidates in that angular separation regime.

We use these remarks to define the angular separation dependence of the common proper motion criterion, $\mu/\Delta \mu_{\text{min}}= \text{x} (\Delta \theta)$, shown in the 3rd row of Figure \ref{fig:subs4A_mu_theta}. After applying this criterion we are left with 10 pairs in the data sample and 3 pairs in the random alignments sample. This constitutes the final search of the {\it Subsample 4 - Branch A}.

\begin{figure*}
\begin{minipage}{\textwidth}
\centering
\subfloat{{\includegraphics[width=0.65\linewidth]{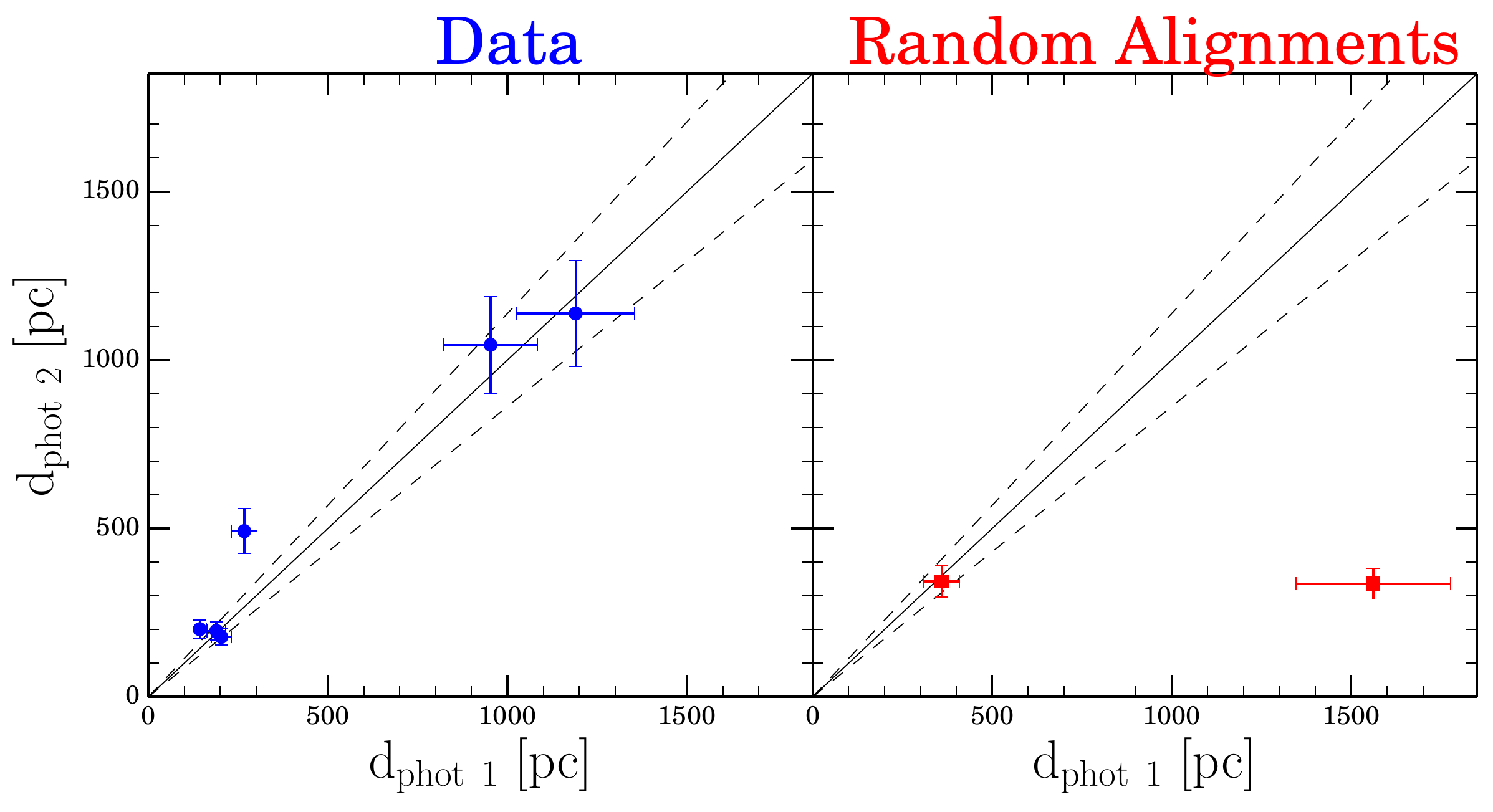}}}
\caption{{\it Subsample 4 - Branch A}: ``PM-only''. Comparison of the photometric distances of the two components of 6 data sample pairs (left column, blue) and 2 random alignments sample pairs (right column, red). The pairs shown are the subset of those obtained in the 3rd row of Figure \ref{fig:subs4A_mu_theta} that have photometric distance estimates for both component stars. The black solid line is the 1:1 relation, and the black dashed lines show the $\pm 1\sigma$ error ($\sim 14 \%$) of the photometric parallax relation used \citep{dhital10,dhital15}.} 
\label{fig:subs4A_dphot}
\end{minipage}
\end{figure*}

\begin{table*}
\begin{minipage}{\textwidth}
\centering
	\caption{List of our wide binary candidates in the {\it Subsample 4 - Branch A}: ``PM-only''. Analogous to Table \ref{tab:subs1_table}, for the subsample of stars with only proper motions (neither parallaxes nor RVs nor metallicities required). For some stars we are able to calculate photometric distance estimates d$_{\text{phot}}$, and we report them in the 7th column. We note, though, that photometric distances were only used for comparison purposes and not required as criteria in this search. Additionally, some of the stars have TGAS parallax values, and we report them as well in the 9th column. We note that the stars of the pair KIC 8909853/ KIC 8909876 (pair \No 8 in the table) have the exact same proper motion in UCAC4, therefore $\mu/\Delta \mu$ is undefined. While we initially thought this could be an artifact of UCAC4, {\gaia} DR2 has confirmed this pair is promising (see \S \ref{sec:gaiadr2_validation}).}
\renewcommand{\arraystretch}{1.0}
\begin{tabular}{cccccccccc}
\hline
Pair \No & KIC ID & $\mu_{\alpha}$ & $\mu_{\delta}$ & $\mu/\Delta \mu$ & $\Delta \theta$ & d$_{\text{phot}}$ & $\sigma_{\text{d}_{\text{phot}}}$ & $\varpi$ & $\sigma_{\varpi}$ \\
  - & - & \multicolumn{2}{c}{[mas yr$^{-1}$] }& - & [$\arcsec$] & \multicolumn{2}{c}{[pc]} &  \multicolumn{2}{c}{[mas]}  \\
\hline
\hline
\multicolumn{10}{c}{List of pairs obtained in \S \ref{subsec:subs4_search_1}} \\
\hline
1 & 11709006 & 16.2 & -39.8 & 8.5 & 6.4 & - & - & 15.25 & 0.44 \\ 
 & 11709022 & 11.4 & -41.3 &  &  & - & - & - & - \\ 
\hline
2 & 7090649 & 16.7 & 13.0 & 24.8 & 9.3 & 203 & 28 & - & - \\ 
 & 7090654 & 17.5 & 13.3 &  &  & 178 & 24 & 6.33 & 0.32 \\ 
\hline
3 & 7871438 & 22.3 & 45.2 & 31.0 & 16.7 & 143 & 19 & - & - \\ 
 & 7871442 & 23.9 & 44.9 &  &  & 201 & 27 & - & - \\ 
\hline
4 & 8674256 & -11.9 & -17.4 & 39.1 & 51.4 & 1190 & 164 & - & - \\ 
 & 8674286 & -11.7 & -17.9 &  &  & 1138 & 157 & - & - \\ 
\hline
5 & 8753617 & 11.0 & -38.0 & 21.3 & 9.0 & 953 & 131 & - & - \\ 
 & 8753619 & 9.9 & -39.5 &  &  & 1045 & 144 & - & - \\ 
\hline
6 & 9788210 & 80.5 & 99.8 & 24.2 & 12.7 & - & - & - & - \\ 
 & 9788227 & 85.0 & 97.0 &  &  & 159 & 21 & - & - \\ 
\hline
7 & 8184075 & 32.8 & -13.7 & 8.1 & 6.0 & 267 & 36 & - & - \\ 
 & 8184081 & 33.9 & -9.5 &  &  & 492 & 67 & - & - \\ 
\hline
8 & 8909853 & 132.0 & 231.0 & - & 17.7 & 189 & 26 & - & - \\ 
 & 8909876 & 132.0 & 231.0 &  &  & 196 & 27 & - & - \\ 
\hline
9 & 4946401 & 58.1 & -91.0 & 39.7 & 40.4 & 745 & 102 & 8.79 & 0.28 \\ 
 & 4946433 & 59.3 & -88.6 &  &  & - & - & - & - \\ 
\hline
10 & 6150118 & -11.1 & -24.2 & 40.6 & 45.5 & 410 & 56 & - & - \\ 
 & 6150124 & -10.7 & -23.7 &  &  & - & - & 2.91 & 0.57 \\ 
\hline
\end{tabular}
\label{tab:subs4_table_1}
\end{minipage}
\end{table*}

One of the 10 candidate pairs of the data sample, KIC 8909853/KIC 8909876, is not shown in the 3rd row of Figure \ref{fig:subs4A_mu_theta}. Its component stars have the exact same proper motion in our data, therefore $\mu / \Delta \mu$ is undefined. This pair has $\Delta \theta \simeq 18 \arcsec$ and $\mu \simeq 266$ mas yr$^{-1}$. While we initially thought this could be an artifact of UCAC4, we have investigated this pair in the Second Data Release (DR2) from {\gaia} \citep{gaia18a} and confirmed it is a promising candidate (see \S \ref{sec:gaiadr2_validation}).

7 out of the 10 data sample pairs have angular separations $\Delta \theta \leq 20 \arcsec$ versus only 1 of the random alignments sample pairs. Given this, we expect the contamination fraction of these {\it close}-separation candidates to be small. On the other hand, for the 3 candidate pairs with wider angular separations ($\Delta \theta \gtrsim 40 \arcsec$), we expect a higher contamination fraction as they overlap in phase space with the remaining 2 random alignments pairs.

\begin{figure*}
\begin{minipage}{\textwidth}
\centering
\subfloat{{\includegraphics[width=0.70\linewidth]{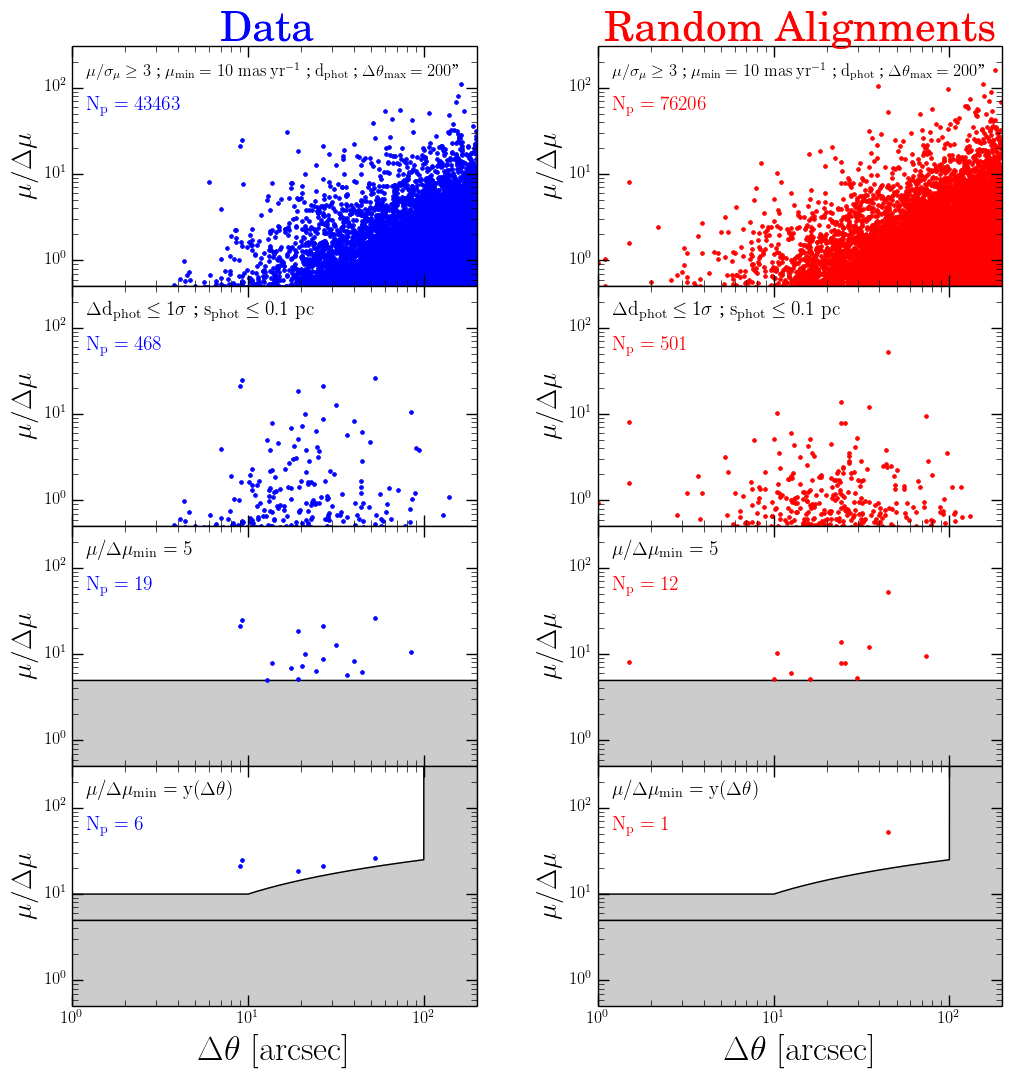}}}
\caption{{\it Subsample 4 - Branch B}: ``PM+photometric distance''. Analogous to Figure \ref{fig:subs1_mu_theta} for the subsample of stars with only proper motions {\it and} photometric distance estimates (neither TGAS nor {\lamost} data required). In the 1st row we only include stars that satisfy the criteria: $\mu/\sigma_{\mu} \geq 3$ and $\mu_{\text{min}}=10$ mas yr$^{-1}$, and we search for pairs up to angular separations of $\Delta \theta_{\text{max}}=200 \arcsec$. In the 2nd row we only keep the pairs whose stars have photometric distances consistent within 1$\sigma$ and with photometric projected physical separations s$_{\text{phot}}\leq 0.1$ pc. In the 3rd row we add the $\mu/\Delta \mu = 5$ common proper motion constraint, and in the 4th row we add an angular separation dependence to it, $\mu/\Delta \mu = \text{y}(\Delta \theta)$.} 
\label{fig:subs4B_mu_theta}
\end{minipage}
\end{figure*}

Although by construction these candidate pairs do not have neither parallaxes nor RVs nor metallicities for both component stars, we can compare their photometric distances, if available (they were not used as criteria in this search). Out of the 10 (3) data (random alignments) sample pairs, 6 (2) of them have photometric distance estimates for both component stars. We show the comparison of these estimates for the data pairs (left column, blue) and random alignments pairs (right column, red) in Figure \ref{fig:subs4A_dphot}.

Out of the 6 data pairs shown, only 1 of them is notably off the 1:1 relation while 5 of them lie close to it (4 have d$_{\text{phot}}$ consistent within $1\sigma$, 1 is within $1.5\sigma$). For the random alignments pairs, 1 of them is clearly off the 1:1 relation and 1 lies close to it, although both of these are the {\it wide} pairs ($\Delta \theta \gtrsim 40 \arcsec$) of the 3rd row of Figure \ref{fig:subs4A_mu_theta}. 

We report these 10 candidate pairs in Table \ref{tab:subs4_table_1}. For this subsample we do not assign qualitative flags as fewer criteria were used when looking for the candidate pairs. A summarized version of the selection criteria used in this subsample can be found in Table \ref{tab:summary_criteria_subsamples}.
\subsection{Branch B: ``PM+photometric distance''}
\label{subsec:subs4_search_2}

\begin{table*}
\begin{minipage}{\textwidth}
\centering
\caption{List of our wide binary candidates in the {\it Subsample 4 - Branch B} (``PM+photometric distance''). Analogous to Table \ref{tab:subs1_table}, for the subsample of stars with only proper motions and photometric distance estimates (neither parallaxes nor RVs nor metallicities required). For this search, in addition to proper motions,  we have used photometric distances and photometric projected physical separations as a criteria, and we report them in the 7th  and 9th columns, respectively. The 3 pairs found in \S \ref{subsec:subs4_search_2} that were also found in {\it Subsample 4 - Branch A} (pairs \No2, \No5, and \No8 of Table \ref{tab:subs4_table_1}) are not reported in this table. Additionally, one of the stars has a TGAS parallax value, and we report it in the 10th column.}
\renewcommand{\arraystretch}{1.0}
\begin{tabular}{ccccccccccc}
\hline
Pair \No & KIC ID & $\mu_{\alpha}$ & $\mu_{\delta}$ & $\mu/\Delta \mu$ & $\Delta \theta$ & d$_{\text{phot}}$ & $\sigma_{\text{d}_{\text{phot}}}$ & s$_{\text{phot}}$ & $\varpi$ & $\sigma_{\varpi}$\\
  - & - & \multicolumn{2}{c}{[mas yr$^{-1}$] }& - & [$\arcsec$] & \multicolumn{2}{c}{[pc]} & [pc] & \multicolumn{2}{c}{[mas]} \\
\hline
\hline
\multicolumn{11}{c}{List of pairs obtained in \S \ref{subsec:subs4_search_2}} \\
\hline
1 & 12507868 & 23.9 & 69.2 & 26.3 & 53.0 & 262 & 36 & 0.07 & 4.87 & 0.29 \\ 
 & 12507882 & 22.2 & 71.4 &  &  & 300 & 41 &  & - & - \\ 
\hline
2 & 12214492 & -6.2 & -37.3 & 18.3 & 19.3 & 635 & 87 & 0.06 & - & - \\ 
 & 12214504 & -7.8 & -38.6 &  &  & 686 & 94 &  & - & - \\ 
\hline
3 & 2709773 & 10.2 & -14.0 & 21.2 & 26.6 & 628 & 86 & 0.09 & - & - \\ 
 & 2709787 & 9.6 & -13.5 &  &  & 801 & 110 &  & - & - \\ 
\hline
\end{tabular}
\label{tab:subs4_table_2}
\end{minipage}
\end{table*}

As a final exercise, we look for wide binary candidates when using proper motions and supplementing them with photometric distances. Even though we know the absolute values of the photometric distance estimates might not be accurate (see Figure \ref{fig:histogram_distance_phot_trig}), we use them as a last resource to exploit all our available data. 

Since photometric distances are required in this subsample, we are introducing a bias against detecting pairs with subgiant/giant components (see \S \ref{subsubsec:distances_photometric}). We note, however, that this is the only search in which we have introduced such a bias.

From the search in the {\it Subsample 4 - Branch A}, we slightly modify the initial pool of stars.  We maintain the proper motion quality cut ($\mu/\sigma_{\mu}\geq$3), but since we are adding new constraints, we relax the total proper motion criterion to $\mu_{\text{min}}=10$ mas yr$^{-1}$. We also require the stars to have photometric distance estimates available. This leaves us with an initial pool 30,682 stars.

For now, we do not require any common proper motion constraint ($\mu/\Delta \mu_{\text{min}}=0$). With these constraints we set the angular separation limit to $\Delta \theta_{\text{max}}=200 \arcsec$ and run our search algorithm, finding 43,463 pairs in the data sample and 76,206 pairs in the random alignments sample. The $\mu/\Delta \mu$ versus $\Delta \theta$ phase space for this search is shown in the top row of Figure \ref{fig:subs4B_mu_theta}.

We now add the photometric distance consistency constraint $\Delta \text{d}_{\text{phot}}\leq 1 \sigma$, meaning we only keep the pairs whose stars have photometric distances consistent within their errorbars. Since distance estimates are available for all of these pairs, we can also calculate photometric physical separations using Equation \ref{eqn:physical_separation_calculation}. Given the uncertainties associated with this estimate, we choose the rather stringent criterion s$_{\text{phot}} \leq 0.1$ pc. 

The 2nd row of Figure \ref{fig:subs4B_mu_theta} shows the pairs left after applying these criteria. We maintain the common proper motion consistency criterion, $\mu / \Delta \mu_{\text{min}}=5$, and discard all pairs located in the shaded region in the 3rd row of Figure \ref{fig:subs4B_mu_theta}. 

At this point we are left with 19 pairs in the data sample and 12 pairs its random alignments counterpart. The distribution of points is similar for both samples, with most of the pairs being located in the regime $10 \arcsec \lesssim \Delta \theta \lesssim 50\arcsec$. Fewer points are found for wider angular separations as a consequence of the stringent projected physical separation criterion.

As in previous subsamples, the 3rd row of Figure \ref{fig:subs4B_mu_theta} shows a number of data sample pairs with high $\mu/\Delta \mu$ values ($\simeq20$) that have no counterpart in the random alignments sample. These pairs constitute promising candidates, and we add an angular dependence to the common proper motion criterion to better select them: $\mu /\Delta \mu_{\text{min}}=\text{y}(\Delta \theta)$. The result of this is shown in the 4th row of Figure \ref{fig:subs4B_mu_theta}.

With this, we are left with 6 pairs in the data sample and 1 pair in its random alignments counterpart. These pairs constitute the final search of the {\it Subsample 4 - Branch B}. We note, however, that 3 out of these 6 candidates were already found in the {\it Subsample 4 - Branch A} (one of them is KIC 8909853/KIC 8909876, with undefined $\mu/\Delta \mu$ and therefore not shown in the 4th row of Figure \ref{fig:subs4B_mu_theta}). 

We only report the 3 new pairs in Table \ref{tab:subs4_table_2}. These are the three widest angular separation pairs plotted in the 4th row of Figure \ref{fig:subs4B_mu_theta}. Similarly as in the previous branch, we do not assign qualitative flags to these pairs. A summarized version of the selection criteria used in this subsample can be found in Table \ref{tab:summary_criteria_subsamples}.

\section{Wide Binary Candidates List}
\label{sec:search_all_subs}

Up to this point, by combining the candidates pairs obtained in all the different subsamples, we have compiled with a list of 55 wide binary candidates. From this list, we have classified 22 of them with the qualitative flag ``{\it a}'', and 20 of them with the flag ``{\it b}''. 13 pairs were not assigned a qualitative flag because they were selected on the basis of proper motions alone, but nonetheless, they seem to be promising candidates as well.

Given that for each subsample we have employed different criteria in the selection process of our candidates, we report them in separate tables (see Tables \ref{tab:subs1_table} to \ref{tab:subs4_table_2}). A summarized version of the selection criteria used in all the subsamples can be found in Table \ref{tab:summary_criteria_subsamples}.

In the following sections we compare our candidates with other studies that have reported wide binaries in the {\kepler} field, validate our candidates using {\gaia} DR2, and look for age and rotation period information for the stars in our candidate list.

\section{Comparison with other works}
\label{sec:comparison_with_others}

Two other studies that have looked for wide binary candidates in the {\kepler} field have been published \citep{deacon16,janes17}. Here we compare our work with theirs and remark the differences in the binary candidates selection processes. Both works also use their candidates to study age-rotation relations, but we compare this part of their work with ours in \S \ref{sec:age_rotation}.

To investigate the differences, we examine their pairs in the light of our data. In other words, for identification we use the KIC IDs of their candidates, look for them in our base catalog, and assign them our proper motion, parallax, and RV data.
\subsection{Comparison with \citet{deacon16}}
\label{subsec:comparison_with_deacon16}

\citet{deacon16} studied the effects of wide binarity on the planet occurrence rate using the stars observed by {\kepler} and Pan-STARRS 1. They constructed their own proper motion catalog by combining archival data from public surveys, as well as UKIRT and Pan-STARRS 1, and found good overall agreement with the UCAC4 proper motions.

They looked for likely binary companions around {\kepler} stars, with potential companions being either another {\kepler} star, or a Pan-STARRS 1 star. Using the Pan-STARRS 1 photometry they fitted spectral energy distributions (SEDs), and derived absolute magnitudes, masses, and distances. Their selection method is based on the calculation of a probability of a given pair being real rather than a chance alignment, which depends on the comparison of masses, distances, and proper motions of the two potential components, as well as their angular separation. Additionally, they only work with stars that satisfy $\mu/\sigma_{\mu}\geq 5$ (in their data), and look for companions in the $6\arcsec \leq \Delta \theta \leq 300\arcsec$ regime.

A caveat of their SED fitting is that it assumes the stars are dwarfs, which propagates to potentially matching unrelated stars. Although they tried to exclude giants by means of color cuts, they warn that giant contamination may still be present. This is an important difference with our work, as we have not excluded pairs containing evolved components. We have only introduce a similar bias in the {\it Subsample 4 - Branch B}, where photometric distances were used.

Out of the 401 pairs reported, 128 of them are {\kepler}-{\kepler} and 273 of them are {\kepler}-Pan-STARRS 1. In our work, however, the candidate pairs are only formed by stars observed by {\kepler}. By matching the KIC IDs, we found that 102 out of their 128 {\kepler}-{\kepler} pairs could, in principle, exist in our catalog (the rest are absent because they are missing in UCAC4; see \S \ref{subsubsec:data_crossmatch}). After removing the pairs that are located within the exclusion areas of the clusters, or that have stars that do not satisfy the $\mu/\sigma_{\mu}\geq 3$ criterion (in our data), there are 50 pairs left. The 52 discarded pairs will naturally be absent in our candidate list, mainly because they do not pass our proper motion quality cut.

\begin{figure}
\centering
\subfloat{{\includegraphics[width=1.0\linewidth]{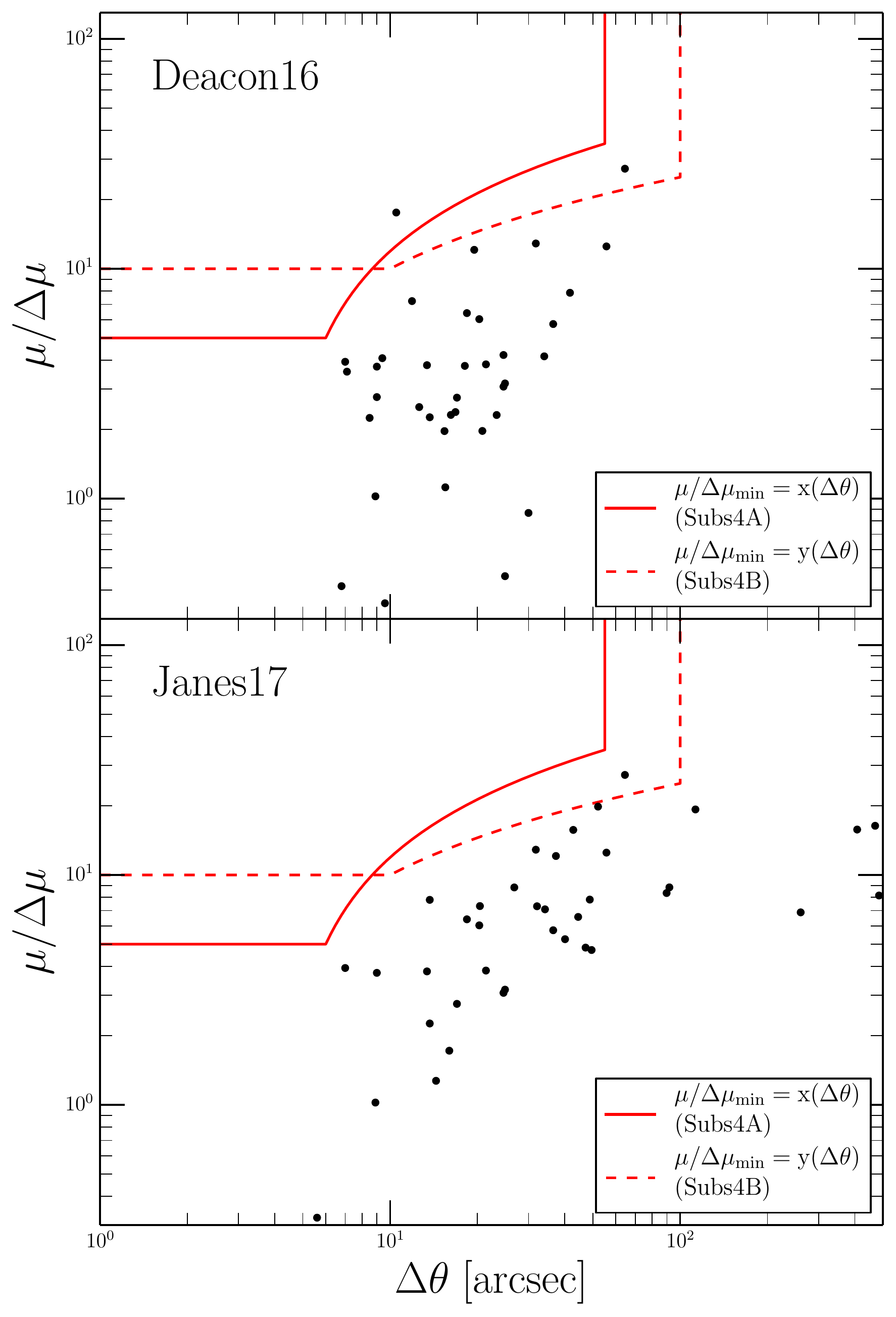}}}
\caption{Top Panel: $\mu/\Delta \mu$ versus $\Delta \theta$ for the 37 pairs reported by \citet{deacon16} {\it not in common} with our candidate list (see text). The $\mu/\Delta \mu_{\text{min}}$ criteria derived in {\it Subsamples 4-A} and {\it 4-B} are shown as the red solid and dashed lines, respectively. Bottom Panel: same as top panel, for the 38 pairs reported by \citet{janes17} {\it not in common} with our candidate list.}
\label{fig:comparison_deacon16_janes17}
\end{figure}

The list of the remaining 50 pairs was then crossmatched with our final candidate list. We found 13 pairs in common with \citet{deacon16}, while 37 of their pairs are not in common with our candidate list (and 42 of ours are missing in their list). We show the position of these 37 pairs in the $\mu/\Delta \mu$ versus $\Delta \theta$ diagram in the top panel of Figure \ref{fig:comparison_deacon16_janes17}.

Out of the 37 {\it not in common} pairs, 16 of them have $\mu/\Delta \mu < 3$, and therefore are naturally absent in our candidate list. According to our data, these pairs do not have similar-enough proper motion to be accepted by our search algorithm. 

Of the 21 remaining pairs, none of them had parallax measurements for both component stars, excluding them from been potentially found in the {\it Subsamples 1} or {\it 2}. 3 of the 21 pairs do have RV measurements for both component stars, meaning they could potentially have been found in the {\it Subsample 3}.

Further investigation of these 3 pairs revealed that 2 of them (KIC 8474682/KIC 8474690 and KIC 8316361/KIC 8316388) were rejected for having too large $\sigma_{\text{RV}}$ values ($>15$ km s$^{-1}$), and one of them (KIC 9050137/KIC 9050190) was rejected for having [Fe/H] values only consistent within 4$\sigma$. We note that these 3 {\it not in common} pairs do have fast and somewhat similar proper motions, but only one of them (KIC 8474682/KIC 8474690) would have been recovered in our search in the {\it Subsamples 4-A} (therefore, shown above the solid line in Figure \ref{fig:comparison_deacon16_janes17}).

For the 18 ($=21-3$) remaining pairs, while they could have been recovered in our searches in the {\it Subsamples 4-A} and/or {\it 4-B}, all of them lie below the derived $\mu/\Delta \mu_{\text{min}}$ lines (one exception, KIC 10557342/KIC 10622511, is located in between the solid and dashed lines, however this pair is rejected for not having a $\Delta \text{d}_{\text{phot}} \leq 1\sigma$ consistency). This explains why these {\it not in common} pairs are missing in our candidates, as they are located in the contamination-dominated region. 
\subsection{Comparison with \citet{janes17}}
\label{subsec:comparison_with_janes17}

\citet{janes17} identified 93 wide binary candidates in the {\kepler} field, derived rotation periods for them, and used them to explore existing age calibrations. They calculated their own proper motions by combining the values from a number of existing catalogs (e.g., UCAC4, the \citealt{deacon16} values, etc).

Their selection method is based on a probability of finding an unrelated star near each target star, which depends on the density of stars with similar positions and proper motions. Additionally, they also impose constraints on the maximum proper motion difference between the stars in pairs, only work with stars that satisfy $\mu/\sigma_{\mu} \geq 5$ (in their data), and look for companions in the $6\arcsec \leq \Delta \theta \leq 600\arcsec$ regime.

Additionally, they discard the pairs that do not follow the correlation seen in the $g$ versus $(g-K)$ diagram (see their Figure 1), expected for MS-MS pairs, therefore excluding pairs with evolved components. This is not our case, as we have not restricted ourselves to look only for MS-MS pairs.

While they do estimate the presence of random alignments in their candidates by a procedure (in principle) similar to ours, the implications of the random alignments samples not following the same angular separation distribution as the data (see \S \ref{subsubsec:random_alignments_angular_separation}) are not considered, and their distribution in a common-proper-motion phase space is not compared with their candidates.

To compare our results with those of \citet{janes17}, we follow the same procedure of \S \ref{subsec:comparison_with_deacon16}. By matching the KIC IDs, we found that 70 of the 93 pairs could exist in our catalog, but only 59 of them had $\Delta \theta <500\arcsec$ (our maximum search radius), were located outside the clusters exclusion area, and had $\mu/\sigma_{\mu} \geq 3$ (in our data).

When crossmatching these 59 pairs of \citet{janes17} with our candidate list, we found 21 pairs in common, while 38 of their pairs are not in common with our list (and 34 of ours are missing in their list). The bottom panel of Figure \ref{fig:comparison_deacon16_janes17} shows the positions of these 38 pairs in the $\mu/\Delta$ versus $\Delta \theta$ diagram. 32 of these 38 {\it not in common} pairs pass the $\mu/\Delta\mu \geq 3$ criterion, and require further examination. 

One of them, KIC 11873181/KIC 11923356, could potentially have been found in the {\it Subsample 1}. While the stars have fast ($\sim$ 50 mas yr$^{-1}$) and similar ($\mu/\Delta \mu \simeq 8$) proper motions, the pair is wide ($\Delta \theta \simeq 485 \arcsec$) and the parallax and RV values are not consistent ($\varpi_1=5.70 \pm 0.27$ and $\varpi_2=9.71 \pm 0.29$ mas; RV$_{1}=-57.3 \pm 17.5$ and RV$_{2}=-27.7\pm1.7$ km s$^{-1}$), justifying its rejection.

Two of them, KIC 11229052/KIC 11229131 and KIC 12218888/KIC 12317678, have parallaxes consistent within $3\sigma$ and could have been found in the {\it Subsample 2}, but fall below the derived $\mu/\Delta \mu_{\text{min}}$ criterion. Three other pairs could have been found in the {\it Subsample 3}, but they all have a component with $\sigma_{\text{RV}}>$15 km s$^{-1}$, making them too uncertain to be included in our search.

For the 26 (=32-6) remaining pairs, their $\mu/\Delta \mu$ values place them below the $\mu/\Delta \mu_{\min}$ lines defined in the {\it Subsamples 4-A} and {\it 4-B}, meaning they are located in the contamination-dominated region. The one exception (KIC 10557342/KIC 10622511, located between the solid and dashed lines), is in common with \citet{deacon16} and was already discussed in \S \ref{subsec:comparison_with_deacon16}.
\subsection{Overall Comparison}
\label{subsec:comparison_with_both}

Compared with the works of \citet{deacon16} and \citet{janes17}, our work is more complete in terms of discussing the underlying selection effects of the stars observed by {\kepler} (see \S \ref{subsubsec:catalog_angular_separation}), in assessing the behavior of the chance alignments in phase space (see \S \ref{sec:search_subs1} to \S \ref{sec:search_subs4}), and in justifying the selection criteria we have used.

It is important to clarify that we are not classifying the \citet{deacon16} and \citet{janes17} pairs not recovered by us as chance alignments. Instead, we argue that, based on our contamination analysis, and according to our base catalog data, these pairs occupy a region of phase space largely dominated by chance alignments. With more precise proper motion (and parallax and RV) measurements they could be confirmed as binary candidates, but the quality of the present data does not allow us to claim that.

We also note that some of the missing pairs of the \citet{deacon16} and \citet{janes17} candidate lists have fast ($\sim$ 40$\textendash$60 mas yr$^{-1}$) and similar ($\mu/\Delta \mu \simeq$ 5-15) proper motions. While these pairs lie below our $\mu/\Delta \mu_{\text{min}}$ criteria derived in the different subsamples, perhaps, had we included a higher $\mu_{\text{min}}$ criterion (particularly in the {\it Subsample 4-A}), a lower $\mu/\Delta \mu_{\text{min}}$ criteria may have been required to discard most of the random alignments pairs, leading us to potentially include these pairs in our list. We leave this further refinement (and extra dimension) of our searches as future work. 

Finally, to the best of our knowledge, and not exclusively to searches in the {\kepler} field, our work is the first to actually use RVs (and metallicities) as a selection criterion rather than just as a confirmation of potential wide binary candidates.

\section{Validation with {\gaia} DR2}
\label{sec:gaiadr2_validation}

\begin{figure*}
\begin{minipage}{\textwidth}
\centering
\subfloat{{\includegraphics[width=0.65\linewidth]{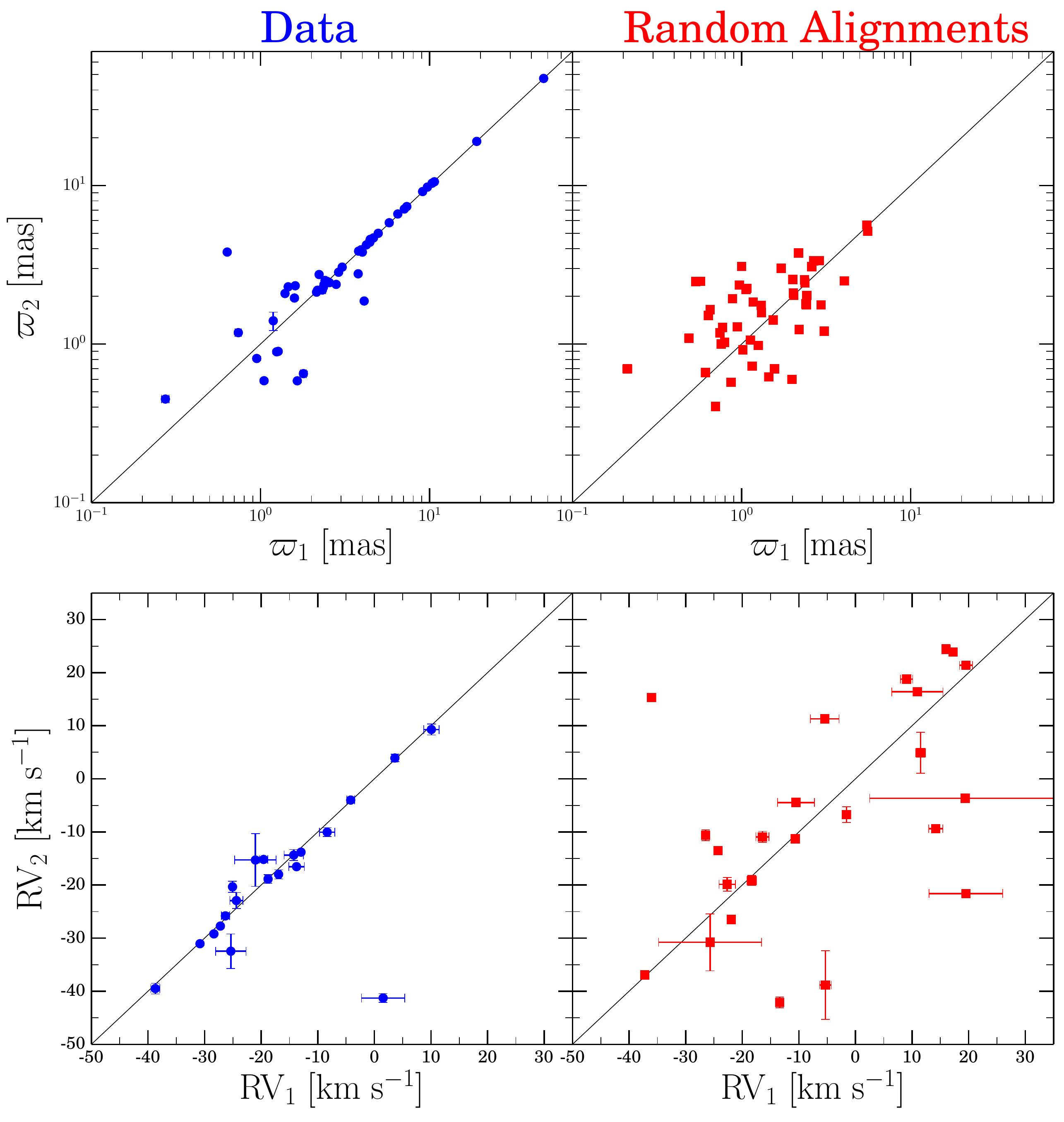}}}
\caption{Comparison of the {\gaia} DR2 parallaxes (top panel) and RVs (bottom panel) of the two components of each pair in the data sample (left column, blue) and random alignments sample (right column, red). The pairs shown are those found after crossmatching our final list of candidates and random alignments (compiled by combining the pairs obtained in all the different subsamples) with {\gaia} DR2. Only pairs with {\gaia} DR2 parallaxes or RVs for both components stars are shown. The black solid line in all panels shows the 1:1 relation. Errorbars are shown for all points, but they can be smaller than the markers.} 
\label{fig:GaiaDR2_candidates_randalign}
\end{minipage}
\end{figure*}

The recently released {\gaia} DR2 \citep{gaia18a} offers the possibility of validating our wide binary candidates in the light of improved astrometric information. Additionally, {\gaia} DR2 reports RV measurements for bright stars ($G <$ 13), which can also be used for validation purposes.

Using VizieR, we have crossmatched our list of pairs with {\gaia} DR2. Figure \ref{fig:GaiaDR2_candidates_randalign} shows the result of this crossmatch in parallax (top) and RV (bottom) space, for both our candidates (left column, blue) and their corresponding random alignments sample (right column, red). In this case, the random alignments sample is simply the sum of all the random alignments counterparts of the individual subsamples. We note that not all the pairs that have parallaxes have RVs measurements, as the availability of the latter in {\gaia} DR2 depends on the brightness of the targets.

The top panel Figure \ref{fig:GaiaDR2_candidates_randalign} shows that the components of most of our candidates with $\varpi \gtrsim 2$ mas have nearly identical parallaxes. For $\varpi \lesssim 2$ mas, while some pairs tend to follow the 1:1 line, there is clear scatter around it and the components of some pairs certainly have different parallaxes, revealing the presence of chance alignments in this regime. For the random alignments sample, it can be seen that they do not follow the 1:1 line, and are mostly concentrated at small parallaxes ($\varpi \sim 1 \textendash 2$). Similar results are shown in the bottom panel of Figure \ref{fig:GaiaDR2_candidates_randalign}, where most of our candidates follow the RV 1:1 line (with only a few outliers), while the random alignments pairs do not.

All of these confirms the reliability of our candidate selection method, as both data and random alignments sample behave as expected. Furthermore, this independently validates the promising status of our candidates, and we highlight that for $\varpi \gtrsim 2$ we expect little contamination in our sample. For $\varpi \lesssim 2$ we expect a higher contamination rate, which is consistent with our results from \S \ref{subsec:subs2_search_1}.

Additionally, we investigated the pair KIC 8909853/KIC 8909876 in the {\gaia} DR2 data. This pair was found in the {\it Subsample 4 - Branch A}, with both stars having the exact same proper motion in UCAC4 (therefore $\mu/\Delta \mu$ is undefined). While we considered the possibility of this being an artifact in UCAC4 (e.g., the same UCAC source matched to two different KIC sources) given the small angular separation of the pair ($\Delta \theta < 20 \arcsec$), according to {\gaia} DR2 these are actually two different stars. Moreover, both stars have similar parallaxes ($\varpi \sim 10$ mas) and proper motions ($\mu \sim 250$ mas yr$^{-1}$), which confirms it as a promising candidate.

\begin{figure*}
\begin{minipage}{\textwidth}
\centering
\subfloat{{\includegraphics[width=1.0\linewidth]{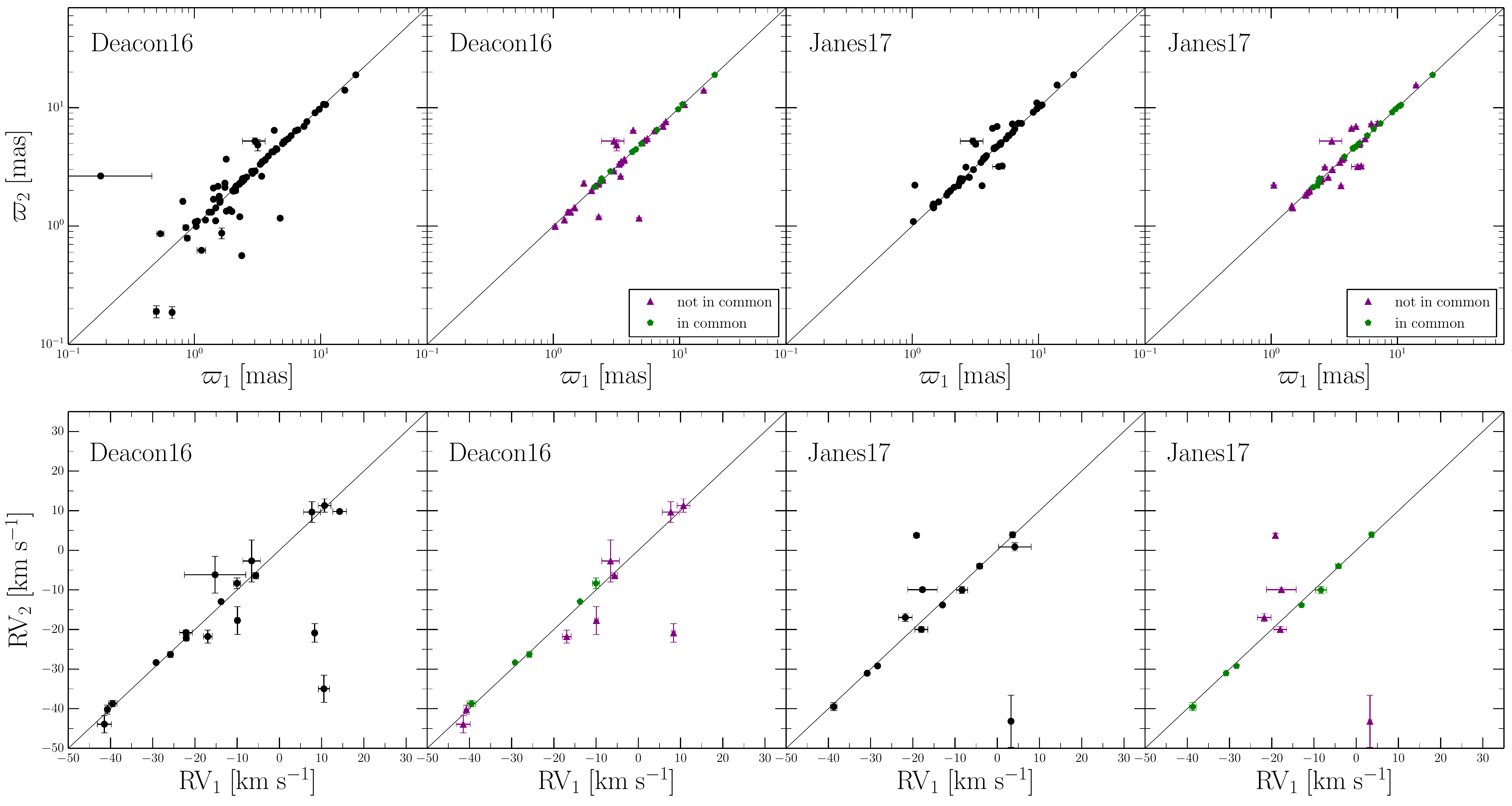}}}
\caption{Comparison of the {\gaia} DR2 parallaxes (top panel) and RVs (bottom panel) for the candidate pairs of \citet{deacon16} (left two columns) and \citet{janes17} (right two columns). The first and third columns show all of their pairs that we have found in {\gaia} DR2 (black points). In the second and fourth columns we have separated their pairs in those {\it in common}/{\it not in common} with us (green/purple points, respectively). For the {\it not in common} subset, we are only including those pairs that satisfy our proper motion quality cut (see \S \ref{sec:comparison_with_others}). Only pairs with {\gaia} DR2 parallaxes or RVs for both components stars are shown. The black solid line in all panels shows the 1:1 relation. Errorbars are shown for all points, but they can be smaller than the markers.} 
\label{fig:GaiaDR2_D16_J17}
\end{minipage}
\end{figure*}

We have also used the {\gaia} DR2 data to further explore our comparison with the works of \citet{deacon16} and \citet{janes17}. The results of crossmatching their candidate lists with {\gaia} DR2 are shown in Figure \ref{fig:GaiaDR2_D16_J17}. Similarly as in Figure \ref{fig:GaiaDR2_candidates_randalign}, the parallax and RV information are shown in the top and bottom panels, respectively. For this comparison we have separated their candidate lists in {\it all} of their pairs (first and third columns; black points), and those that {\it could exist} in our candidate list (second and fourth columns). This latter category consists of those pairs that satisfy $\mu/\sigma_{\mu} \geq 3$ (in our data) and $\Delta \theta \leq 500\arcsec$ (see \S \ref{sec:comparison_with_others}), and we separate it in those that are {\it in common} or {\it not in common} with us (green or purple points, respectively). The {\it not in common} subsets are the same pairs previously shown in Figure \ref{fig:comparison_deacon16_janes17}.

Regarding the parallax comparison, the first and third columns of Figure \ref{fig:GaiaDR2_D16_J17} ({\it all} pairs) show that both \citet{deacon16} and \citet{janes17} have, to first order, candidates with components of similar parallaxes. Nonetheless, both works show scatter around the 1:1 line for $\varpi \lesssim 5$ mas, and this is more pronounced in the \citet{deacon16} sample for $\varpi \lesssim 2$ mas. In the RV comparison both works tend to follow the 1:1 line too, although they have some pairs located far from it.

The second and fourth columns of Figure  \ref{fig:GaiaDR2_D16_J17} show that practically the entire {\it in common} subsets (green points) simultaneously follow the 1:1 in both parallax and RV. For the {\it not in common} subsets (purple points), however, we notice a different behavior. Virtually all the pairs that do not follow the parallax 1:1 line belong to the {\it not in common} subsets, which confirms that our selection method is properly rejecting pairs that do not seem to be consistent. The RV comparison strengthens this point, as many of the rejected pairs lie far from the 1:1 line (particularly in the \citet{janes17} sample).

Nonetheless, we note that there are {\it not in common} pairs that do lie along the parallax 1:1 line that our method is rejecting. We remind the reader that these pairs have been analyzed using our base catalog data, and that we do not classify them as chance alignments but we rather argue that we cannot reliably distinguish them from chance alignments (see \S \ref{subsec:comparison_with_both}).

Altogether the crossmatch of our candidate pairs with {\gaia} DR2 has given us a confirmation that our search method is selecting promising candidates, particularly for stars with $\varpi \gtrsim 2$ mas. Moreover, when comparing previous works with {\gaia} DR2, we seem to be recovering many of their promising candidates, and rejecting many pairs that lie off the parallax and/or RV 1:1 line. We note that the {\gaia} DR2 data has only been used for validation purposes and was not employed in our candidate selection.
\section{Age-Rotation Results}
\label{sec:age_rotation}

\subsection{Data Compilation}
\label{subsec:age_prot_compilation}

In order to study our pairs in the context of age-rotation relations, we have looked for information of our stars in existing catalogs of rotation period and asteroseismic ages for stars observed by {\kepler}.

For rotation periods we have crossmatched our base catalog with those of \citet{garcia14}, \citet{mcquillan14}, \citet{janes17}, \citet{reinhold13}, and \citet{nielsen13}. From the catalog of \citet{janes17} we have excluded the stars with autocorrelation functions classified as ``indeterminate'' or ``complex''. For cases where a star was found in more than one catalog, we have adopted the rotation period coming from the earliest cited work (e.g., if a star was found in both \citet{garcia14} and \citet{janes17}, we have adopted the \citet{garcia14} rotation period). The adopted prioritization reflects the robustness of the method used.

For ages we have crossmatched our base catalog with those of \citet{metcalfe14}, \citet{serenelli17}, and \citet{chaplin14} (we also crossmatched our candidate list with that of \citet{wu18} but did not find any matches). Again, for cases where a star was found in more than one catalog, we have adopted the age coming from the earliest cited work.

Additionally, 16 Cygni (KIC 12069424/KIC 12069449), a well known wide binary in the {\kepler} field that we have recovered in {\it Subsample 2}, has been the subject of dedicated asteroseismic studies \citep{davies15,metcalfe15}. This pair has age and rotation period estimates available, and we have included them in our crossmatch.

Since the literature on age-rotation relations uses the $(B-V)$ color as a proxy for mass, we needed to compile $B$ and $V$ magnitudes for our candidates. One source of this was already in our base catalog, as UCAC4 includes photometry from APASS \citep{henden14}. Additionally, we compiled $B$ and $V$ magnitudes from the MAST data archive\footnote{http://archive.stsci.edu/kepler/}, which come from the survey of \citet{everett12}. 

Following \S \ref{subsubsec:kic_photometry} we corrected the $B$ and $V$ magnitudes by extinction and calculated $(B-V)$ colors. A comparison between the APASS and \citet{everett12} photometry showed a relatively good agreement (standard deviation of $\sigma \simeq 0.1$ mag). For pairs with photometry from both catalogs, we prioritized the APASS values, as they are more widely available for our candidates.

After crossmatching our list of 55 candidates with the aforementioned catalogs, we are left with 19 pairs with rotation periods and $(B-V)$ color for both stars. In order to provide the most meaningful constraints on age-rotation relations, we focus on the subset of pairs that are validated by {\gaia} DR2. 15 of these 19 pairs have {\gaia} DR2 parallaxes (and radial velocities, when available) consistent within 3$\sigma$, and we use them in the subsequent analysis. Of the remaining 4 pairs, 2 of them were not found in {\gaia} DR2, and 2 other pairs have parallaxes inconsistent within 3$\sigma$. Additionally, only 6 stars were found to have age estimates (all of them belong to validated pairs). We show the results of this crossmatch in Table \ref{tab:info_prot_bv_age}.

\subsection{Binary Candidates and Age-Rotation Relations}
\label{subsec:gyro_results}

\begin{figure}
\centering
\subfloat{{\includegraphics[width=1.0\linewidth]{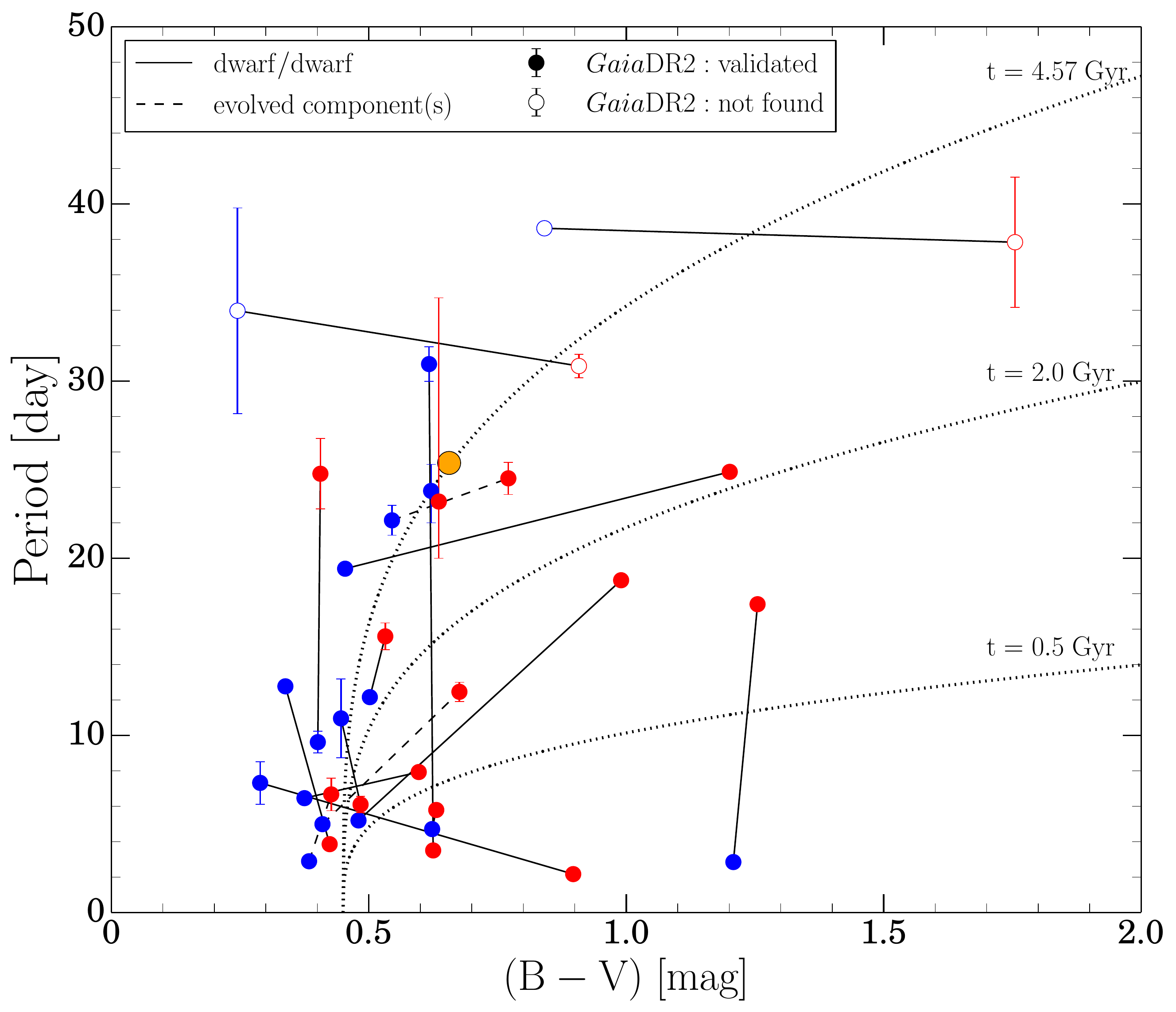}}}
\caption{Rotation period versus $(B-V)$ color for the candidate pairs for which both quantities are available for both component stars. The 15 pairs validated by {\gaia} DR2 are shown as filled circles. Additionally, the 2 pairs not found in {\gaia} DR2, but that seem promising according to UCAC4, are shown as open circles. In each pair the star with the bluer/redder color is shown as a blue/red circle. The components of the pairs are connected by a solid line if they are both classified as dwarfs ({\logg} $\geq$ 4), or by a dashed line if otherwise (see text). For reference we show the Sun as the orange circle, and the age-rotation relation of \citet{angus15} as dotted lines, evaluated at ages of 0.5 Gyr, 2 Gyr, and 4.57 Gyr.}
\label{fig:gyro_Prot_BV_GaiaDR2}
\end{figure}

Figure \ref{fig:gyro_Prot_BV_GaiaDR2} shows the 15 pairs validated by {\gaia} DR2 for which rotation period and $(B-V)$ color are available for both component stars (filled circles). Since the 2 pairs not found in {\gaia} DR2 seem to be promising candidates according to UCAC4, they are also included in the figure (open circles; these pairs are KIC 4043389/KIC 4142913 and KIC 4946401/KIC 4946433). Following the layout of Figure 11 of \citet{janes17}, the star with the bluer/redder color in the pair is plotted as a blue/red circle. The Sun is shown as the orange circle at $(B-V)=0.656$ (\citealt{gray92}; but see also \citealt{ramirez12}). We follow \citet{janes17} and references therein and plot it at a period of 25.38 days (age of 4.57 Gyr). For reference, the dotted lines show the age-rotation relation derived by \citet{angus15}, evaluated at ages of 0.5 Gyr, 2 Gyr, and 4.57 Gyr.

The components of pairs are connected by lines. If both components are classified as dwarfs ({\logg} $\geq$ 4) the connecting line is solid, and it is dashed if otherwise. We have made this distinction as stars that have evolved off the MS are not expected to follow the same age-rotation relations \citep{garcia14,angus15}. Further investigation of the 3 pairs with evolved component(s) of Figure \ref{fig:gyro_Prot_BV_GaiaDR2} revealed that for only one of them (the pair in which both stars have a period $< 10$ day) the {\logg} value of an evolved star is precisely measured from asteroseismic studies. For the other two pairs, their {\logg} values come from the KIC and \citet{huber14}, meaning they only come from photometric studies and making their classification not as reliable.

\begin{table*}
\begin{minipage}{\textwidth}
\centering
\caption{Crossmatch of our wide binary candidates with catalogs of $BV$ photometry, rotation periods, and asteroseismic ages. In the first block we report the 15 pairs validated by {\gaia} DR2 for which period and color information is available for both stars. Additionally, in the second block we report the 2 pairs not found in {\gaia} DR2 that seem to be promising according to UCAC4. The source catalog of the data reported is indicated in the corresponding column, with the values meaning: for color source=1 for \citet{henden14}, 2 for \citet{everett12}; for rotation periods source=1 for \citet{garcia14}, 2 for \citet{mcquillan14}, 3 for \citet{janes17}, 4 for \citet{reinhold13}, 5 for \citet{nielsen13}, 6 for \citet{davies15}; for asteroseismic ages source=1 for \citet{metcalfe14}, 2 for \citet{serenelli17}, 3 for \citet{chaplin14}, 4 for \citet{metcalfe12}. The $(B-V)$ color reported has been corrected by extinction (see \S \ref{subsec:age_prot_compilation}). For cases where the asteroseismic catalogs report systematic and statistical age uncertainties separately, they have been added in quadrature.}
\renewcommand{\arraystretch}{1.0}
\begin{tabular}{ccccccc}
\hline
KIC ID & $(B-V)$ & source$_{(B-V)}$ & Period & source$_{\text{Period}}$ & Age & source$_{\text{Age}}$  \\
\hline
- & [mag] & - & [day] & - & [Gyr] & -  \\
\hline
\hline
\multicolumn{7}{c}{Pairs with period and $(B-V)$ color information validated by {\gaia} DR2} \\
\hline
11069655 & 0.63 & 1 & 3.51$\pm$0.01 & 2 & - & - \\ 
11069662 & 0.62 & 1 & 30.96$\pm$0.97 & 3 & - & - \\ 
\hline
12156630 & 0.38 & 1 & 6.46$\pm$0.11 & 2 & - & - \\ 
12156742 & 0.60 & 1 & 7.93$\pm$0.04 & 2 & - & - \\ 
\hline
12507868 & 0.45 & 2 & 19.41$\pm$0.28 & 2 & - & - \\ 
12507882 & 1.20 & 2 & 24.88$\pm$0.20 & 2 & - & - \\ 
\hline
7871438 & 1.21 & 2 & 2.85$\pm$0.01 & 2 & - & - \\ 
7871442 & 1.25 & 2 & 17.41$\pm$0.03 & 2 & - & - \\ 
\hline
9139151 & 0.45 & 1 & 10.96$\pm$2.22 & 1 & $1.71 \pm 0.28$ & 1 \\ 
9139163 & 0.48 & 1 & 6.10$\pm$0.47 & 1 & $1.07 \pm 0.21$ & 1 \\ 
\hline
9944337 & 0.50 & 1 & 12.16$\pm$0.15 & 2 & - & - \\ 
9944356 & 0.53 & 1 & 15.59$\pm$0.76 & 3 & - & - \\ 
\hline
12069424 & 0.62 & 1 & $23.8^{+1.5}_{-1.8}$ & 6 & $7.0 \pm 0.3$ & 4 \\ 
12069449 & 0.64 & 1 & $23.2^{+11.5}_{-3.2}$ & 6 & $7.0 \pm 0.3$ & 4 \\ 
\hline
12366681 & 0.43 & 1 & 6.67$\pm$0.92 & 1 & $2.1^{+0.9}_{-0.6}$ & 2 \\ 
12366719 & 0.38 & 1 & 2.90$\pm$0.01 & 2 & - & - \\ 
\hline
7013635 & 0.68 & 1 & 12.46$\pm$0.54 & 3 & - & - \\ 
7013649 & 0.41 & 1 & 4.99$\pm$0.01 & 3 & - & - \\ 
\hline
8293539 & 0.48 & 1 & 5.20$\pm$0.11 & 3 & - & - \\ 
8293571 & 0.99 & 1 & 18.76$\pm$0.10 & 2 & - & - \\ 
\hline
8174654 & 0.62 & 1 & 4.71$\pm$0.12 & 2 & - & - \\ 
8242135 & 0.63 & 1 & 5.79$\pm$0.01 & 2 & - & - \\ 
\hline
8241071 & 0.42 & 1 & 3.86$\pm$0.09 & 3 & - & - \\ 
8241074 & 0.34 & 1 & 12.77$\pm$0.09 & 3 & - & - \\ 
\hline
2992956 & 0.55 & 1 & 22.14$\pm$0.84 & 3 & - & - \\ 
2992960 & 0.77 & 1 & 24.51$\pm$0.91 & 3 & - & - \\ 
\hline
4386086 & 0.41 & 1 & 24.77$\pm$1.99 & 3 & - & - \\ 
4484238 & 0.40 & 1 & 9.62$\pm$0.61 & 3 & - & - \\ 
\hline
6225718 & 0.29 & 1 & 7.32$\pm$1.20 & 3 & $2.6^{+0.9}_{-0.8}$ & 3 \\ 
6225816 & 0.90 & 1 & 2.17$\pm$0.00 & 3 & - & - \\ 
\hline
\multicolumn{7}{c}{Pairs with period and $(B-V)$ color information not found in {\gaia} DR2} \\
\hline
4043389 & 0.84 & 1 & 38.63$\pm$0.38 & 2 & - & - \\  
4142913 & 1.75 & 1 & 37.84$\pm$3.68 & 3 & - & - \\ 
\hline
4946401 & 0.24 & 1 & 33.97$\pm$5.81 & 2 & - & - \\   
4946433 & 0.91 & 1 & 30.85$\pm$0.67 & 3 & - & - \\ 
\hline
\end{tabular}
\label{tab:info_prot_bv_age}
\end{minipage}
\end{table*}

The location of most of our candidates in Figure \ref{fig:gyro_Prot_BV_GaiaDR2} is at odds with expectations from age-rotation relations for coeval stars. We think of the component stars of a binary as being coeval, and would therefore expect them to lie along the same period-color line, with the redder star having a longer period. The connecting lines of our pairs, however, show a wide variety of slopes.

Furthermore, if we only focus on the pairs confirmed by {\gaia} DR2 in Figure \ref{fig:gyro_Prot_BV_GaiaDR2}, in many cases the blue star in a pair has a longer rotation period than its redder, lower mass companion (5 out of 15 cases). When only considering dwarf/dwarf pairs, we find that in 7 out of 12 cases the redder component has a longer period than its bluer companion.

Although not shown, we have re-plotted Figure \ref{fig:gyro_Prot_BV_GaiaDR2} but only keeping the highest-quality pairs from our sample (i.e., those classified with flag$=$``{\it a}'). This exercise, however, did not preferentially select pairs that have slopes mostly aligned with the expectations, meaning that the situation depicted in Figure \ref{fig:gyro_Prot_BV_GaiaDR2} remains even when only considering the higher confidence pairs.

Somewhat similar results, nonetheless, were previously obtained by \citet{janes17}. Although their period-color plot is more populated than ours (see their Figure 11), and most of their pairs have the redder star with a longer period than the bluer star, they do obtain a wide variety of slopes as well. This result was also seen when they only considered pairs of stars with similar colors (see their Figure 14), where the rotation periods were expected to be nearly identical. One of their conclusions is that their binary candidates do not follow a simple period-color relation.

To first order, no simple period-color relation is shown by our binary candidates either. We do not venture to further speculate on the implications of our result, as we are limited by a small sample size, and leave a more thorough discussion of this as future work.

\citet{deacon16} also relate their binary candidates with gyrochronology studies, but their approach is different. They use the \citet{mamajek08} relations to derive gyro-ages for their stars with measured periods, and then use the similarities and discrepancies of these ages to validate their contamination rate. This, however, starts from the presumption that available age-rotation relations work well for the components of wide binaries.

If one assumes that current age-rotation relations must work for the components of wide binaries, then we need to discuss a number of factors that could be influencing our results of Figure \ref{fig:gyro_Prot_BV_GaiaDR2}. Although we have tried to minimize the presence of chance alignments in our sample, contamination could still be present. Additionally, as only a fraction of our candidates have multiple RV epochs, the presence of potentially triple/multiple systems (instead of just binaries) cannot be ruled out. If at the appropriate distance, these close companions could produce tidal interactions, affecting the rotation periods we observe. Lastly, perhaps the periods we use do not actually correspond to the true rotation periods, and effects like differential rotation or spot evolution are important for some stars.

\subsection{Interesting Pairs}
\label{subsec:gyro_interesting_pairs}

From our list of 55 candidates, we have found 6 stars with asteroseismic ages, all of them belonging to pairs validated by {\gaia} DR2. Two of them correspond to the members of the aforementioned pair 16 Cygni. This pair has been thoroughly discussed in the literature \citep{davies15,metcalfe12,metcalfe15}, and has already been included in recent age-rotation relation studies \citep{angus15,vansaders16}. Thus, we simply note that both component stars have the same asteroseismic age, in agreement with the expectations, besides almost identical colors and rotation periods, as expected for approximately equal mass stars.

Of the remaining 4 (=6-2) stars, two of them, KIC 9139151 and KIC 9139163, are actually in the same pair, so it is interesting to compare their age estimates. This pair was classified with flag$=$``{\it a}'' in the {\it Subsample 2}. \citet{metcalfe14} reports the ages of KIC 9139151 and KIC 9139163 to be 1.71$\pm$0.28 Gyr and 1.07$\pm$0.21 Gyr, respectively (where we have added the systematic and formal uncertainties in quadrature). While not strictly consistent within the errorbars, the ages of these stars are certainly similar. \citet{chaplin14} and \citet{serenelli17} have also derived ages for these two stars and obtained values consistent with each other given the uncertainties, but we prioritize the ages reported by \citet{metcalfe14} as they have fitted individual oscillation frequencies.

We note that this pair has already been identified as a binary star in the literature \citep{garcia14}. Therefore, the fact that the asteroseismic ages derived by \citet{metcalfe14} are similar but not strictly consistent, could be an indication of underestimated errorbars. We highlight the potential of using stars in wide binaries as tests of consistency in asteroseismic studies.

For two other pairs, we have an age estimate for one component star, and a rotation period for its companion. For KIC 12366681, \citet{serenelli17} reports an age of 2.1$^{+0.9}_{-0.6}$ Gyr (systematic and statistical uncertainties added in quadrature). Its companion, KIC 12366719, has a period of 2.9 day. Their {\logg} values classify both of them as evolved stars, although for KIC 12366719 this is only based on photometry (for KIC 12366681 it is based on asteroseismology).

For KIC 6225718, \citet{chaplin14} reports an age of 2.6$^{+0.9}_{-0.8}$ Gyr. Its companion, KIC 6225816, has a period of 2.2 day. Their {\logg} values, both coming from asteroseismology, classify them as a dwarf/dwarf pair.

These two pairs are cases where, according to the expectations, and using the idea of \citet{chaname12}, we can assign the age of one star to the entire system and use them to obtain new gyrochronology constraints. We leave this task as future work, but encourage the community to consider these pairs in future age-rotation studies.
\section{Conclusions} 
\label{sec:conclusions}

We have compiled a catalog of 55 wide binary candidates formed by stars observed by {\kepler}. We assembled our base catalog by compiling positions and proper motions (UCAC4) for the $\sim$ 200,000 stars observed by {\kepler}, and supplement it with parallaxes (TGAS), and RVs and metallicities ({\lamost}). 

We then mined our base catalog for binary candidates by means of matching the stars' positions, proper motions, parallaxes, RVs, and metallicities, depending on the availability of the data. In order to select promising candidates, we have characterized the behavior of random alignments in phase space using the empirical procedure of \citet{lepine07}.

To the best of our knowledge, this is the first work to have used (at least partially) RVs and metallicities as a search criteria in a binary search, rather than just using them for confirmation purposes.

Given the selection function of the targets observed by {\kepler}, as well as its pixel size, our base catalog is incomplete as a function of angular separation for $\Delta \theta \lesssim 20\arcsec$. We are therefore biased against the detection of close separation pairs, precisely the regime where most of the binaries would be expected to appear. A similar effect, though more severe, can be expected from the Transiting Exoplanet Survey Satellite mission (TESS; \citealt{ricker14})

Because of the presence of a variety of selection effects in the base catalog, we did not attempt any degree of completeness in our search for wide binaries. Instead, and based on our random alignments analysis, we have selected the pairs more likely to be real, gravitationally bound wide binaries, to the limits of the available data.

As an independent validation of our binary candidates, we have crossmatched our pairs with the recently released {\gaia} DR2. This exercise has confirmed the reliability of our search method, as well as reinforced our claim that little contamination is expected for pairs with $\varpi \gtrsim 2$ mas.

When crossmatching our list of pairs with catalogs of rotation periods, we observed that our candidates do not follow a simple period-color relation. Additionally, only a small number of our candidates have asteroseismic age estimates. Two of them are the components of the well known binary 16 Cygni, for which asteroseismic studies derive the same age. For the other pair with ages for both component stars, although not strictly consistent, they have similar estimates. For two other pairs, an age estimate is available for one component while its companion has a measured period, making them cases that can be used to obtain new gyrochronology constraints.

Our search method is adaptable to current and future data releases of astrometric (e.g., {\gaia} DR2 and DR3) and spectroscopic (e.g., {\lamost} DR4) surveys. Including these higher quality data will allow not only to verify the nature of existing candidates, but also to perform a more thorough, multi-dimensional search using stars that, until very recently, only had proper motion measurements available.
 
We nonetheless strengthen that proper motions (and parallaxes, when available) remain as the fundamental basis of our searches. As an example of a potential improvement, we have performed a quick crossmatch of our base catalog with the recently released {\gaia} DR2 catalog \citep{gaia18a}. While $\approx$ 27\% of the stars in our base catalog have $\mu/\sigma_{\mu}\geq 3$, this fraction could potentially rise to near 100\% for both $\mu/\sigma_{\mu}$ {\it and} $\varpi/\sigma_{\varpi}$, increasing the number of possible pairs by over an order of magnitude. We leave this exercise, as well as others including forthcoming {\gaia} and {\lamost} data releases, as future work.
\section*{Acknowledgments}
We thank the referee for his/her comments and suggestions, which have improved the clarity and consistency of our work. We thank Marc Pinsonneault and Andy Gould for useful comments and discussions. We thank Jeff Andrews for useful discussions and for helping us with the construction of the random alignment samples algorithm. We also thank Antonio Frasca for providing us with the catalog of {\lamost}-{\kepler} parameters.

JC acknowledges support from Proyecto FONDECYT Regular 1130373; BASAL PFB-06 Centro de Astronomía y Tecnologías Afines; and by the Chilean Ministry for the Economy, Development, and Tourism’s Programa Iniciativa Científica Milenio grant IC 120009, awarded to the Millennium Institute of Astrophysics.

This work has made use of data from the European Space Agency (ESA) mission {\it Gaia} (\url{https://www.cosmos.esa.int/gaia}), processed by the {\it Gaia} Data Processing and Analysis Consortium (DPAC, \url{https://www.cosmos.esa.int/web/gaia/dpac/consortium}). Funding for the DPAC has been provided by national institutions, in particular the institutions participating in the {\it Gaia} Multilateral Agreement.

Guoshoujing Telescope (the Large Sky Area Multi-Object Fiber Spectroscopic Telescope LAMOST) is a National Major Scientific Project built by the Chinese Academy of Sciences. Funding for the project has been provided by the National Development and Reform Commission. LAMOST is operated and managed by the National Astronomical Observatories, Chinese Academy of Sciences.

This research was made possible through the use of the AAVSO Photometric All-Sky Survey (APASS), funded by the Robert Martin Ayers Sciences Fund.

\bibliographystyle{mn2e}
\bibliography{bibliography}

\end{document}